\documentclass[aps,prd,10pt,twocolumn,superscriptaddress,floatfix,nofootinbib,amsmath,amssymb,altaffilletter,preprintnumbers]{revtex4-1}
\usepackage[colorlinks=true,pdfstartview=FitV,breaklinks=true]{hyperref}

\usepackage{bm}
\usepackage{graphicx}
\usepackage{multirow}
\usepackage{soul}
\usepackage{amsmath}
\usepackage{fontawesome}
\usepackage{mathrsfs}
\usepackage{amssymb}
\usepackage{yfonts}
\usepackage{makecell}
\usepackage{color}
\usepackage{xspace}
\usepackage{url}
\usepackage{verbatim}
\usepackage{mathtools}
\usepackage{upgreek}
\usepackage{units}
\usepackage{siunitx}
\usepackage{amstext}
\usepackage[sc]{mathpazo}
\usepackage{booktabs}
\usepackage{tabulary}
\usepackage{tabularx}
\usepackage{etoolbox}
\usepackage[version=4]{mhchem}
\usepackage[utf8]{inputenc}
\usepackage[dvipsnames,table]{xcolor}
\newcolumntype{Y}{>{\centering\arraybackslash}X}
\hypersetup{urlcolor=BlueViolet,
	    citecolor=Plum,
	    linkcolor=PineGreen}
\usepackage[official]{eurosym}
\definecolor{lightgray}{rgb}{0.9,0.9,0.9}	    
\definecolor{green}{rgb}{0,0.5,0}
\definecolor{red}{rgb}{1,0,0}
\definecolor{blue}{rgb}{0,0,0.5}

\long\def\symbolfootnote[#1]#2{\begingroup%
\def\thefootnote{\fnsymbol{footnote}}\footnotetext[#1]{#2}\footnotemark[#1]\endgroup}

\newcommand{\dbd}[2]{\ifmmode \frac{\textrm{d}#1}{\textrm{d}#2}\else $\textrm{d}#1/\textrm{d}#2$\fi}
\newcommand{\pbp}[2]{\ifmmode \frac{\partial#1}{\partial#2}\else $\partial#1/\partial#2$\fi}

\newcommand{\ra}[1]{\renewcommand{\arraystretch}{#1}}

\newcommand{\drm}{\mathrm{d}}

\DeclareMathAlphabet{\mathpzc}{OT1}{pzc}{m}{it}
 \newcommand{\eV}{\text{e\kern-0.15ex V}\xspace}

 \newcommand{\TeV}{\text{T\kern-0.1ex \eV}\xspace}

\DeclareMathAlphabet{\mathpzc}{OT1}{pzc}{m}{it}

\newcommand{\North}{\hat{\mathcal{N}}}
\newcommand{\West}{\hat{\mathcal{W}}}
\newcommand{\Zenith}{\hat{\mathcal{Z}}}
\newcommand{\bfhat}[1]{\hat{\mathbf{#1}}}
\newcommand{\lat}{\lambda_{\rm lab}}

\newcommand{\be}{\begin{equation}}
\newcommand{\ee}{\end{equation}}
\newcommand{\bea}{\begin{eqnarray}}
\newcommand{\eea}{\end{eqnarray}}

\newcommand{\excl}{\langle \cos^2{\theta}\rangle^{\rm excl.}_T}
\newcommand{\disc}{\langle \cos^2{\theta}\rangle^{\rm disc.}_T}

\begin{document}

\title{Dark photon limits: a handbook}

\author{Andrea Caputo}\email{andrea.caputo@uv.es}
\affiliation{School  of  Physics  and  Astronomy,  Tel-Aviv  University,  Tel-Aviv  69978,  Israel}
\affiliation{Department  of  Particle  Physics  and  Astrophysics,Weizmann  Institute  of  Science,  Rehovot  7610001,  Israel}%

\author{Alexander J. Millar}\email{alexander.millar@fysik.su.se}\affiliation{The Oskar Klein Centre, Department of Physics, Stockholm University, AlbaNova, SE-10691 Stockholm, Sweden}
\affiliation{Nordita, KTH Royal Institute of Technology and
Stockholm
  University, Roslagstullsbacken 23, 10691 Stockholm, Sweden}
  
  \author{Ciaran A. J. O'Hare}\email{ciaran.ohare@sydney.edu.au}
\affiliation{ARC Centre of Excellence for Dark Matter Particle Physics, Sydney, NSW, Australia}
\affiliation{School of Physics, Physics Road, The University of Sydney, NSW 2006 Camperdown, Australia}

\author{Edoardo Vitagliano}\email{edoardo@physics.ucla.edu}\affiliation{ Department  of  Physics  and  Astronomy,  University  of  California,  Los  Angeles,  California,  90095-1547,  USA}

\preprint{NORDITA-2021-036}

%\date{\today}
\smallskip
\begin{abstract}
The dark photon is a massive hypothetical particle that interacts with the Standard Model by kinetically mixing with the visible photon. For small values of the mixing parameter, dark photons can evade cosmological bounds to be a viable dark matter candidate. Due to the similarities with the electromagnetic signals generated by axions, several bounds on dark photon signals are simply reinterpretations of historical bounds set by axion haloscopes. However, the dark photon has a property that the axion does not: an intrinsic polarisation. Due to the rotation of the Earth, accurately accounting for this polarisation is nontrivial, highly experiment-dependent, and depends upon assumptions about the dark photon's production mechanism. We show that if one does account for the DP polarisation, and the rotation of the Earth, an experiment's discovery reach can be enhanced by over an order of magnitude. We detail the strategies that would need to be taken to properly optimise a dark photon search. These include judiciously choosing the location and orientation of the experiment, as well as strategically timing any repeated measurements. Experiments located at $\pm$35$^\circ$ or $\pm$55$^\circ$ latitude, making three observations at different times of the sidereal day, can achieve a sensitivity that is fully optimised and insensitive to the dark photon's polarisation state, and hence its production mechanism. We also point out that several well-known searches for axions employ techniques for testing signals that preclude their ability to set exclusion limits on dark photons, and hence should not be reinterpreted as such.  \smallskip \href{https://github.com/cajohare/DarkPhotonCookbook}{\large\faGithub}
\end{abstract}

\maketitle

\section{Introduction} 
\label{sec:intro}
The hypothesis that galactic dark matter (DM) halos are comprised of a cold population of bosons is accumulating substantial interest in both theoretical and experimental circles~\cite{Essig:2013lka,Battaglieri:2017aum,Agrawal:2021dbo}. Examples of these DM candidates include the pseudoscalars--- like the widely popular QCD axion~\cite{DiLuzio:2020wdo}, or its generalisation, the axion-like particle~\cite{Masso:1995tw, Masso:2002ip, Ringwald:2012hr, Ringwald:2012cu, Arvanitaki:2009fg, Cicoli:2012sz, Jaeckel:2010ni}---as well as light scalars~\cite{Taylor:1988nw,Hu:2000ke,Magana:2012ph,Lesgourgues:2002hk,Hui:2016ltb,Damour:2010rp,Damour:2010rm,Piazza:2010ye}, and vector particles~\cite{Holdom:1985ag,Okun:1982xi,Fayet:1980rr,Georgi:1983sy}. From an experimental standpoint, one of the primary appeals of these ideas is that many of them possess a coupling to electromagnetism that, while usually suppressed by a high energy scale, is generically nonzero. This permits a diverse array of laboratory experiments to directly detect them as galactic DM. Much of the experimental activity has been driven towards the detection of the axion. However many of the axion's experimental signatures are shared by another DM candidate---the dark photon (DP)\footnote{aka hidden photon, or paraphoton.}~\cite{Jaeckel:2010ni,Jaeckel:2013ija,Fabbrichesi:2020wbt}---which can be searched for with very similar techniques~\cite{Jaeckel:2007ch,Horns:2012jf,Suzuki:2015sza,Jaeckel:2015kea,Knirck:2018ojz,Brun:2019kak,Nguyen:2019xuh,Andrianavalomahefa:2020ucg,Tomita:2020usq,Godfrey:2021tvs,Baryakhtar:2018doz,Chaudhuri:2018rqn,Gelmini:2020kcu,Arias:2012az}.

The DP is the gauge boson of a new dark U(1) added to the Standard Model (SM) gauge group, under which the SM fields are uncharged. This makes the DP nearly unobservable, save for a small kinetic mixing with the visible photon that is left in the theory at low energies~\cite{Holdom:1985ag}. The kinetic mixing leads to photon-DP oscillations, reminiscent of neutrino oscillations, or the axion-photon mixing relied upon by DM axion detectors known as haloscopes (see e.g.~Ref.~\cite{Irastorza:2018dyq} for a review of experimental techniques). The primary practical difference between the two is that axion-photon conversion requires an applied magnetic field, whereas DP-photon mixing is an inherent feature of the model---making the latter in some scenarios less demanding to search for.

While DPs lack as compelling a theoretical backing as the QCD axion~\cite{Peccei:1977hh, Peccei:1977ur, Weinberg:1977ma, Wilczek:1977pj, Kim:2008hd}, they can, at the very least, function as a viable cold DM candidates~\cite{Nelson:2011sf,Arias:2012az}, and several production mechanisms have been proposed to generate a sufficient abundance of them in the early Universe~\cite{AlonsoAlvarez:2019cgw,Graham:2015rva,Kolb:2020fwh,Ema:2019yrd,Ahmed:2020fhc,Nakai:2020cfw,Agrawal:2018vin,Co:2018lka, Bastero-Gil:2018uel,Co:2021rhi,Long:2019lwl}. So as a minimal extension of the SM, the DP is therefore just as worthwhile a target for experimental investigation as, say, an axion-like particle.\footnote{Indeed, DPs can also play the role of a $Z^\prime$-like mediator to a dark sector, hence there is substantial interest in searching for DPs in accelerators~\cite{Hewett:2012ns,Beacham:2019nyx,Alemany:2019vsk,Graham:2021,Lanfranchi:2020crw} and via searches for the millicharged particles they may couple to~\cite{Prinz:1998ua,Ball:2016zrp,Magill:2018tbb,Kelly:2018brz,Harnik:2019zee,Ball:2020dnx}.} See Ref.~\cite{Fabbrichesi:2020wbt} for a recent review of DPs that extends to their role in contexts well outside of the scope of our work.

While many haloscopes designed to search for axions are also sensitive to DPs, there are few, if any, dedicated exclusion analyses made by their respective collaborations. The non-observation of axions of a particular mass is simply reinterpreted as a constraint on DPs of the same mass, without fully taking into account the subtle differences between the ways the two particles couple to experiments. Most notably, the DP possesses a polarisation, and therefore the direction of the $E$-field it generates will have nothing to do with the experiment. In contrast, the $E$-field generated by axions has a preferred direction provided by the applied magnetic field. Most experiments designed to detect DM-induced $E$ or $B$-fields are sensitive to the directions of those fields. The parallels and differences between the behaviours of axions and DPs in haloscopes were first pointed out in 2012 by Ref.~\cite{Arias:2012az}.\footnote{The haloscope limits from Ref.~\cite{Arias:2012az} were updated very recently in Ref.~\cite{Ghosh:2021ard}.} Since then, some subtly erroneous results of the recasting exercise detailed in that work have propagated through the literature.

However, even dedicated DP searches must reckon with the DP polarisation. There are several reasons why this is fraught with difficulty. Firstly, the DP polarisation distribution around the Earth will depends upon assumptions about its production mechanism. Secondly, experiments operate in a rest frame that rotates with respect to the DP field, meaning the preferred direction of detection will vary over the day if one assumes any degree of coherence to the underlying polarisation distribution. Since many experiments, especially those designed to detect axions, are only sensitive to $E$-fields aligning with one axis, this will make any DP interpretation of those experimental analyses strongly orientation- and time-dependent. Approaches taken to address these issues in past analyses have been overly simplistic at best, and plainly incorrect at worst.

A large number of new DP searches, and an even larger number of proposals, have appeared since 2012. It is therefore high time that the steps be laid out to perform a more careful treatment of DP polarisation. We will show that with some simple changes to the data-taking procedure (without changing the total measurement time, and therefore cost, of the experiment), an experiment can obtain vastly improved limits on DPs in the absence of detection, and would be in a much more robust position to study the signal if in the presence of one.

In what follows, we will derive a set of mathematical formulae to compute relevant quantities needed to account for the Earth's rotation with respect to the DP. These results take the form of conversion factors that quantify how much an exclusion limit is impacted by the range of possible angles between the DP polarisation and the axis or plane that the experiment is sensitive to. We will see that the dependencies of these conversion factors entail certain experimental configurations being less impacted by this uncertainty than others. The fully optimised scenario is one in which the experimental location, running time, and orientation all conspire to give a signal that is effectively insensitive to the DP polarisation. For the location, we will see that latitudes of $\lambda_{\rm lat} = \pm 35^\circ$ or $\pm 55^\circ$ are optimal for experiments that are sensitive along the North-South or the Zenith-Nadir axes respectively. The optimal measurement time is always one sidereal day, since in this case the experiment samples all possible DP polarisation angles available to it. However, we will explain that a continuous measurement lasting the entire day is not necessary, and a close-to-fully optimised discovery reach can be achieved if the experiment splits a short $\mathcal{O}({\rm min})$-long measurement into three measurements spaced evenly over the day. As mentioned above, we aim for all of our recommendations to be incorporated relatively unintrusively via simple scheduling changes. So we emphasise that we do not require these measurements to be conducted on the same date---three measurements at different times of the sidereal day, spaced months apart, would achieve the same result.

To begin, in Sec.~\ref{sec:fig1explanation}, we set the stage by giving a brief overview of existing constraints on dark photons, all of which are displayed in Fig.~\ref{fig:bounds}. Then in Sec.~\ref{sec:DPcosmology} we discuss ways in which DPs can be produced in the early Universe with sufficient abundance to explain the DM, and in the process what these production mechanisms imply about the present day DP polarisation distribution around the Earth. Then in Sec.~\ref{sec:DPelectrodynamics} we walk through the electrodynamics associated with detecting DPs, and in Sec.~\ref{sec:DPdetection} we discuss various experimental techniques. In Sec.~\ref{sec:DPscanning}
 we discuss the direction and time dependence inherent in DP detection, and how this dictates the optimum scanning strategy. Finally, in Sec.~\ref{sec:postdetection} we determine how an experiment could use the daily modulation signal to \emph{measure} the DP polarisation following a positive detection. We summarise and conclude in Sec.~\ref{sec:summary}. All the figures created for this paper can be reproduced using the code available at~\url{https://github.com/cajohare/DarkPhotonCookbook} with the exception of the DP constraint plots and their associated data which are available at~\url{https://cajohare.github.io/AxionLimits}.
 
\section{Existing constraints on the dark photon}\label{sec:fig1explanation}
\begin{figure*}[t]
\centering
% 	%trim option's parameter order: left bottom right top
  \includegraphics[width=0.98\textwidth]{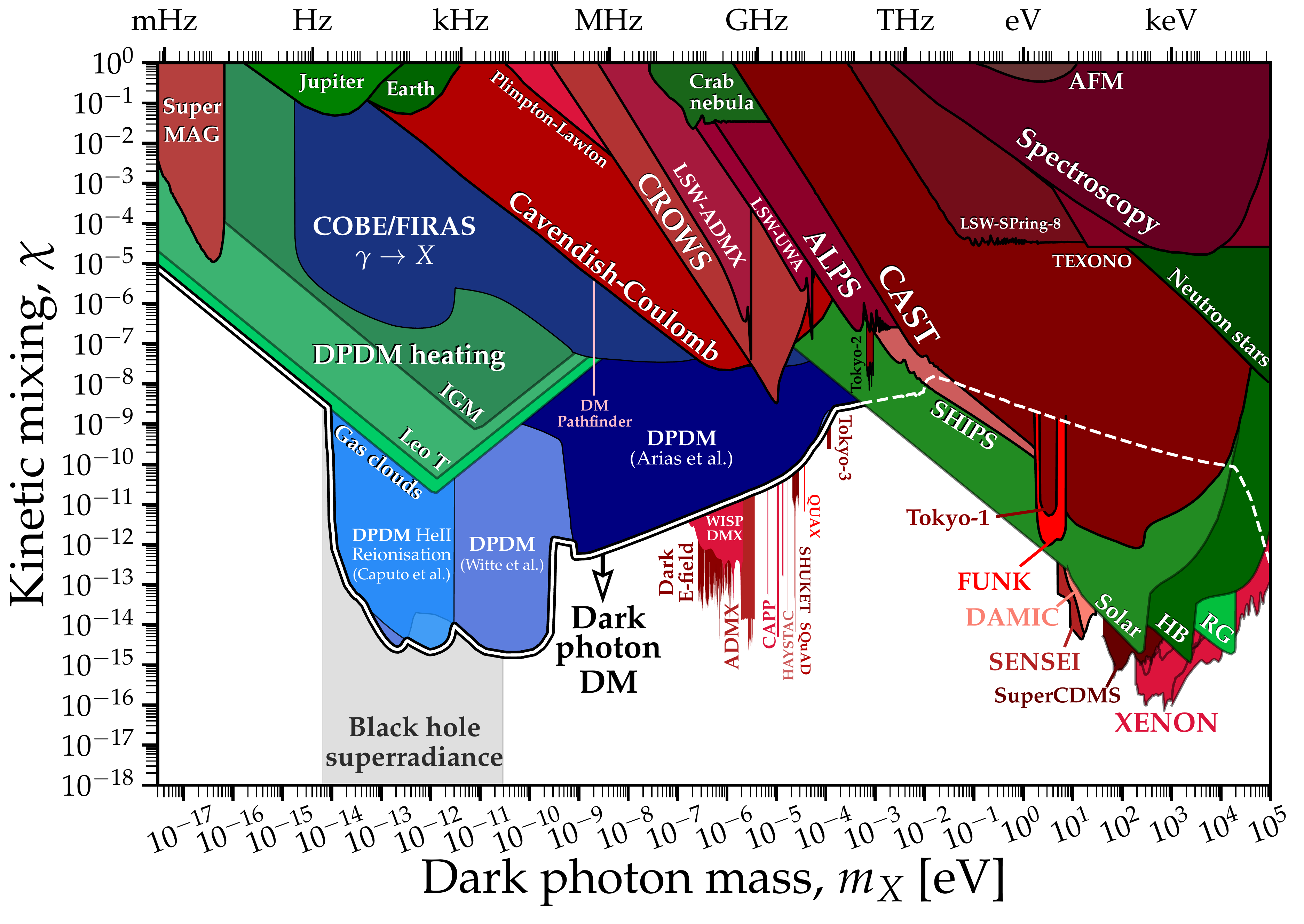}
\caption{Current constraints on the DP's mass, $m_X$, and kinetic mixing parameter with the SM photon, $\chi$. The general colour-scheme is: cosmological bounds in blue, experimental bounds in red, and astrophysical bounds in green. The thick white line that divides the parameter space in two is the upper limit for which DPs are a viable candidate for 100\% of the DM. The focus of this work are the experimental bounds that reach below this line. Descriptions of each bound are given in Sec.~\ref{sec:fig1explanation}.} 	\label{fig:bounds}
\end{figure*}
In Fig.~\ref{fig:bounds} we show the current landscape of bounds on DPs lighter than 0.1~MeV.\footnote{Accelerator bounds on heavier DPs have been purposefully ignored because the focus here is on DPs which can constitute DM in and of themselves.} We have chosen to show any bound set by a physical laboratory experiment in red, those set using astrophysical data in green, and those set using cosmological data in blue. We now briefly run through the sources of each bound.

Many model-independent bounds on the existence of the DP in Nature have been obtained through tests of the Coulomb $1/r^2$ force law, or, equivalently, via bounds on the photon mass~\cite{Goldhaber:2008xy}. The ones we have shown here are from Cavendish-like experiments~\cite{Williams:1971ms,Bartlett:1988yy,Tu:2005ge,Kroff:2020zhp}, Plimpton \& Lawton's experiment~\cite{Plimpton:1936ont,Kroff:2020zhp}, atomic spectroscopy~\cite{Jaeckel:2010xx}, atomic force microscopy (AFM)~\cite{Kroff:2020zhp}, and, at the lightest masses displayed here, from the static magnetic fields of the Earth~\cite{Goldhaber:1971mr} and Jupiter~\cite{Davis:1975mn}. Similarly, there are purely laboratory bounds on DPs set using light-shining-through-walls (LSW) experiments, e.g.~those run at ALPs~\cite{Ehret:2010mh}, SPring-8~\cite{Inada:2013tx} and UWA~\cite{Povey:2010hs,Parker:2013fxa}, as well as the microwave LSW experiments performed by ADMX~\cite{Wagner:2010mi} and CROWS~\cite{Betz:2013dza}. CAST~\cite{Redondo:2008aa} and SHIP~\cite{Schwarz:2015lqa} are both helioscopes, setting bounds on DPs emitted by the Sun. Finally, TEXONO~\cite{Soma:2014zgm} is a reactor neutrino experiment, for which a low mass DP limit was derived in Ref.~\cite{Danilov:2018bks}.

Dedicated direct detection bounds on the DP, specifically as a DM candidate, are set by the following experiments: DAMIC~\cite{Aguilar-Arevalo:2019wdi}, Dark E-field Radio~\cite{Godfrey:2021tvs}, DM Pathfinder~\cite{Phipps:2019cqy}, FUNK~\cite{Andrianavalomahefa:2020ucg}, SENSEI~\cite{Barak:2020fql}, SHUKET~\cite{Brun:2019kak}, SuperCDMS~\cite{Aralis:2019nfa}, SQuAD~\cite{Dixit:2020ymh}, three Tokyo dish antennae experiments~\cite{Suzuki:2015sza,Knirck:2018ojz,Tomita:2020usq}, WISPDMX~\cite{Nguyen:2019xuh}, and XENON1T/XENON100~\cite{Bloch:2016sjj,Aprile:2019xxb,Aprile:2020tmw, Bloch:2020uzh, Alonso-Alvarez:2020cdv,An:2020bxd}. Several other underground DM detectors sensitive to keV-mass DPs have also set limits~\cite{Abgrall:2016tnn,Armengaud:2018cuy,She:2019skm,Sato:2020ebe,Lasenby:2020goo,GERDA:2020emj} that are less sensitive than XENON's---we have neglected these to reduce clutter.

One of the focuses of this work is on reinterpreting haloscope limits on axions in the context of DPs. Those shown are ADMX~\cite{Asztalos:2001jk,Asztalos:2009yp,Du:2018uak,Boutan:2018uoc,Braine:2019fqb}, HAYSTAC~\cite{Zhong:2018rsr,Backes:2020ajv}, CAPP~\cite{Lee:2020cfj}, and QUAX~\cite{Alesini:2020vny}. Results from several well-known axion haloscopes~\cite{DePanfilis:1987dk,Hagmann:1990tj,McAllister:2017lkb,Ouellet:2018beu,Gramolin:2020ict} are not shown because they used their $B$-field to test for potential (axion) signals. In other words, a DP could have been observed, but its signal would have been vetoed. 

The upper limit of viable dark photon dark matter (DPDM), shown by a thick white line, is taken from various references. Although we run the risk of being overly-stringent, we adopt the most democratic approach of taking the lower envelope of all published analyses, including: Arias et al.~\cite{Arias:2012az}, Witte et al.~\cite{McDermott:2019lch,Witte:2020rvb}, and Caputo et al.~\cite{Caputo:2020rnx,Caputo:2020bdy}, though we note that there are some substantive disagreements between these analyses. Three astrophysical limits also require DPDM: those based on the heating of the intergalactic medium (IGM)~\cite{Dubovsky:2015cca}, the gas in the Leo T dwarf~\cite{Wadekar:2019mpc}, and the gas cloud at the galactic centre G357.8-4.7-55~\cite{Bhoonah:2019eyo}, and again, there are also disagreements between these analyses. Also at these lightest masses, a recent experimental bound was set on DPDM from an analysis~\cite{Fedderke:2021rrm,Fedderke:2021aqo} of SuperMAG data---a global network of magnetometers studying the geomagnetic field.

The astrophysical bounds at higher masses are those based on stellar cooling arguments applied to the Sun, horizontal branch (HB) stars, and red giant (RG) stars in Ref.~\cite{Redondo:2013lna}, and neutron stars in Ref.~\cite{Hong:2020bxo}. Note that for the straight part of the solar bound below 10 eV we use the improved limit from the solar global fit performed in Ref.~\cite{Vinyoles:2015aba}. These bounds assume a non-dynamical generation of the DP mass: the Stueckelberg case. However, if the DP mass originated via a Higgs mechanism, the stellar bounds would be much stronger---plateauing at $\chi\sim 10^{-13}$ for $m_X~\lesssim~100$~eV, down to arbitrarily small masses~\cite{An:2013yua,An:2020bxd}.

Another astrophysical bound was set using gamma rays from the Crab nebula~\cite{Zechlin:2008tj}. The final cosmological bound is on $\gamma\rightarrow X$ happening in the early universe to the degree that it would generate spectral distortions to the CMB, which are tightly constrained by COBE and FIRAS~\cite{Fixsen:1996nj}. Several groups have derived these constraints in the past~\cite{Mirizzi:2009iz,Caputo:2020rnx,Caputo:2020bdy,Garcia:2020qrp}, with broad, but not perfect, agreement. The one shown in Fig.~\ref{fig:bounds} is from Ref.~\cite{Caputo:2020bdy}. Lastly, we shade in grey the mass window $6.5\times 10^{-15}\,{\rm eV}<m_X<2.9\times10^{-11}\, {\rm eV}$. If a DP existed in that range, the field would spin down stellar mass black holes due to superradiance~\cite{Stott:2020gjj,Ghosh:2021zuf,Cardoso:2018tly}. 

Data for every bound shown in this figure can be downloaded individually at \url{https://cajohare.github.io/AxionLimits/docs/dp.html}.

\section{Dark photon cosmology}\label{sec:DPcosmology}
Dark photons are a compelling candidate for new physics. As they are simply the gauge boson of an additional U(1), they represent a very minimal extension to the SM. Recently, several novel production mechanisms for DPDM have been proposed, igniting interest even further. In this section we summarise some of these mechanisms, focusing on the degree of polarisation they leave the relic DPDM with. This point is often overlooked in the literature but is extremely relevant for their subsequent detection. We also stress that, contrary to previous claims found in the literature, the DP polarisation is different from the isotropy of the stress-energy tensor associated to the field, and depends on the production mechanism.

Arguably one of the simplest ways to produce DPDM is the misalignment mechanism, which is the most popular mechanism used to generate axions~\cite{Preskill:1982cy,Abbott:1982af,Dine:1982ah}. However, unlike axions, a minimal coupling to gravity does not lead to the correct relic abundance---a nonminimal coupling to the Ricci scalar needs to be invoked~\cite{Arias:2012az,Graham:2015rva, AlonsoAlvarez:2019cgw}. This typically comes at the cost of introducing instabilities in the longitudinal DP mode~\cite{Himmetoglu:2008zp,Himmetoglu:2009qi, Karciauskas:2010as}. Therefore, some extra work is required to make the theory consistent (though finding such a UV theory is not the goal of this work). For our purposes, the most notable consequence of the misalignment mechanism, however it may be constructed, is that it naturally leads to relic DPDM with a fixed polarisation within the cosmological horizon. We refer to this scenario later on as the \emph{fixed polarisation scenario}, and it will have the most dramatic consequences for direct detection. A scenario similar to the misalignment mechanism consists of the DPDM production via quantum fluctuations during inflation~\cite{Graham:2015rva} (see also Refs.~\cite{Kolb:2020fwh,Ema:2019yrd,Ahmed:2020fhc,Nakai:2020cfw}). In contrast to scalars and tensors, the vector is produced with a power spectrum peaked at intermediate wavelengths, evading bounds from long-wavelength isocurvature perturbations. Furthermore, this mechanism does not require a non-minimal coupling to gravity.

Another scenario is one based on tachyonic instabilities\footnote{A similar mechanism, based on a different instability, is through parametric resonance. See e.g.~Ref.~\cite{Dror:2018pdh}.} which arise when the DP couples to a misaligned axion~\cite{Agrawal:2018vin,Co:2018lka, Bastero-Gil:2018uel}. The energy density is initially stored in the axion field, and then the axion's zero mode transfers to both transverse and longitudinal components of the DP. Any production mechanism involving tachyonic instabilities dominantly produces a specific DP helicity, and we would expect that the final relic will also carry the same helicity---although later scatterings can deplete the degree of polarisation~\cite{Ratzinger:2020oct}. However, there exist other scenarios~\cite{Co:2021rhi} where the degree of polarisation may be even more pronounced and likely surviving scatterings. Lattice simulations of these models would be very relevant to the experimental campaign. 

DPDM could also be produced from the decay of topological defects, such as a network of near-global, Abelian-Higgs cosmic strings~\cite{Long:2019lwl}. In this scenario, the transversely polarised DP interaction is suppressed and the radiation is dominated by the emission of longitudinally polarised DPs. These modes would then come to constitute the DM. The evolution of the network is complicated, and consists of both short loops and infinite strings. An educated guess would be that long strings, with lengths of the order of the Hubble horizon, may lead to some degree of polarisation in the DP field by identifying a preferred direction. On the other hand, DPs would be produced also from the collapse of smaller closed loops, so eventually the polarisation alignment may be washed out. We refer to the case where the DP has no single polarisation as the \emph{random polarisation scenario}. 

To summarise, there are many interesting DP production mechanisms that can successfully produce the correct abundance of DM, several of them leading to some level of coherence in the polarisation distribution of the relic DP field, possibly over the entire Universe. Unfortunately, a more precise statement than this is not possible at the moment. In fact, the situation becomes even less clear when considering structure formation. It is not readily apparent what effect the formation of DM halos would have, if any, on the distribution of DP polarisations on the mpc-scales probed by an experimental campaign. Although dedicated simulations will be needed to resolve this issue, we can at least try to appreciate what impact gravity will have on the DP polarisation with a simple back-of-the-envelope calculation. 

Consider a particle with four-velocity $u^\alpha$ and polarisation $S^\alpha$; we know that $u_\alpha S^{\alpha} = 0$ should hold in any frame. One can thus derive the precession of the polarisation according to the equation of parallel transport,
\begin{equation}
\frac{\textrm{d}S_\alpha}{\textrm{d}\tau} = \Gamma^\lambda_{\alpha\nu} S_\lambda \frac{\textrm{d}x^\nu}{\textrm{d}\tau} \, .
\end{equation}
where $\Gamma^\lambda_{\alpha \nu}$ are the Christoffel symbols and $\tau$ is proper time. From here we can specialise to the motion in a gravitational potential $\phi$. Following e.g. Ref.~\cite{Weinberg:1972kfs}, we can write the time variation of the polarisation as,
\begin{align}
  \frac{\textrm{d}\mathbf{S}}{\textrm{d}t} = - \mathbf{S}\frac{\partial \phi}{\partial t} - 2 \mathbf{v}\cdot \mathbf{S}\nabla\phi-\mathbf{S}(\mathbf{v}\cdot\nabla\phi) \nonumber \\
  +\mathbf{v}(\mathbf{S}\cdot\nabla\phi)  + \frac{1}{2}\mathbf{S}\times(\nabla\times \mathbf{S}) \, .
\end{align}
It is then useful to define a new vector
\begin{equation}
    \mathbf{S}_1 =(1+\phi)\mathbf{S}-\frac{1}{2}\mathbf{v}(\mathbf{v}\cdot \mathbf{S}) \, ,
\end{equation}
whose evolution is governed by a spin-orbit equation,
\begin{equation}
\frac{\textrm{d}\mathbf{S}_1}{\textrm{d}t} = \boldsymbol{\Omega} \times \mathbf{S}_1,
\end{equation}
where $\boldsymbol{\Omega}=-\frac{1}{2}\nabla\times \boldsymbol{\zeta}-\frac{3}{2}\mathbf{v}\times \nabla \phi$, with $\boldsymbol{\zeta}$ the vector potential.
We can then estimate the relative polarization variation as 
\begin{equation}
   \frac{\delta S}{S} \sim T v \frac{\phi}{R c^2}  
\end{equation}
where $T$ and $R$ are the typical time and length scales of the problem, and we have restored the factor of $c$ for clarity. We can then use the virial theorem, $\phi \sim v^2$,
%and assume the following rough values: $T_{\rm gal} \simeq 13.6 $ billion years, $R_{\odot} \simeq 8$ kpc, $M \simeq 10^{12}\, M_{\odot}$.
and normalize everything to typical values of our galaxy. We thus find
$$
 \frac{\delta S}{S} \sim 4 \times 10^{-3} \Big(\frac{v}{2\times 10^{-3}}\Big)^3\frac{T}{13 \times 10^9 \, {\rm yr}}\frac{8 \, \rm kpc}{R},
$$
which shows how dark matter can easily preserve some degrees of its initial polarisation over the lifetime of the galaxy.
While a dedicated study is needed to understand how a real halo of dark matter would be affected, this estimate shows, at the very least, that a fixed polarisation over laboratory scales is a plausible scenario.

In this work we will take a phenomenological approach and consider the two extreme cases: fixed polarisation, and totally randomised polarisation. Our results are such that any real scenarios will be bounded within these two limits. The fixed polarisation scenario will be the focus of the majority of our results since it leads to signals that require the most care to describe. We emphasise though that the details of this study do not solely apply to the extreme case where the DP polarisation is fixed over the entire Universe. They also apply to cases where only a fraction of the field is coherently polarised, or if the polarisation varies slowly relative to our measurements.

\section{Dark photon electrodynamics}\label{sec:DPelectrodynamics}
Working with the assumption that the DM distribution around the Earth is comprised of a cold population of DPs, we now discuss how to detect them in the laboratory. This requires us to first explain some aspects of electrodynamics in the presence of DPs.

The low-energy effective Lagrangian due to the presence of a gauge boson $X$ of a dark U(1) that kinetically mixes~\cite{Holdom:1985ag,Arias:2012az,An:2014twa,Fabbrichesi:2020wbt} with the visible photon $A$ reads,
\begin{align}
\mathcal{L} \supset & -\dfrac{1}{4} {F}_{\mu \nu} {F}^{\mu \nu} - \dfrac{1}{4}{X}_{\mu \nu} {X}^{ \mu \nu} +\frac{\sin{\alpha}}{2}F^{\mu\nu}{X}
_{\mu\nu}  \notag \\
& + e J_{\rm EM}^{\mu}{A}_{\mu}+ \dfrac{m_X^2 \cos^2{\alpha}}{2}{X}^{\mu} {X}_{\mu}\, ,
\end{align}
where $F_{\mu \nu}$, $X_{\mu \nu}$ are the field strengths of the SM photon and the DP, $J_{\rm EM}^{\mu}$ is the electromagnetic current, $m_X$ is the DP mass and $\sin \alpha$ is the kinetic mixing parameter. Note that we neglect terms $\mathcal{O}(\alpha^2)$.

We can remove the kinetic mixing term by diagonalisation through $\tilde{A}={A}\cos{\alpha}, \tilde{X}=X-\sin{\alpha} A$. In the so-called interaction basis, $(\tilde{A},\tilde{X})$, the effective Lagrangian is,
\begin{align}\label{Eq:Lagrangian1}
\mathcal{L} \supset & -\dfrac{1}{4} {\tilde{F}}_{\mu \nu} \tilde{F}^{\mu \nu} - \dfrac{1}{4}\tilde{X}_{\mu \nu} \tilde{X}^{ \mu \nu} + \dfrac{e}{\cos{\alpha}}  J_{\rm EM}^{\mu}\tilde{A}_{\mu} \notag
  \\
& + \dfrac{m_X^2 \cos^2{\alpha}}{2}\left(\tilde{X}^{\mu} \tilde{X}_{\mu} +2\chi\tilde{X}_\mu \tilde{A}^\mu+ \chi^2\tilde{A}^\mu \tilde{A}_\mu\right)~, 
\end{align}
where $\chi \equiv \tan {\alpha}$ and $\tilde{A}$, $\tilde{X}$ are interpreted as the photon produced in electromagnetic interactions and the DP sterile state respectively. These interaction states are the most relevant when discussing DP detection. 

In the interaction basis the electromagnetic coupling is renormalised to $(e/\cos \alpha)$ and there are $\tilde{A}$-$\tilde{X}$ oscillations due to the mass-mixing term. Due to decoherence, the DPDM will be in the massive propagation eigenstate found instead by diagonalising the mass term in the DP Lagrangian, Eq.~\eqref{Eq:Lagrangian1}. Assuming the kinetic mixing $\chi$ to be small, the DM is mostly aligned with the sterile eigenstate $\tilde{X}$.

Neglecting $\mathcal{O}(\chi^2)$ terms and dropping the tildes, we obtain the Lagrangian
\begin{align}\label{Eq:Lagrangian}
\mathcal{L} \supset & -\dfrac{1}{4} {{F}}_{\mu \nu} {F}^{\mu \nu} - \dfrac{1}{4}{X}_{\mu \nu} {X}^{ \mu \nu} + e  J_{\rm EM}^{\mu}{A}_{\mu}
\notag \\
& + \dfrac{m_{X}^2}{2}\left({X}^{\mu} {X}_{\mu} +2\chi{X}_\mu {A}^\mu\right) \, ,
\end{align}
from which one finds the wave equation in momentum space
\begin{align}
    -K^2 {A}^\nu= \chi m_X^2{X}^\nu \, .
    \label{eq:wave_equation}
\end{align}
Here, we defined the four momentum $K=(\omega,\mathbf{k})$ and used the Fourier expansion for a free field with the energy $\omega=+\sqrt{|\mathbf{k}|^2 +m_X^2}$. In the following, we will treat the fields as complex, $X^\mu_c(t, \mathbf{x})$, of which the actual fields constitute the real part $X^\mu=$ Re$\{X^\mu_c\}$. Following Refs.~\cite{Knirck:2018knd,Gelmini:2020kcu}, we include a volume $V$ in the definition of the Fourier transform,
\begin{align}\label{eq:Fourier}
    {X_c^\mu}(t,\mathbf{x})= \sqrt{V} \int \frac{\drm^3 \mathbf{k}}{(2\pi)^3} &X^\mu (\mathbf{k}) e^{-i(\omega t -\mathbf{k}\mathbf{x}+\delta({\bf k}))}\, .
\end{align}
In principle, one can have non-trivial information such as the clumpiness of DM in the phases $\delta({\bf k})$~\cite{Knirck:2018knd}, however for our purposes we are neglecting such issues and assuming the density of DM remains constant throughout measurements.
Our use of a classical field description is justified as the state occupation number required to make up a local DM density of $\rho_{\rm DM} \sim 0.3$--$0.5$~GeV~cm$^{-3}$~\cite{deSalas:2020hbh} out of sub-eV DPs must be very large. As the classical calculation gives the expectation value of the measurement for bosonic two-level mixing~\cite{Raffelt:1991ck}, the large occupation number means a large number of DPs will be involved, so that the overall measurement simply gives the expectation value.\footnote{As with axion haloscopes, there is no Bose-enhancement if the final state is occupied~\cite{Ioannisian:2017srr} .} Thus one would only need to worry about a non-classical state if higher order correlations are measured.

We will now show that most of the DM energy is stored in the zero mode of the DP field. The energy density of the DP field is given by 
\begin{equation}\label{density}
    \rho = \frac{1}{V}\int \frac{d^3\textbf{k}}{(2\pi)^3}\frac{\omega(\textbf{k})^2}{2}|\textbf{X}(\textbf{k})|^2.
\end{equation}
The space-averaged DP field can be written as
\begin{align}
\langle X_c^{\mu}(t) \rangle = \frac{1}{V}&\int \drm^3\textbf{x} \,X^\mu_c(t,\textbf{x}) \nonumber\\&= \frac{X^\mu(\textbf{k}=0)}{\sqrt{V}}e^{- i m_X t} \equiv X^\mu_0 e^{-i m_X t}\, ,
\end{align}
where we have defined the amplitude of the plane wave $X^\mu_0 = \left(X^0_0, \textbf{X}_0\right) = X^{\mu}(\textbf{k}= 0)/\sqrt{V}$. At this point, we can make contact between the field and the  particle description. Given the local velocity distribution in the laboratory rest frame, $f_{\rm lab}(\textbf{v})$, the DM density is 
\begin{equation}
    \rho = \rho \int \drm^3\textbf{v} \, f_{\rm lab}(\textbf{v})\, ,
\end{equation}
so the DM velocity distribution can be identified with
\begin{equation}
    f_{\rm lab}(\textbf{v}) = \frac{m_X^3 \omega^2}{2(2\pi)^3\rho}|\textbf{X}(\textbf{k})|^2 \, ,
\end{equation}
where we assumed the DM to be non-relativistic, i.e.~$\textbf{k} =m_X \textbf{v}$.
Neglecting the DM kinetic energy, $\omega(\textbf{k})^2 =m_X^2$, we can  write 
\begin{equation}
    \rho = \frac{m_X^2}{2} \langle| \textbf{X}_c(t)|^2 \rangle\, ,
\end{equation}
by taking a spatial average of the squared amplitude of the DP field. Finally, if we also neglect velocity in the Fourier transform Eq.~\eqref{eq:Fourier} (i.e.~we assume the velocity of the DM to be zero), we can write
\begin{equation}
    \rho \simeq \frac{m_X^2}{2} |\langle \textbf{X}_c(t)| \rangle |^2 = \frac{m_X^2}{2}|\textbf{X}_0|^2 \, .
\end{equation}

The quantity relevant to experiments is the ordinary electric field induced by the DP field which acts like a source in the wave equation, Eq.~\eqref{eq:wave_equation}, for the interaction eigenstate. In the limit of classically oscillating fields, Gauss's law in a homogeneous, isotropic medium with no external sources reads $\nabla\cdot \mathbf{D}=0$. The electric displacement field is related to the electric field via $\mathbf{D}=\epsilon \mathbf{E}$ where $\epsilon$ is the dielectric function of the medium ($\epsilon_{\rm vacuum}=1$).

Therefore, the electric field produced by the DP is,
\begin{equation}
    \left|\textbf{E}_0 \right| = |\frac{\chi m_X}{\epsilon}\textbf{X}_0| \, .
\end{equation}
But if the experiment is sensitive to a particular component of the produced electric field, for example in the direction $\bfhat{z}$, then the resulting electric field shall be written as 
\begin{equation}
   \left|\textbf{E}_0 \right| = \left|\frac{\chi m_X}{\epsilon}\textbf{X}_0\cos\theta\right|\, ,
    \label{eq:electric_direction}
\end{equation}
where $\cos\theta = \bfhat{z}\cdot \textbf{X}_0/|\textbf{X}_0| \equiv \bfhat{z}\cdot \bfhat{X}$. In conclusion, we see that the DP signal depends upon the orientation of the experiment with respect to the DP's polarisation.

\section{Dark Photon detection}\label{sec:DPdetection}

 \begin{table*}[t]\centering
\ra{1.3}
\begin{tabularx}{1\textwidth}{llr|Y|Y|Y|Y|Y}
\hline\hline
\multicolumn{3}{c|}{\bf Experiment}  & {\bf Magnetic field} [T] & {\bf Latitude} [$^\circ$] & {\bf Measurement time}, $T$ &  {\bf Directionality} & $\langle \cos^2{\theta}\rangle^{\rm excl.}_T$ \\ 
\hline
\multirow{15}{*}{\bf Cavities} & ADMX-1 & ~\cite{Asztalos:2009yp} & 7.6 & 47.66 & $\mathcal{O}({\rm min})$ & $\Zenith$-pointing  & $\sim$0.025\\
& ADMX-2 & ~\cite{Du:2018uak} & 6.8 & 47.66 & $\mathcal{O}({\rm min})$ & $\Zenith$-pointing  & $\sim$0.019\\
& ADMX-3 & ~\cite{Braine:2019fqb} & 7.6 & 47.66 & $\mathcal{O}({\rm min})$ & $\Zenith$-pointing  & $\sim$0.019\\
& ADMX Sidecar & ~\cite{Boutan:2018uoc} & 3.11\footnote{Run B of Ref.~\cite{Boutan:2018uoc} started at 0.78 T before the magnetic field was ramped up to 2.55~T, so it cannot naively be reinterpreted as a DP limit.} & 47.66 & $\mathcal{O}({\rm min})$ & $\Zenith$-pointing  & $\sim$0.019\\
& HAYSTAC-1 & ~\cite{Zhong:2018rsr} & 9 & 41.32 & $\mathcal{O}({\rm min})$ & $\Zenith$-pointing  & $\sim$0.019\\
& HAYSTAC-2 & ~\cite{Backes:2020ajv} & 9 & 41.32 & $\mathcal{O}({\rm min})$ & $\Zenith$-pointing  & $\sim$0.019\\
& CAPP-1 & ~\cite{Lee:2020cfj} & 7.3 & 36.35 & $\mathcal{O}({\rm min})$ & $\Zenith$-pointing  & $\sim$0.019 \\
& CAPP-2 & ~\cite{Jeong:2020cwz} & 7.8 & 36.35 & $\mathcal{O}({\rm min})$ & $\Zenith$-pointing  & $\sim$0.019 \\
& CAPP-3 & ~\cite{Kwon:2020sav} & 7.2 and 7.9 & 36.35 & $90~{\rm s}$ & $\Zenith$-pointing  & $\sim$0.019 \\
& CAPP-3 [KSVZ] & ~\cite{Kwon:2020sav} & 7.2 & 36.35 & $15~{\rm hr}$ & $\Zenith$-pointing  & 0.20 \\
& QUAX-$\alpha\gamma$ & ~\cite{Alesini:2020vny} & 8.1 & 45.35 & 4203 s & $\Zenith$-pointing  & 0.023 \\
& $^\dagger$KLASH & ~\cite{Alesini:2017ifp} & 0.6 & 41.80 & $\mathcal{O}({\rm min})$ & $\Zenith$-pointing  & $\sim$0.019 \\

\cline{4-8} & RBF & ~\cite{DePanfilis:1987dk} & \multicolumn{5}{c}{Magnetic field veto}\\
& UF & ~\cite{Hagmann:1990tj} & \multicolumn{5}{c}{Magnetic field veto} \\
& ORGAN & ~\cite{McAllister:2017lkb} & \multicolumn{5}{c}{Magnetic field veto} \\
& RADES & ~\cite{Melcon:2021dyi} & \multicolumn{5}{c}{Magnetic field veto} \\
\hline
\multirow{5}{*}{\bf LC-circuits} & ADMX SLIC-1 & ~\cite{Crisosto:2019fcj} & 4.5 & 29.64 & $\mathcal{O}({\rm min})$ & $\North/\West$-facing  & $\sim$0.19\\
 & ADMX SLIC-2 & ~\cite{Crisosto:2019fcj} & 5 & 29.64 & $\mathcal{O}({\rm min})$ & $\North/\West$-facing  & $\sim$0.19\\
  & ADMX SLIC-3 & ~\cite{Crisosto:2019fcj} & 7 & 29.64 & $\mathcal{O}({\rm min})$ & $\North/\West$-facing  & $\sim$0.19\\

\cline{4-8}
& ABRACADABRA & ~\cite{Ouellet:2018beu} & \multicolumn{5}{c}{Magnetic field veto\footnote{The pickup geometry also does not allow for linearly polarised DP to be detected.}}\\
& SHAFT & ~\cite{Gramolin:2020ict} & \multicolumn{5}{c}{Magnetic field veto}\\
\hline
\multirow{1}{*}{\bf Plasmas}& $^\dagger$ALPHA  & ~\cite{Lawson:2019brd} & 10 & Unknown & $\mathcal{O}$(week) & $\Zenith$-pointing  & 0.28--0.33\\
\hline
\multirow{3}{*}{\bf Dielectrics} & $^\dagger$MADMAX & ~\cite{TheMADMAXWorkingGroup:2016hpc} & 10 & 53.57 & $\mathcal{O}$(week)  & $\Zenith$-pointing or $\North/\West$-facing & 0.26 or 0.62--0.66\footnote{MADMAX can sense the $E$-field parallel to its disks, which will be vertical. But whether MADMAX will be axial or planar depends on its antenna. The first number is if the antenna can sense only a single polarisation, and the second is if it can sense arbitrary polarisation, in which case the range reflects the unknown final horizontal orientation of the experiment.}\\
& $^\dagger$LAMPOST & ~\cite{Baryakhtar:2018doz} & 10 & Unknown & $\mathcal{O}$(week) & Any-facing & 0.61--0.66\\
& $^\dagger$DALI & ~\cite{DeMiguel-Hernandez:2020mon} & 9 & 28.49 & $\mathcal{O}$(month) &  Any-facing\footnote{The DALI experiment is designed to be on an altazimuth mount and so can be aligned as required.} & 0.61--0.66\\
\hline
\multirow{1}{*}{\bf Dish antenna} & $^\dagger$BRASS & ~\cite{BRASS} & 1 & 53.57 & $\mathcal{O}$(100 days) & Any-facing & 0.61--0.66\\
\hline
\makecell{ {\bf Topological} \\ {\bf insulators}} & $^\dagger$TOORAD & ~\cite{Schutte-Engel:2021bqm} & 10\footnote{The $B$-field would be varied for tuning purposes} & Unknown & $\mathcal{O}$(day) & Any-pointing & 0.18--0.33\\
\hline \hline
\end{tabularx}
\caption{Summary of axion haloscopes and their parameters relevant for recasting exclusion limits to DPs. We denote planned experiments with a ``$\dagger$''. Note that some axion exclusions cannot be reinterpreted as DP exclusions if the experiment used a magnetic field to veto potential signals~\cite{Gelmini:2020kcu}. We assumed that, unless stated otherwise, all experiments did not use some form of magnetic field veto, though it is possible that one was used without explicit acknowledgement. In the final column we have put our estimate of $\excl$ which is a geometric factor used to convert axion exclusion limits (with varying C.Ls) to 95\% C.L. DP exclusion limits, accounting for the unknown polarisation. The factor $\excl$ is defined in Eq.(\ref{eq:excl}), and its calculation is the subject of Sec.~\ref{sec:DPscanning}. For experiments with unknown locations, we have taken the largest and smallest values in the range $\lat \in [35^\circ,55^\circ]$. For experiments with unknown orientations, we have taken the largest and smallest values over the range of possible orientations.\label{tab:axionhaloscopes}}
\end{table*}

\begin{table*}[t]\centering
\ra{1.3}
\begin{tabularx}{1\textwidth}{llr|c|Y|c|Y|Y}
\hline\hline
\multicolumn{3}{c|}{\bf Experiment}  & {\bf Latitude} & {\bf Measurement} &  {\bf Directionality} & {\bf Assumed} & $\langle \cos^2{\theta}\rangle^{\rm excl.}_T$ \\ 
 &  &  &  [$^\circ$] & {\bf time}, $T$ &   & $\langle \cos^2{\theta}\rangle_T$ & \\ 
\hline
\multirow{2}{*}{\bf Cavities} & WISPDMX & ~\cite{Nguyen:2019xuh} & 46.14 & $\mathcal{O}({\rm day})$ & ($0.92\North + 0.38 \West$)-pointing & 1/3 & 0.23 \\
& SQuAD & ~\cite{Dixit:2020ymh} & 41.88 & 12.81 s & Unspecified & 1/3 & 0.019 \\
\hline
\multirow{1}{*}{\bf Dielectrics} & $\dagger$NYU Abu Dhabi & ~\cite{Manenti}  & 24.45 & ${\cal O}({\rm day})$ & $\Zenith$-facing & N/A & 0.65\\
\hline
\multirow{5}{*}{\bf Dish}\vspace{.5ex}\hspace{-2.25em}\multirow{6.5}{*}{\bf antennae} & Tokyo-1 & ~\cite{Suzuki:2015sza} & 35.68 & 29 days\footnote{This measurement alternated signal and background every 30s} & $\West$-facing & 2/3 & 0.62 \\
& Tokyo-2 & ~\cite{Knirck:2018ojz} & 36.06 & $\mathcal{O}$(week) & Axial, $\North$/$\West$-pointing & 1/3 & 0.15--0.2\\
& Tokyo-3 & ~\cite{Tomita:2020usq} & 36.13 & 12~hr & $\North$/$\West$-pointing or $\Zenith$-facing  & Unspecified & 0.15 or 0.62\\
& SHUKET & ~\cite{Brun:2019kak} & 48.86 & 8000~s & $\Zenith$-pointing & 1/3 & 0.04\\
& FUNK & ~\cite{Andrianavalomahefa:2020ucg} & 49.10 & $\mathcal O$(month) & ($-0.5\North - 0.87 \West+0.28\Zenith$)-facing & 2/3 & 0.56\\
\hline
\multirow{3}{*}{\bf LC-circuits} & DM Pathfinder & ~\cite{Phipps:2019cqy} & 37.42 & 5.14~{\rm hr}& $\Zenith$-pointing & 1\footnote{Rather than assume a randomised angle, the DP current was assumed to be aligned with the device} & 0.075\\
& Dark E-field & ~\cite{Godfrey:2021tvs} & 38.54 & 3.8\,{\rm hr} \footnote{These numbers include dead time, so should not be taken as a single continuous measurement} & $\West$-pointing & 1/3 &  0.29 \\
& Dark E-field spots & ~\cite{Godfrey:2021tvs} & 38.54 & 5.8\,days\footnote{These numbers include dead time, so should not be taken as a single continuous measurement} & $\West$-pointing & 1/3 &  0.58 \\
\hline \hline
\end{tabularx}
\caption{Summary of dedicated DPDM experiments and some of their relevant parameters. We mark planned experiments with a ``$\dagger$''. The penultimate column refers to the value of $\langle \cos^2{\theta}\rangle_T$ assumed by the corresponding reference, whereas the last column is our best attempt to estimate the value of $\langle \cos^2{\theta}\rangle^{\rm excl.}_T$ for the fixed polarisation scenario, based on the published measurement times, locations, and experimental orientations.\label{tab:DPexperiments} This factor is defined in Eq.(\ref{eq:excl}) and describes how much a 95\% C.L.~exclusion limit on the DP is impacted by the unknown DP polarisation angle with respect to the experiment. The calculation of $\excl$ factor is the subject of Sec.~\ref{sec:DPscanning}.}
\end{table*}

\begin{figure*}
\centering
% 	%trim option's parameter order: left bottom right top
  \includegraphics[trim = 0mm 0mm 0mm 0mm, clip, width=0.95\textwidth]{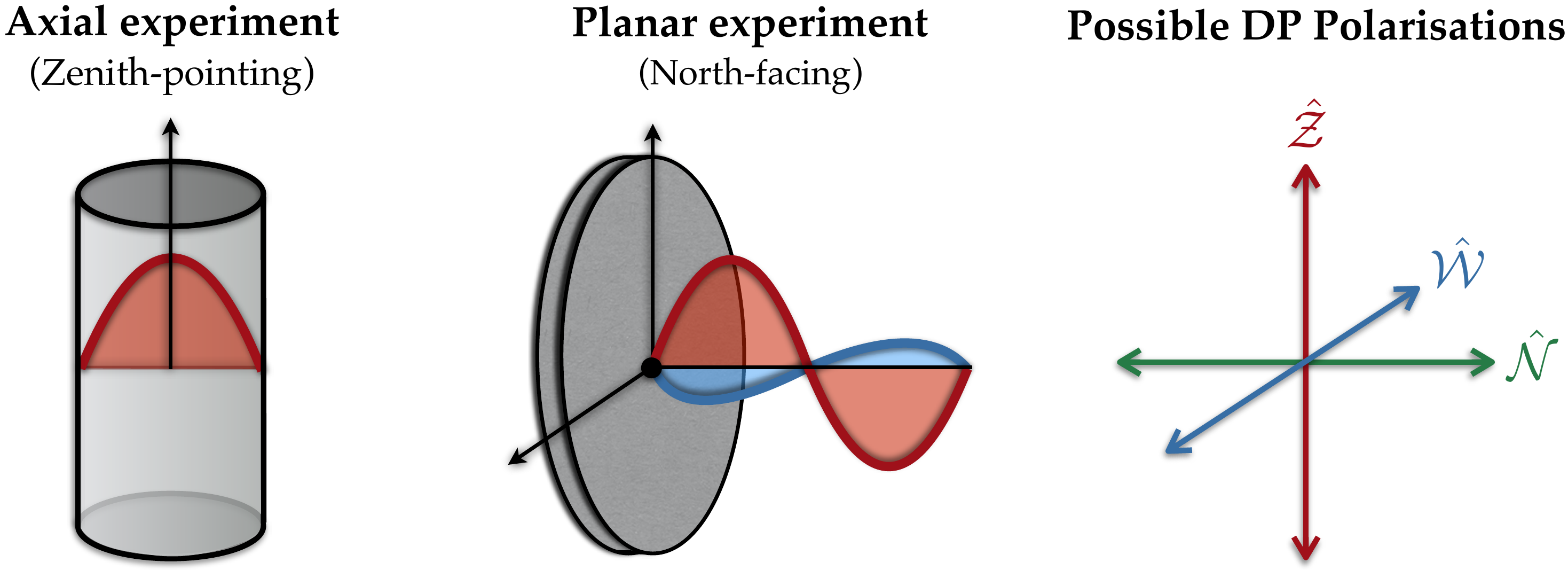}
\caption{Schematic of the two categories of experiment that we structure this study around. They are distinguished by their sensitivity to the direction of the DP polarisation. On the left, ``axial'' experiments sensitive to a single direction of polarisation are represented via a cylindrical cavity. In the middle, ``planar'' experiments, sensitive to a plane of polarisation, are shown using a simple sketch of a dish antenna. We express the DP polarisation here in the lab-centered North-West-Zenith basis shown in the far right image. To introduce some nomenclature that will be important later: the cavity in this example would be sensitive to electric fields along the $\Zenith$ axis, so we refer to this as ``Zenith-pointing''. On the other hand, the dish antenna is facing North, so it is sensitive polarisations in the $\Zenith$-$\West$ plane, which we refer to as ``North-facing''.}
 	\label{fig:ExperimentPolarisation}
\end{figure*}

Most searches for DPs as dark matter use the mixing of the DP and visible photon to induce $E$ or $B$ fields, which are then coupled to a detector. This is similar to searches for the axion-photon coupling, which typically use an external $B$-field to mix axions with photons. Hence there is substantial overlap between experiments that can search for axions and those that can search for DPs. All experiments that we will discuss are listed in Tables~\ref{tab:axionhaloscopes} and~\ref{tab:DPexperiments}.

Types of experiment exploiting electromagnetic mixing include: cavities \cite{Sikivie:1983ip,Rybka:2014xca,Woohyun:2016,Goryachev:2017wpw,Alesini:2017ifp,Melcon:2018dba,Melcon:2020xvj}, dielectric disks~\cite{TheMADMAXWorkingGroup:2016hpc,Baryakhtar:2018doz}, dish antennae~\cite{Horns:2012jf,Jaeckel:2013sqa,Suzuki:2015sza,Experiment:2017icw,BRASS}, plasmas~\cite{Lawson:2019brd}, LC circuits~\cite{Sikivie:2013laa,Chaudhuri:2014dla,Kahn:2016aff,Silva-Feaver:2016qhh,Crisosto:2018div}, and electric-field radios~\cite{Godfrey:2021tvs}. However, there are two key differences between DPs and axions. The first is that the axion-induced $E$-field is always aligned with the external magnetic field due to the ${\bf E}\cdot {\bf B} $ coupling of the axion, whereas the DP can be polarised in any direction. The second is that the mixing does not ``turn off" in the absence of a $B$-field, which as pointed out in Ref.~\cite{Gelmini:2020kcu} is often not checked when reinterpreting axion limits in the context of DPs. This behaviour both changes potential noise vetos (such as demanding the signal vary as $B^2$) and the requirements of shielding~\cite{Arias:2014ela,Chaudhuri:2014dla}. 

By design, the majority of these experiments are only sensitive to fields induced in specific directions. Most commonly a single direction, e.g.~a single polarisation of light as in Eq.~\eqref{eq:electric_direction}, but for some designs all polarisations lying along a two-dimensional plane may be measurable. A schematic of these two classes is shown in Fig.~\ref{fig:ExperimentPolarisation}. We refer to experiments that are sensitive to a single polarisation direction as \emph{axial}, and those that can sense polarisations along a plane as \emph{planar}. In the following subsections we will describe which types of experiment fall into each of these two categories. In this figure we also define our lab-centered coordinate system basis that will become important when we discuss daily modulation: $(\North,\West,\Zenith)$, which point towards the North, West and Zenith respectively.

\subsection{Cavity haloscopes}
First introduced by Sikivie in Ref.~\cite{Sikivie:1983ip}, cavity haloscopes are the prototypical axion (and DP) direct detection experiment. A cavity haloscope uses the resonant enhancement of a cavity mode to increase the probability of DM converting to photons over a narrow frequency range (given by the width of the resonant mode, or quality factor, of the cavity). However, resonant enhancement is only achieved if the overlap of the photon wavefunction and DM wavefunction is nonzero~\cite{Sikivie:1983ip,Knirck:2018knd}. Which cavity modes achieve a nonzero overlap depends on the geometry of the cavity. For the popular cylindrical cavity, only transverse magnetic (TM) modes have a nonzero overlap---transverse electric (TE) modes, on the other hand, do not couple~\cite{Stern:2015kzo}.\footnote{More precisely, only TM$_{0n0}$ modes have a significant nonzero overlap, with the largest being the TM$_{010}$ mode~\cite{Stern:2015kzo}.} Thus, cylindrical cavities are only sensitive to the component of the DP polarised along the axis parallel to the magnetic field, which is usually aligned vertically. 

For a more complicated cavity, such as multicavity arrays~\cite{Goryachev:2017wpw,Melcon:2018dba,Melcon:2020xvj,Jeong:2020cwz}, dedicated mode analyses are required to determine how the DP couples. Even if the DP couples to modes in multiple directions, each mode will have a preferred $E$-field direction. Depending on how the cavity is read out, as long as these modes are not degenerate and have sufficient frequency separation to be resolved, then each mode will provide information on a specific polarisation. If the modes are not similar in their overall signal power, then the sensitivity will be directionally sensitive in the same way as a cylindrical cavity. On the other hand, if each mode is similar in power (for example, a cubic cavity), then the experiment will be sensitive regardless of the DP polarisation. Cavities designed to detect axions are usually not optimised for multiple polarisation directions since they do not need to be.

So we consider cavities to be ``axial'' in the sense that they are sensitive to a single polarisation direction. In these cases, the signal power is suppressed by the angular factor introduced in Eq.(\ref{eq:electric_direction}), $\cos^2{\theta}$, where $\theta$ the angle between $\bfhat{X}$ and that direction. The average of this factor over all possible DP polarisation axes is $\langle \cos^2{\theta} \rangle = 1/3$.

\subsection{Dish antennae}
In stark contrast to cavity experiments, dish antennae allow a broadband DM search, relying on the nonresonant breaking of translation invariance~\cite{Horns:2012jf,Jaeckel:2013sqa}.
As shown in Eq.\eqref{eq:electric_direction} the mixing between dark and ordinary photons depends on the medium. So by changing medium, the DP-induced $E$-field also changes. Because parallel $E$ and $B$-fields must be conserved across changing media, propagating photon waves are emitted to compensate, which could then be detected. In other words, when the DPDM passes through the dish the small electric field of the DP makes the electrons in the dish oscillate, emitting an (almost) ordinary electromagnetic wave perpendicular to the surface. Dish antennae can, at most, only be sensitive to the component of $E$-field parallel to the interface (usually a metallic mirror)~\cite{Jaeckel:2013sqa}. In contrast to cavities, we classify this kind of experiment as ``planar'', as displayed by the middle panel of Fig.~\ref{fig:ExperimentPolarisation}. As a consequence, the emitted $E$-field is suppressed by $\cos\theta$, where $\theta$ is the angle between $\bfhat{X}$ and the plane of the interface.

Complicating matters is the issue that dish antennae are often, as the name suggests, rounded. If the polarisation is spatially constant over scales comparable to the radius of curvature, $\cos\theta$ will depend upon the position on the dish. Because of this, the total power detected is proportional to $\int \drm A \cos^2\theta({\bf x)}$~\cite{Horns:2012jf}, only being trivial for planar or completely spherical systems. The details depend on the exact geometry of the dish antenna, which can vary considerably between experiments, so we leave the details for dedicated analyses. To provide a rough guide, we make an estimate assuming a planar dish geometry. When averaged over all possible DP polarisations, $\langle \cos^2\theta\rangle=2/3$, for any dish antenna with a sufficiently large radius of curvature.

Furthermore, depending on the antenna technology, some experiments may only be sensitive to certain polarisations of light, e.g.~some stages of the Tokyo dish antennae~\cite{Knirck:2018ojz,Tomita:2020usq}. This would introduce an additional factor of $\cos\theta_a$, where $\theta_a$ is the angle between the light emitted from the dish antenna and detector polarisation. For our purposes we will consider dish antennae either planar or axial depending on whether the detector is sensitive to two polarisations or just one. 

\subsection{Dielectric haloscopes}
Dielectric haloscopes use the same principles as a dish antenna, but employ many semi-transparent dielectric layers, arranged so that the waves emitted at each add constructively. The constructive interference of the emitted waves enhances the sensitivity at the expense of reduced bandwidth. This is the core design principle of MADMAX~\cite{TheMADMAXWorkingGroup:2016hpc,Millar:2016cjp}, but the concept can be implemented on much smaller scales/higher frequencies, as proposed for LAMPOST~\cite{Baryakhtar:2018doz}. Dielectric haloscopes have the same polarisation dependence as a planar dish antenna. Depending on whether the antenna is polarisation-specific, a dielectric haloscope could be sensitive to $E$-fields aligned with either the plane of the dielectric disks or the polarisation of the antenna~\cite{Jaeckel:2013eha,Baryakhtar:2018doz,Brun:2019lyf}.

The recent proposal of using a topological insulator as an axion detector, TOORAD, is mathematically equivalent to a single-layer dielectric haloscope with near-zero refractive index~\cite{Schutte-Engel:2021bqm}. However, the device would only be sensitive to a single polarisation due to the very strong anisotropy of the medium.

\subsection{Plasma haloscopes} 
A more recent proposal, ALPHA, belongs to a distinct class of experiment known as plasma haloscopes~\cite{Lawson:2019brd}. These experiments allow DM axions or DPs to convert to photons by matching the photon mass (plasma frequency) to the DM mass. While the principle holds for any plasma, a more specific proposal allowing for tuneable, cryogenic plasmas at the GHz regime is to use thin wire metamaterials, aligned in a single direction~\cite{Lawson:2019brd}. As pointed out in Ref.~\cite{Gelmini:2020kcu}, the matching of the dispersion relation only occurs in the direction of wire alignment. If the boundary of the plasma is a conducting cylinder, the experiment behaves like a resonant cavity for non-axial electric fields. Thus for transverse polarisations the experiment behaves like a cylindrical cavity, that is, with transverse electric modes not coupling to DM. Thus a plasma haloscope consisting of a cylindrical cavity with wires in only one direction will only be sensitive to $E$-fields aligned with the wires.\footnote{However, this is a geometry-dependent statement. In principle, one could design an isotropic plasma inside a geometry that led to equal sensitivity to any polarisation.}

\subsection{LC circuits}
Rather than measuring the $E$-field induced by DPDM, LC circuits (also referred to as lumped element circuits) attempt to inductively measure a $B$-field. This field can be generated directly from the DM~\cite{Sikivie:2013laa}, or more commonly, indirectly via the $E$-field causing a current in a conductor~\cite{Chaudhuri:2014dla,Kahn:2016aff,Silva-Feaver:2016qhh,Crisosto:2018div,Ouellet:2018beu}. In fact, the (tiny) electric field generated by the DP will create a displacement current, $\textbf{J}_{X}$. This in turn creates a magnetic field and an induced electric field, defined respectively via
\begin{subequations}
\begin{align}
    \nabla \times \textbf{B}_{X} = \textbf{J}_{X}  \, , \\
    \nabla \times \textbf{E}_{X} = -\frac{\partial \textbf{B}_{X}}{\partial t} \, .
\end{align}
    \end{subequations}
If the system size, $r$, is much smaller than a Compton wavelength, $m^{-1}_X$, the $E$-field will be suppressed, however the $B$-field will not, $|\textbf{E}_{X}| \sim m_X r |\textbf{B}_{X}| $~\cite{Chaudhuri:2014dla,Arias:2014ela}. 

As the induced current is in the direction of the DP-induced $E$-field, only polarisations parallel to the conductor can induce a $B$-field~\cite{Arias:2014ela}. As this $B$-field is read out by an inductive loop, the directionality is further constrained to polarisations that generate a $B$-field correctly aligned with the readout loop. For an example, DM Radio~\cite{Silva-Feaver:2016qhh} has a closed toroidal conducting sheath and reads out azimuthal $B$-fields. With such a geometry the read-out $B$-field is proportional to $\cos\theta = \bfhat{z}\cdot\bfhat{X}$. Whereas in other geometries, such as ADMX SLIC~\cite{Crisosto:2019fcj}, the sensitivity can instead lie in a plane~\cite{Arias:2014ela}. 

A similar idea (in fact, an electromagnetic dual) is to have a shielded room much larger than the Compton wavelength, and simply place an antenna to read out any DP-induced $E$-fields. Though there was an early proposal for such an experiment to search for axions~\cite{Smith:1987kz}, it has only recently been revived~\cite{Godfrey:2021tvs}. While in principle all polarisations could be read out, the experiment operating currently, referred to as Dark $E$-field Radio~\cite{Godfrey:2021tvs}, uses a polarisation sensitive antenna. 

Thus we can see that a large class of the most developed and promising light wave-like DM experiments are sensitive to the polarisation of DPs.

\subsection{Signal-to-noise}\label{sec:S2N}
While the exact methods for reading out a signal depend on the design of an experiment, measurement always boils down to increasing the signal-to-noise ratio ($S/N$) by integrating the signal over some measurement time, $T$. Measurements at different times are often combined, either to increase the signal-to-noise, or to test an excess for signal veracity (often referred to as a ``rescan''). This combination usually relies on the assumption that the signal will remain constant with time, i.e.~if a $3\sigma$ excess is measured during an initial campaign, it should persist with subsequent interrogation. However, this is not necessarily a safe assumption for DP signals.

For the experiments we are exploring, there are two main methods of detection: linear amplification and single photon counting. Regardless of whether the $E$ or $B$-field is being measured, the fundamental sensitivity comes from the power generated by the conversion of DPs, relative to the noise of the system. In all cases, the power in the measuring device is determined by the dominant field, either $E^2$ or $B^2$. As discussed above, projecting a DP polarisation onto the experiment also results in a factor of $\cos\theta$, where from now on $\theta$ is taken to be the angle between the DP polarisation and the experimentally relevant geometry: either a plane or an axis. This means we can assume that the DP signal power can be written as,
\begin{equation}
   P(t)=P_{\rm X}\cos^2\theta(t)
\end{equation}
While real experiments will design dedicated statistical analyses to test for signals, we can understand what the generic consequences of this kind of temporal variation would be by writing down some simple formulae for S/N.

\begin{itemize}
    \item {\bf Linear Amplification:} The signal-to-noise of a linear amplifier with constant signal and noise temperatures is easily written via Dicke's radiometery equation~\cite{Dicke}
    \begin{equation} \label{eq:signalnoise}
\dfrac{S}{N} = \dfrac{P}{T_{\rm sys}}\sqrt{\dfrac{T}{\Delta \nu_{\rm DP}}}\,,
\end{equation}
where $T_{\rm sys}$ is the system noise temperature and $\Delta \nu_{\rm DP}$ is the DP linewidth. Since most experiments relying on this formula are not analysing the variation in the power within the measurement time, $T$, the relevant power in this expression is, $P\to\int \drm t P(t)/T$.
\item {\bf Photon Counting:} At higher frequencies, rather than amplifying a measured voltage, it is often more practical to use a photon counting device. Instead of a noise temperature, the background is often better characterised by a dark count (false counts). The signal-to-noise in such a system is
\begin{equation}
    \dfrac{S}{N}=2(\sqrt{n_s+n_d}-\sqrt{n_d})\,,
\end{equation}
where $n_s$ is the number of signal events and $n_d$ the number of dark counts~\cite{Bityukov:1998ju,Bityukov:2000tt,Arias:2010bh}. The integrated signal over a continuous measurement is given by 
\begin{equation}
    n_s=\eta \int \drm t \frac{P(t)}{\omega}\,,
\end{equation}
where $\eta$ is the detector efficiency. 
\end{itemize}
So we can see quite generically that a simple analysis of the total $S/N$ for a continuous measurement is determined by 
\begin{equation}
    \frac{1}{T} \int \drm t P(t) \equiv P_X \langle \cos^2{\theta}\rangle_T \, .
\end{equation}
Crucially, the quantity $\langle \cos^2{\theta} \rangle_T$ will depend upon the distribution of $\bfhat{X}$ around the Earth. Recall that we classified two scenarios in Sec.~\ref{sec:DPcosmology}: the random polarisation scenario, where $\bfhat{X}$ is randomly drawn in every coherence time; and the fixed polarisation scenario where there is only a single $\bfhat{X}$ for the entire experimental campaign.  

In the randomly polarised scenario, as long as the measurement covers many coherence times, i.e.
\begin{equation}
    T\gg \tau = \frac{2 \pi}{m_{X} v^{2}} \simeq 400 \, \upmu \mathrm{s}\left(\frac{10\,\upmu \mathrm{eV}}{m_{X}}\right) \, ,
\end{equation}
then $\langle \cos^2\theta\rangle_{T}$ is simply given by the average of $1/3$ for a single polarisation, or $2/3$ for a plane. In this case the only thing the experiment needs to do to account for the DP polarisation is to apply this factor to their expected signal, e.g. $P = P_X/3$

The situation for the fixed polarisation scenario is more involved because $\theta(t)$ varies over course of the measurement (or measurements). This can have important consequences for the signal-to-noise. Take the simple case of two measurements, each of equal length $T$ but separated by some time $T_{\rm wait}$, with signals-to-noise $S_1/N_1$ and $S_2/N_2$. For simplicity, we will assume the backgrounds are the same for both ($N_1=N_2$), and that the measurement is not in a background-free regime. In this case, the combined $S/N$ is given by,
\begin{align}
    \frac{S}{N}&\simeq\frac{S_1+S_2}{\sqrt{2N_1}}  \\
    &\propto \frac{P_{\rm X}}{T} \int_0^T \drm t \cos^2\theta(t) + \frac{P_{\rm X}}{T} \int_{T_{\rm wait}}^{T_{\rm wait}+T} \drm t \cos^2\theta(t)\,. \nonumber
\end{align}
Assuming the first time average of $\cos^2\theta(t)$ is equal to the second, then when the original signal is real there would be a $\sqrt{2}$ increase in $S/N$. But if the first result was a fluke, the second measurement would reduce $S/N$ by $\sqrt{2}$. However, imagine if the case were,
\begin{equation}
     \int_0^T \drm t \cos^2\theta(t) \gg \int_{T_{\rm wait}}^{T_{\rm wait}+T} \drm t \cos^2\theta(t) \, ,
\end{equation}
i.e.~the DP field was well aligned for the first measurement, but poorly aligned for the second. In such a scenario, an analysis assuming a constant $P(t)$ would then see a reduction in $S/N$ by a factor of $\sqrt{2}$, and potentially conclude that the signal was a statistical fluke. While this is an oversimplified case, it serves to illustrate that it can be dangerous to reinterpret analyses that did not consider time varying signals. However, we are not completely without handles on this time variation. The Earth rotates in a predictable and unambiguous way. As we will discuss in the next section, simply accounting for---or, even better, planning the experiment around---the Earth's rotation, one can greatly reduce the chances of encountering the sorts of bad luck that led to the example described above.

\begin{figure*}
\centering
% 	%trim option's parameter order: left bottom right top
  \includegraphics[trim = 0mm 0mm 0mm 0mm, clip, width=.99\textwidth]{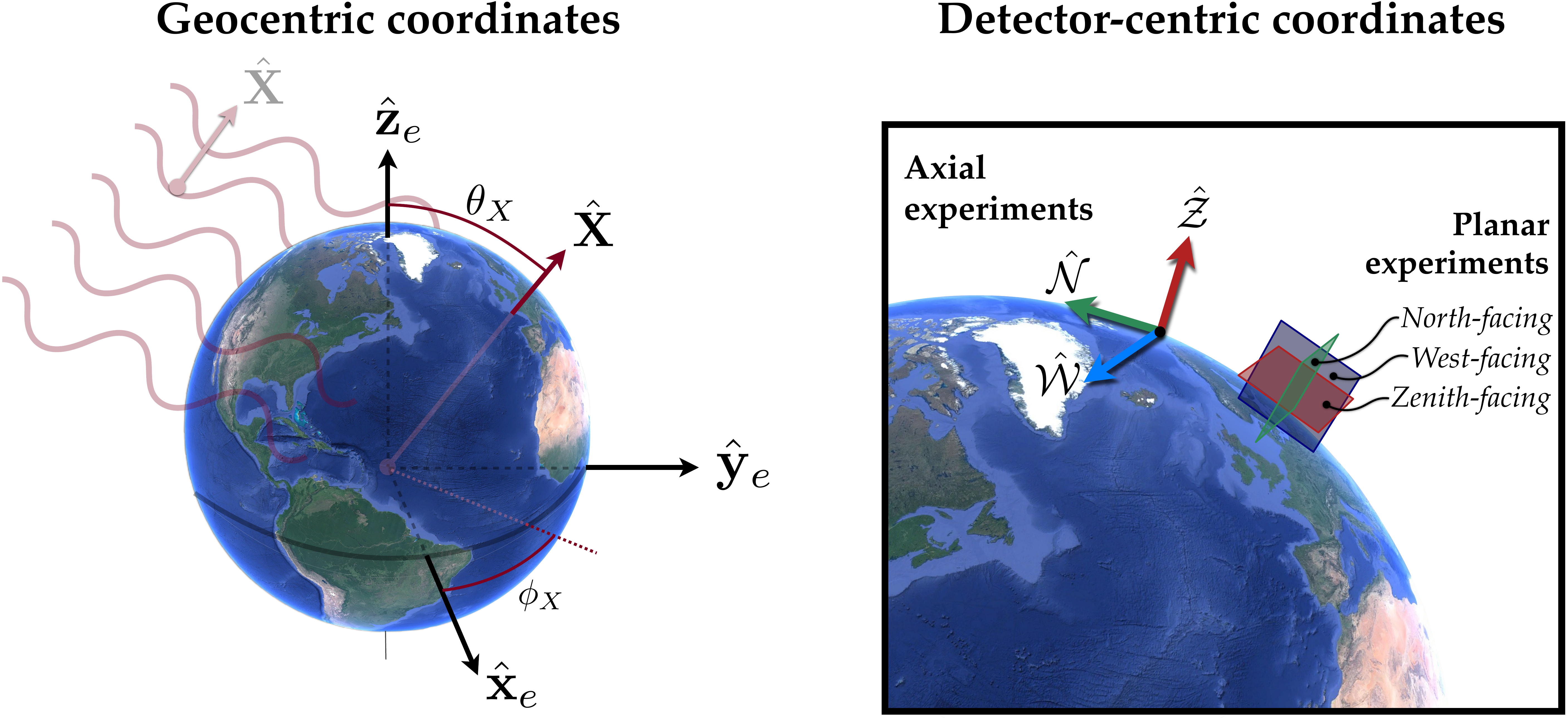}
\caption{A diagram of the coordinate systems used in this paper. On the left, a geocentric equatorial coordinate system in which we fix the DP polarisation vector, $\bfhat{X}$, defined by angles $(\theta_X,\phi_X)$. The $\bfhat{x}_e$ direction is conventionally chosen to point towards the vernal equinox, with $\bfhat{y}_e$ pointing 90$^\circ$ of right ascension to the East. The Earth rotates anticlockwise in the $\bfhat{x}_e$-$\bfhat{y}_e$ plane. On the right, the detector-centric Cartesian coordinate system using axes pointing towards the North, West, and Zenith. We also show three planes in the coordinate system which we will need when describing those experiments that are sensitive to the component of $\bfhat{X}$ projected onto a two-dimensional plane. The Zenith-facing plane is horizontal, whereas the North-facing and West-facing planes are both vertical in the lab.}
 	\label{fig:coordinates}
\end{figure*}

\subsection{Deriving limits on dark photons}\label{sec:derivinglimits}
We have listed all past, current, and planned axion haloscopes searching for the axion-photon coupling in Table~\ref{tab:axionhaloscopes}. This table includes all the relevant information needed to convert their limits on axions to limits on DPs.
 
The crucial differences between axions and DPs come down to two factors: the absence of $B$ in the DP case, and the non-trivial polarisation. The former implies that vetos demanding that the signal vanish in the absence of a magnetic field, as often used in axion searches, would also veto any DP signals and so make an experiment incapable of discovering them. For clarity, we have also listed the experiments for which this caveat applies in Table~\ref{tab:axionhaloscopes}. 

As for the DP polarisation, this issue requires more care and we must think about what an exclusion limit means. Most haloscope analyses take the form of a hypothesis test--- testing for the presence of an axion signal above some noise. In the absence of any convincing signal candidates, they can set an exclusion limit on the coupling---the convention usually being at the 90 or 95\% C.L. In the axion case the signal is a fixed number and the exclusion is straightforward, however in the DP case, there is an unknown parameter which influences the signal strength: $\langle \cos^2{\theta}\rangle_T$. Therefore, since the measured $\langle \cos^2{\theta}\rangle_T$ will be drawn from some (known) distribution, to preserve the statistical meaning of an axion limit when translating it to DPs, we should calculate some additional factor $\excl$ that accounts for this. We define this factor to be ratio of the axion and DP power thresholds that can be excluded at 95\% C.L.~in the absence of a signal. The quantity $\excl$ (or more precisely, its square root since $P\propto \chi^2$) will therefore encode how much the exclusion limit on $\chi$ is weakened by the distribution of $\langle \cos^2{\theta}\rangle_T$.

Doing this conversion accurately would require additional information we cannot easily obtain. For example, we would need to know the noise distribution of each experiment, as well as the local significance of the measured power at each mass point. Nevertheless, we can attempt an estimate of the limit one would obtain from a full data analysis by 1) assuming the noise follows Gaussian distribution, and 2) assuming the measured signal equals the median expected noise. The former assumption is very often the case for most experiments, since they often employ a technique of data stacking which will render most noise distributions Gaussian by the central limit theorem. The latter assumption is a practical one that will make the resulting limit slightly inaccurate and only an estimate of the median exclusion limit. We emphasise here that our intention is not to set definitive limits on the DP using axion haloscope data, but to lay out the recipe for doing so, and predict what would be found.

Next, we adopt a stripped down toy model of the signal and noise distributions to explain how we can relate the axion and DP cases. Let us define the signal and noise in the axion case as,
\begin{equation}
    P = P_a + N \, 
\end{equation}
where $P_a$ is the axion signal power and $N$ is the noise which follows a Gaussian distribution. For simplicity we assume the noise has a mean of $0$ and variance $\sigma_N$. The expected (i.e. median) 95\% C.L.~exclusion limit will therefore be an axion coupling that gives,
\begin{equation}
    \frac{P_a}{\sigma_N} > \Phi^{-1}[0.95] = 1.64 \, ,
\end{equation}
where $\Phi^{-1}$ is the inverse cumulative distribution function of the Gaussian noise.\footnote{For experiments that quote 90\% C.L. exclusions we use 1.28 here instead.} 

On the other hand, in the DP case, the signal is not a single value, but follows a distribution given by,
\begin{equation}
    P = P_X + N \equiv P^0_X \langle \cos^2{\theta}\rangle_T + N \, .
\end{equation}
To get an equivalent 95\% C.L.~exclusion limit on $P^0_X$ we must evaluate an inverse cumulative distribution function, but this time not just of the normally distributed noise, but the joint distribution of the noise and our nuisance parameter $\langle \cos^2{\theta}\rangle_T$. The cumulative distribution function for a variable $z = x+y$ can be written in terms of the distributions on $x$ and $y$, which, assuming they are independent variables, is the following,
\begin{equation}
    \Phi(Z)=\mathbb{P}(z \leq Z)=\int_{-\infty}^{\infty}  \mathrm{~d} x \int_{-\infty}^{Z-x} \mathrm{d} y \, f(x) f(y) \, .
\end{equation}
For our problem we must therefore solve,
\begin{equation}\label{eq:cumPX}
    \Phi[0] \equiv \int_{-\infty}^{+\infty} \textrm{d}P_X  \int_{-\infty}^{0-P_X} \textrm{d}N \, f(P_X) f(N)= 1 - 0.95 \, ,
\end{equation}
where we integrate the noise up to $N = 0 - P_X$, because we want the cumulative distribution function for $P<0$, which is the expected noise level---i.e.~we are demanding that 95\% of measured signals would give a signal greater than the observed noise. We can perform the integral over $f(N)$, which is just a Gaussian, to find,
\begin{align}\label{eq:solvePX}
    \int_{0}^{1} \textrm{d}\langle \cos^2{\theta}\rangle_T \frac{f(\langle \cos^2{\theta}\rangle_T)}{2} \bigg[1+\text{erf}\bigg( \frac{-P^0_X\langle \cos^2{\theta}\rangle_T}{\sqrt{2}\sigma_N} \bigg)\bigg] \nonumber \\ 
     = 1-0.95 \, .
\end{align}
where $f(\langle \cos^2{\theta}\rangle_T) = f(P_X)/P^0_X$. We then solve this equation for $P^0_X/\sigma_N$, giving us the power threshold required for 95\% of all possible signals to be above the median noise. Then, since we are matching this threshold to the axion's one, $P_a/\sigma_N = 1.64$, we can take the ratio of these thresholds to define,
\begin{equation}\label{eq:excl}
    \excl = \frac{1.64\,\sigma_N}{P^0_X} \, ,
\end{equation}
which effectively describes how much the DP power threshold for a 95\% C.L.~exclusion needs to be enhanced over the axion case. We will see in the next section that this results in a value of $\excl$ that is typically in the 15--30th percentile of the distribution $f(\langle \cos^2{\theta}\rangle_T)$.% {\color {blue} To check, what we are doing here is putting $P^0_X=1.64\excl$ int equation 28, and then solving for $\excl$?} \Ciaran{I think it is better to think of it in terms of finding $P^0_X$ by solving equation 28, and then defining $\excl$, but yes the process is the same}

As a sanity check we can imagine what would happen if the distribution of $\langle \cos^2{\theta}\rangle_T$ was a delta function: $\delta(1-\langle \cos^2{\theta}\rangle_T)$. In that case, the solution to Eq.(\ref{eq:solvePX}) is 
$P^0_X/\sigma_N = 1.64$, giving $\excl = 1$. So the power threshold for the DP exclusion limit is identical to the axion, as expected. This case is similar to what we expect in the randomised polarisation scenario. If the DP has a random $\cos^2\theta$ in every coherence time---and typically a single measurement will be over a very large number of coherence times---the central limit theorem will dictate that the resulting $f(\langle \cos^2{\theta}\rangle_T)$ for the stacked data will be very tightly contained around its average of $1/3$ (axial) or $2/3$ (planar). This means we can assume that $\excl$ will be equal to either $1/3$ or $2/3$ for the random polarisation scenario.
 
For the fixed polarisation scenario on the other hand, there is no single factor we can use since $f(\langle \cos^2{\theta}\rangle_T)$ depends upon $T$, as well as the location, orientation and readout of the experiment. In general though, we will find that for very short measurement times, the value of $\excl$ is $\sim$0.025 for axial experiments and $\sim$0.37 for planar experiments\footnote{Note that if the original published exclusion limit was at 90\% C.L. rather than our chosen benchmark of 95\%, then these factors are 0.019 and 0.29}. For longer $T$, the rotation of the Earth will restrict the distribution of possible $\langle \cos^2{\theta}\rangle_T$ that are available to the experiment and therefore $\excl$ will increase. We present our estimate of $\excl$ for axion haloscopes in the final column of Table~\ref{tab:axionhaloscopes}, but the full mathematical recipe for its calculation is the subject of the following section.

We note in passing that previous treatments of the fixed polarisation scenario~\cite{Arias:2012az,Ghosh:2021ard} defined their conversion factor as the 5th percentile of $f(\langle \cos^2{\theta}\rangle_T)$, which gives a value of $0.0025$ instead of 0.025 for very short $T$ axial experiments. Such a prescription is more demanding of the signal and the resulting limit using that conversion factor would turn out to be suppressed over a limit that was truly at the 95\% C.L.

For the sake of completion, we have also made a similar table for experiments that are dedicated to searching only for DPs: Table~\ref{tab:DPexperiments}. Despite the fact that the directionality of the DP polarisation and its relationship to the device's geometry and orientation is of central importance to computing the signal in these experiments, it is surprisingly not always explicitly stated how this information was taken into account. In the cases where it was not possible to glean the orientation or location of the experiment from the manuscript cited, the authors were contacted for further information.

Finally, as well as the expected exclusion limits, we can use the same technique as described above to define the conversion factor needed to find the DP's \emph{discovery limit}, $\disc$. We define ``discovery'' in this context to be the case when 95\% of experiments could reject the noise-only hypothesis at $5\sigma$ or more.
%{\color{blue} Why is it insufficient to just reject the noise-only hypothesis at 5 sigma?} \Ciaran{It would be fine to take the median experiment, i.e. putting 0.5 on the RHS below and then $\excl = 5/P^0_X$, however imposing 95\% of experiments is sort of like saying, "this is the parameter space where you are guaranteed a discovery". it is not uncommon to define discovery limits this way.} 
To calculate this we simply need to take Eq.(\ref{eq:cumPX}) and replace the upper limit of the integral over $N$ with $5\sigma_N-P_X$, so that we are imposing that 95\% of the signal's distribution is $5\sigma_N$ away from the median noise. This leaves us with solving,
\begin{align}\label{eq:solvePX_disc}
    \int_{0}^{1} \frac{f(\langle \cos^2{\theta}\rangle_T)}{2} \bigg[1+\text{erf}\bigg( \frac{5\sigma_N-P^0_X\langle \cos^2{\theta}\rangle_T}{\sqrt{2}\sigma_N} \bigg)\bigg] \textrm{d}\langle \cos^2{\theta}\rangle_T \nonumber \\ 
     = 1-0.95 \, .
\end{align}
If the solution to this equation is some value of $P^0_X/\sigma_N$ then
\begin{equation}\label{eq:disc}
    \disc = \frac{(5+1.64)\sigma_N}{P^0_X} \, .
\end{equation}
We will use $\disc$ when we want to discuss ways of optimising the experiment to have the greatest chance of \emph{discovering} the DP signal, not just setting exclusion limits based on its absence.

\section{Accounting for the rotation of the Earth}\label{sec:DPscanning}
So far we have seen that because the DP possesses a polarisation, induced electromagnetic signals of the DP will inherent the features of that polarisation. In particular, since we observe the DM signal from a rotating reference frame, any fixed DP polarisation on scales much larger than the mpc-scales probed by experimental campaigns, will generate signals with a strong preference on the experimental orientation, and consequently, a daily modulation. In this section we will derive these signals.

We first need to lay down some coordinate systems. There are two relevant ones: we need a coordinate system to define the DP polarisation which is fixed relative to the rotating Earth, and a coordinate system to define our detector that rotates with the surface of the Earth. These are both shown in Fig.~\ref{fig:coordinates}.

Following the left-hand panel of Fig.~\ref{fig:coordinates}, we set the first of these coordinate systems to be the conventional geocentric equatorial frame defined by $(\bfhat{x}_e,\bfhat{y}_e,\bfhat{z}_e)$. These axes point, respectively, towards the vernal equinox, 90 degrees of right ascension East from vernal equinox, and parallel to the Earth's spin axis.\footnote{We are adopting the conventional definition of equatorial coordinates but the orientation of $\bfhat{x}_e$ and $\bfhat{y}_e$ is irrelevant when are assuming we have no knowledge of the polarisation axis of the DP and we are only looking at differences in time. We refer the reader to the Appendix of Ref.~\cite{Mayet:2016zxu} for the full rotation matrices needed to convert between laboratory and equatorial coordinates at a precise local time.} Next we define the detector-centric coordinate system, as displayed by the three arrows in the right-hand panel of Fig.~\ref{fig:coordinates}. This system is defined by axes $(\North,\West,\Zenith)$ which point towards the North, West, and the zenith respectively.

\begin{figure*}
\centering
% 	%trim option's parameter order: left bottom right top
  \includegraphics[trim = 0mm 0mm 0mm 0mm, clip, height=0.44\textwidth]{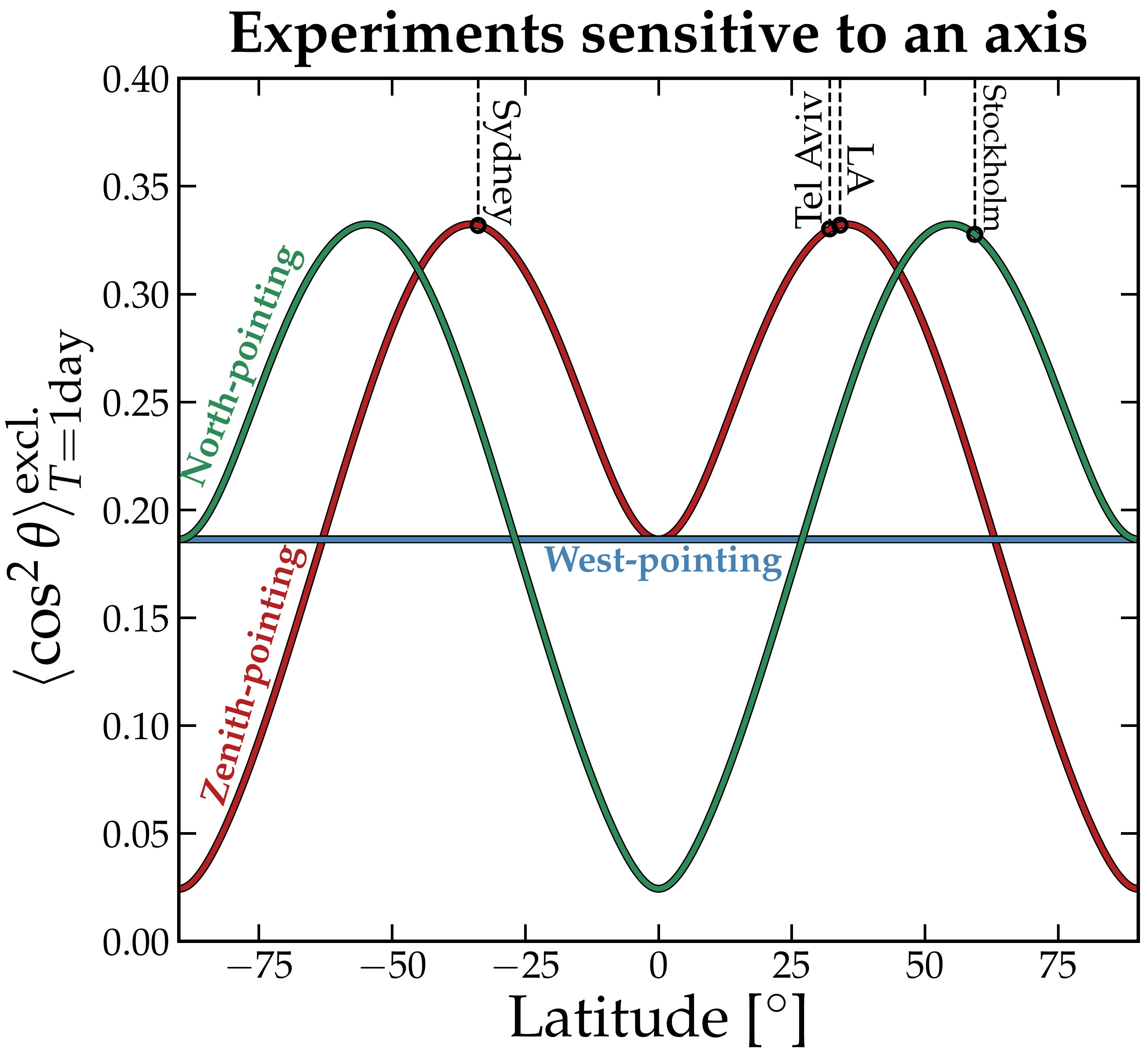}\hspace{1em}
  \includegraphics[trim = 0mm 0mm 0mm 0mm, clip, height=0.44\textwidth]{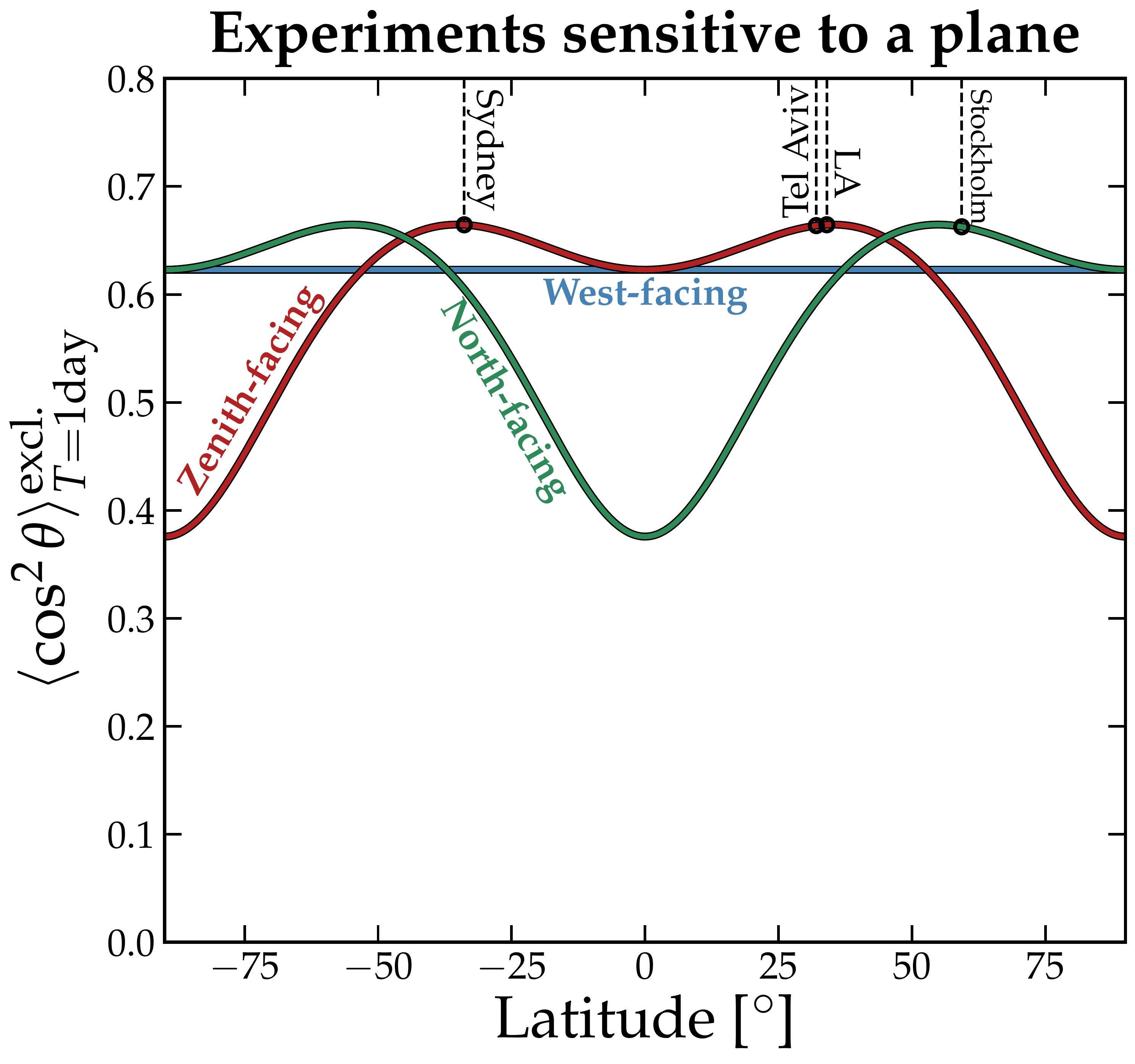}\hspace{1em}

\caption{We define $\excl$ in Eq.(\ref{eq:excl}) to parameterise how much a 95\% C.L. DP exclusion limit is impacted by the distribution of possible polarisation angles. This parameter is an effective conversion factor for recasting prior exclusion limits into the fixed polarisation scenario. Here, we show the latitude dependence of this factor when $T$ is an integer number of sidereal days. The left-hand panel is for axial (1D) experiments, and the right-hand panel is for planar (2D) experiments. Each panel has three lines for the three possible orientations of those experiments. This figure is to demonstrate the preferential latitudes derived in Eqs.(\ref{eq:locationdependence}) and (\ref{eq:locationdependence2}) for integer-day-long measurements. The factor plotted here depends upon the shape of the distribution of $\langle \cos^2{\theta} \rangle_T$ and is typically at the 15-30th percentile.}
 	\label{fig:LocationDependence}
\end{figure*}

\begin{figure*}
\centering
% 	%trim option's parameter order: left bottom right top
  \includegraphics[trim = 0mm 0mm 0mm 0mm, clip, width=0.9\textwidth]{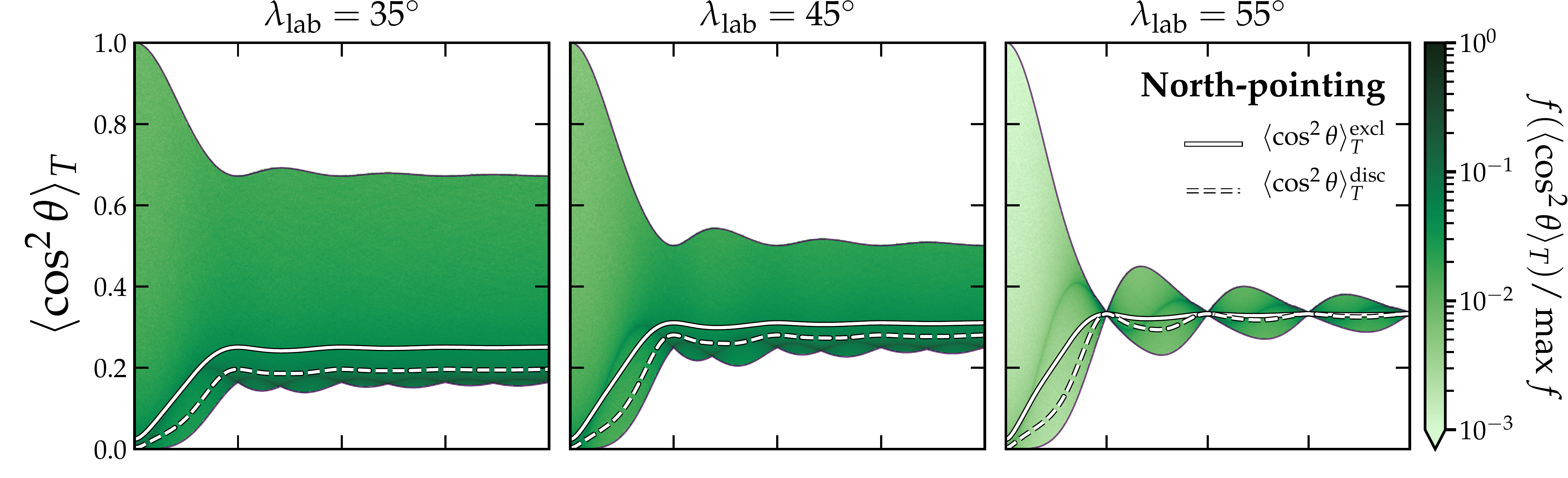}
    \includegraphics[trim = 0mm 0mm 0mm 0mm, clip, width=0.9\textwidth]{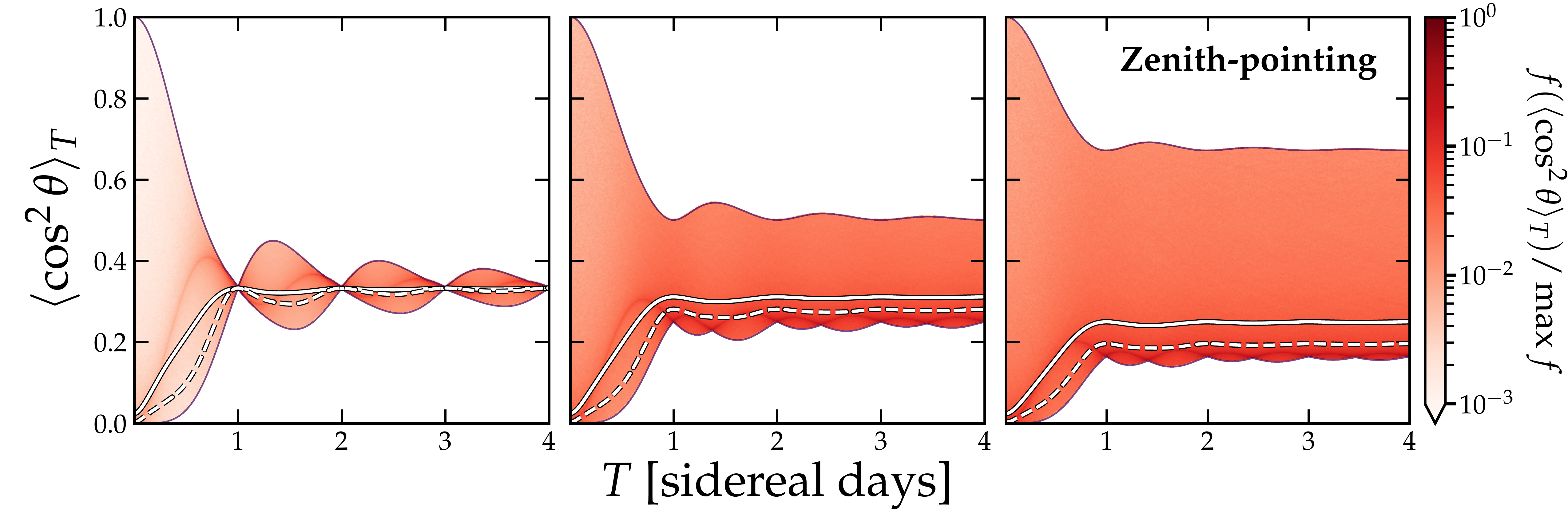}
\caption{Distributions of $\langle \cos^2{\theta(t)} \rangle_T$ after sampling the DP polarisation $(\theta_X, \phi_X)$ isotropically across the sky. For each value of $T$ we use the logarithmic colour-scale to show the value of the distribution, normalised by its maximum value. The three columns correspond to three different latitudes, $\lat = 35^\circ, 45^\circ$, and $55^\circ $, from left to right. The two rows correspond to North-pointing experiments (upper panels) and Zenith-pointing experiments (lower panels). Notice that at integer values of $T$ in sidereal days, the distribution approaches a single point at a value of $1/3$ in the upper right-most panel, and the lower left-most panel. These correspond to the peaks of the North-pointing and Zenith-pointing lines shown in the left-hand panel of Fig.~\ref{fig:LocationDependence}. The white lines correspond to our two conversion factors which we use to parameterise how much DP limits are impacted by the distribution of $\langle \cos^2{\theta}\rangle_T$ in the fixed polarisation scenario, as described in Sec.~\ref{sec:S2N}. The dashed line is $\excl$, as defined in Eq.(\ref{eq:excl}), and can be used to rescale \emph{exclusion} limits. Whereas the dot-dashed line is defined in Eq.(\ref{eq:disc}), and can be used to rescale \emph{discovery} limits.}
 	\label{fig:costh_axis}
\end{figure*}

\begin{figure*}
\centering
% 	%trim option's parameter order: left bottom right top
  \includegraphics[trim = 0mm 0mm 0mm 0mm, clip, width=0.9\textwidth]{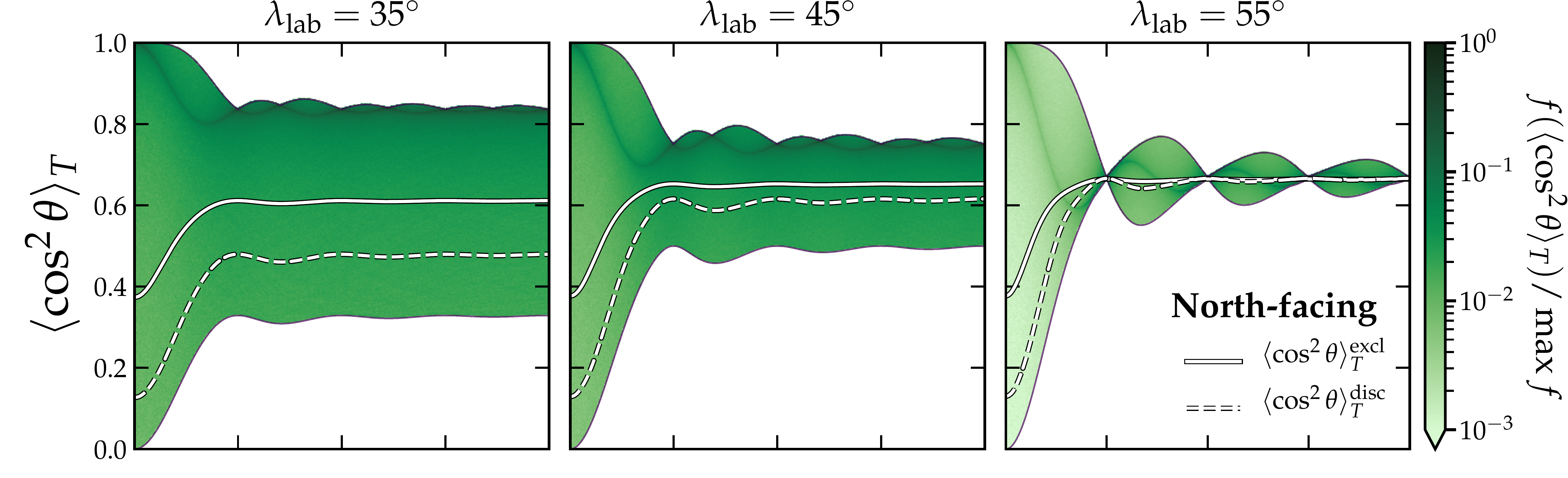}
    \includegraphics[trim = 0mm 0mm 0mm 0mm, clip, width=0.9\textwidth]{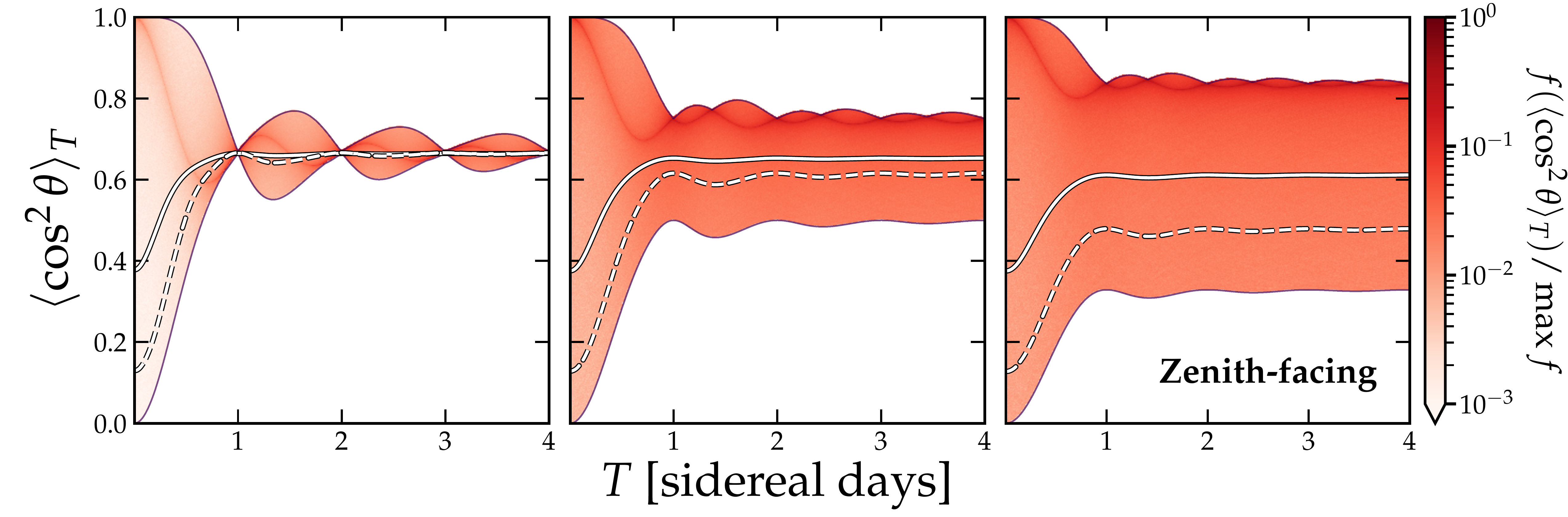}
\caption{As in Fig.~\ref{fig:costh_axis}, but now for planar experiments, i.e. those where the DP signal depends upon the angle between the DP polarisation and a plane. We show the distributions of $\langle \cos^2{\theta(t)} \rangle_T$ after sampling the DP polarisation $(\theta_X, \phi_X)$ isotropically across the sky. For each value of $T$ we use the colour-scale to display the distribution, which we have normalised by its maximum value. The three columns correspond to three different latitudes, $\lat = 35^\circ, 45^\circ$, and $55^\circ $, from left to right. The two rows correspond to North-facing (upper panels) and Zenith-facing (lower panels) experiments. Notice that at integer values of $T$ in sidereal days, the distribution approaches a single point at a value of $2/3$ in the upper right-most panel, and the lower left-most panel. These correspond to the peaks of the North-facing and Zenith-facing lines in the right-hand panel of Fig.~\ref{fig:LocationDependence}.}
 	\label{fig:costh_plane}
\end{figure*}

\subsection{Axial experiments}
To see one example of how the daily modulation plays out, we first consider an experiment in which the DP signal is proportional to the cosine squared of the angle with respect to the vertical (as is the case for all experiments in Tables~\ref{tab:axionhaloscopes} and~\ref{tab:DPexperiments} where the direction in the Directionality column is listed as ``Zenith-pointing''). This direction, expressed in our equatorial coordinate system $(\bfhat{x}_e,\bfhat{y}_e,\bfhat{z}_e)$ is, as a function of time,
\begin{equation}
\Zenith(t)=\left(\begin{array}{c}
\cos{\lat}\cos{\omega_\oplus t} \\
\cos{\lat}\sin{\omega_\oplus t} \\
\sin{\lat}
\end{array}\right) \, 
\end{equation}
where $\omega_\oplus = 2\pi/{\rm 1 \, day}$ is the angular frequency of the Earth's rotation.\footnote{Note that these signals will have a period equal to one sidereal day, which is 23 hours and 56 minutes long, as opposed to the solar day which is 4 minutes longer. This nuance is unimportant for our analysis, but it would be an important part of testing potential signals.} Since we are only going to take time-averages, we have neglected the temporal phase that describes the longitude and the local time of observation.

If we now define a direction for the DP's polarisation in equatorial coordinates,
\begin{equation}
\bfhat{X}=\left(\begin{array}{c}
\sin{\theta_X}\cos{\phi_X} \\
\sin{\theta_X}\sin{\phi_X} \\
\cos{\theta_X}
\end{array}\right) \, 
\end{equation}
then the angle we are interested in is,
\begin{equation}
    \cos^2{\theta(t)} = \big(\bfhat{X}\cdot\Zenith(t)\big)^2 \, .
\end{equation}
The DP signal power accumulated over a measurement time, $T$, is proportional to the time-average,
\begin{equation}
    \langle \cos^2{\theta(t)} \rangle_T \equiv \frac{1}{T}\int_0^T \cos^2{\theta(t)} \mathrm{d}t \, .
\end{equation}
As discussed in the previous section, the statistics of this quantity when sampling ($\theta_X,\phi_X)$ across the sky will be what influences the sensitivity of the experiment. It can be shown that,
\begin{equation}
   \frac{1}{4\pi}\int \langle \cos^2{\theta(t)} \rangle_T \, \drm \cos{\theta_X} \, \drm \phi = \frac{1}{3}
\end{equation}
for all $T$, meaning the squared polarisation component when averaged over the sky is always $1/3$ regardless of how long the observation time is. 

This result has implications for the exclusion and discovery thresholds that we defined in Sec.~\ref{sec:derivinglimits}. The factors $\excl$ and $\disc$ must be below the mean of the distribution, as they describe how much the power threshold needs to be enhanced due to the distribution of $\langle \cos^2{\theta(t)} \rangle_T$. Therefore the maximum that $\excl$ and $\disc$ can ever reach will be when the distribution is very tightly contained around the mean. This means that for axial experiments, $\excl$ will be at most 1/3, which happens to be the value for the random polarisation scenario. So when we are optimising our experiment \emph{the best we could hope to do is to make the limit the same for both DP polarisation scenarios.}

However, although the all-sky mean of $\langle \cos^2{\theta(t)} \rangle_T$ is independent of the location and $T$, its \emph{distribution} will vary greatly, as we will now show. To simplify things further, let us consider $\langle \cos^2{\theta(t)} \rangle_T$ when $T = 1$~sidereal day (or any integer number). Since in this case the detector axis has spun around $2\pi$ in the $(\bfhat{x}_e,\bfhat{y}_e)$ plane, we expect any dependence on $\phi_X$ (which is the angle defined in this plane) to drop out,
\begin{align}
   \langle \cos^2{\theta(t)} &\rangle_{T = n\,{\rm days}} = \nonumber \\
   \frac{1}{8}& \big(3 + \cos{2\theta_X} - (1 + 3 \cos{2\theta_X}) \cos{2\lat}\big) \, .
\end{align}
So $\langle \cos^2{\theta(t)} \rangle_{T = n\,{\rm days}}$ is bounded between $(1~+~\cos{2\lat})/4$ and $(1~-~\cos{2\lat})/2$ with some distribution. This distribution is straightforward to compute numerically since $\cos{\theta_X}$ is drawn from a uniform distribution between $[-1,1]$.

We can do the same thing if we were interested in West or North-pointing experiments. For these cases we write the detector axes in equatorial coordinates in a similar fashion,
\begin{equation}
    \West(t)=\left(\begin{array}{c}
\sin{\omega_\oplus t} \\
-\cos{\omega_\oplus t} \\
0
\end{array}\right) \, ,
\end{equation}
and,
\begin{equation}
 \North(t)=\left(\begin{array}{c}
\sin{\lat}\cos{\omega_\oplus t} \\
-\sin{\lat}\sin{\omega_\oplus t} \\
\cos{\lat}
\end{array}\right) \, .
\end{equation}
Repeating the calculation for these two directions leaves us with three expressions for $\langle \cos^2{\theta(t)} \rangle_{n-\mathrm{days}}$ that hold whenever $T$ is an integer number of sidereal days:
\begin{widetext}
\begin{equation}\label{eq:costh_day_axis}
    \langle \cos^2{\theta(t)} \rangle_{n-\mathrm{days}} =\left\{\begin{array}{ll}
\frac{1}{8} \big(3 + \cos{2\lat} + (1 + 3 \cos{2\lat}) \cos{2\theta_X}\big) & \text {\bf North} \\
\frac{\sin^2{\theta_X}}{2} & \text {\bf West} \\
\frac{1}{8} \big(3 + \cos{2\theta_X} - (1 + 3 \cos{2\theta_X}) \cos{2\lat}\big) & \text {\bf Zenith}
\end{array}\right. \, .
\end{equation}
\end{widetext}

Interestingly, we notice that one can choose the value of $\lat$ such that the $\theta_X$ dependence drops out. In these instances, the value of $\langle \cos^2{\theta(t)} \rangle_{n-\mathrm{days}}$ is the same for every possible DP polarisation. The distribution $f(\langle \cos^2{\theta}\rangle_T)$ is a delta function, so these cases will be when $\excl$ and $\disc$ are the largest possible. Inspecting Eq.(\ref{eq:costh_day_axis}), we can see that this occurs when,
\begin{align}\label{eq:locationdependence}
   {\bf North:} &\quad \lat = \frac{1}{2} \cos^{-1}{\bigg(-\frac{1}{3}\bigg)} \approx \pm 54.74^\circ \, , \nonumber \\
       {\bf West:} &\quad \text{no latitude dependence,}   \\
  {\bf Zenith:} &\quad \lat = \frac{1}{2} \cos^{-1}{\bigg(\frac{1}{3}\bigg)} \approx \pm 35.26^\circ \, .  \nonumber
\end{align}
For experiments that use measurements that are on the order of days in length, these latitudes will clearly be preferential.\footnote{Conveniently, all the experiments from Tables~\ref{tab:axionhaloscopes} and~\ref{tab:DPexperiments} are located within $\sim10^\circ$ of one of the latitudes that maximise the North and Zenith-pointing sensitivities} Since a West-pointing experiment always points in the direction that the Earth is rotating, it will always do a $2\pi$ loop in space once every sidereal day, independent of the latitude. It is worth remarking that the effect of being at a different latitude can be replicated by tilting the device by the relevant angle, however many experiments are unlikely to have this level of freedom.

We show the full latitude-dependence of the distribution of $\langle \cos^2{\theta(t)} \rangle^{\rm excl.}_{n-\mathrm{days}}$ in the left-hand panel of Fig.~\ref{fig:LocationDependence}. We see that the preferred latitudes from Eq.(\ref{eq:locationdependence}) are the latitudes that maximise $\excl$. Also note that when a Zenith-pointing experiment is located at the North/South poles, or when a North-pointing experiment is located at the Equator, the resulting value of $\langle \cos^2{\theta(t)} \rangle^{\rm excl.}_T$ converges the instantaneous value of 0.025, since those instruments will not rotate with respect to $\bfhat{X}$.

Next, we show the full distribution, $f(\langle \cos^2{\theta(t)} \rangle_T)$, as a function of $T$ by sampling over all DP polarisations. In Fig.~\ref{fig:costh_axis}, we show these distributions for the North and Zenith cases, and for three example latitudes. We can see that for $\lat\approx 55^\circ$ in the North-pointing case, and $\lat\approx35^\circ$ in the Zenith-pointing case, the distribution converges on a single value ($1/3$) for integer-day-long experiments. This coincides with the preferential locations observed in Fig.~\ref{fig:LocationDependence}.

\subsection{Planar experiments}
We also wish to calculate the signal for experiments that are sensitive to any polarisation component lying along a plane. So for example, to calculate the angle $\theta$ with respect to the plane defined by the $\Zenith$-$\West$ axes, we simply take the complement of the angle with respect to the axis perpendicular to that plane i.e.,
\begin{equation}
\cos{\theta(t)} = \sqrt{1-\big(\bfhat{X}\cdot\North(t)\big)^2} \, ,
\end{equation}
To inspect how this angle behaves further we simplify in the same way as before by taking the average over an integer number of sidereal days,
\begin{widetext}
\begin{equation}
    \langle \cos^2{\theta(t)} \rangle_{n-\mathrm{days}} =\left\{\begin{array}{ll}
\frac{1}{8} \big(5 - \cos{2\lat} - (1 + 3 \cos{2\lat}) \cos{2\theta_X}\big) &  \text {Zenith-West \quad\, (perpendicular to floor)} \\
\frac{1}{4}(3+\cos{2\theta_X}) & \text {North-Zenith \quad (perpendicular to floor)} \\
\frac{1}{8} \big(5 + \cos{2\theta_X} + (3 \cos{2\lat}-1) \cos{2\theta_X}\big) & \text {North-West \quad\,\,\, (parallel to floor)}
\end{array}\right.
\end{equation}
\end{widetext}
To make discussing these planes less confusing, we refer to them as ``North-facing'', ``West-facing'', and ``Zenith-facing'', respectively, and the axial experiments as ``North-pointing'' etc. We find that we have an identical latitude preference as for the angles with respect to an axis, 
\begin{align}\label{eq:locationdependence2}
   \text{\bf North-facing} &\quad \lat = \frac{1}{2} \cos^{-1}{\bigg(-\frac{1}{3}\bigg)} \approx \pm 54.74^\circ \, , \nonumber \\
       \text{\bf West-facing:} &\quad \text{no preferred latitude,}   \\
 \text{\bf Zenith-facing:} &\quad \lat = \frac{1}{2} \cos^{-1}{\bigg(\frac{1}{3}\bigg)} \approx \pm 35.26^\circ \, ,  \nonumber
\end{align}
We show the full latitude dependence of $\excl$ in the right-hand panel of Fig.~\ref{fig:LocationDependence}. Notice that in this case the average angle with respect to a plane at any one time is $2/3$, meaning that at the preferential latitudes the entire distribution of $\langle \cos^2{\theta(t)} \rangle_{n-\mathrm{days}}$ converges on this value.

As we did for the axial case, we also show the full distribution of $\langle \cos^2{\theta(t)} \rangle_T$ as a function of $T$, and for three latitudes, in Fig.~\ref{fig:costh_plane}. The behaviour is very similar, with the 35$^\circ$ and $55^\circ$ latitude cases having singular points at integer values of $T$ for the Zenith-facing and North-facing planes respectively.

\subsection{Reinterpreting dark photon limits}
\begin{figure*}
\centering
% 	%trim option's parameter order: left bottom right top
  \includegraphics[trim = 0mm 0mm 0mm 0mm, clip, width=1\textwidth]{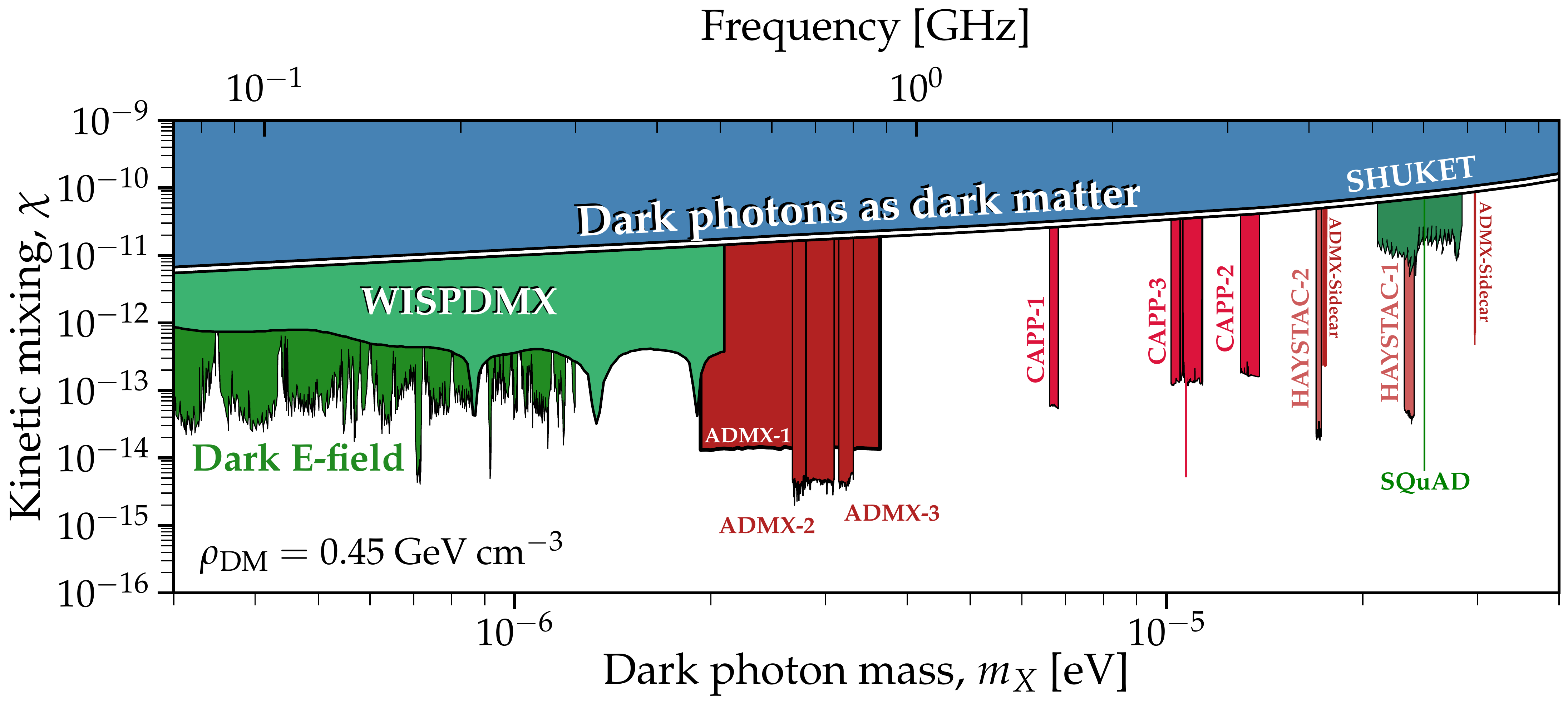}
\caption{Closeup of bounds on DPs in the radio-microwave frequency regime. The limits set by axion cavity haloscopes are in shades of red, whereas the limits from devoted DP experiments operating in the same range of frequencies are shown in shades of green. We impose an upper limit on DPs as DM from the bounds of Refs.~\cite{Arias:2012az,Witte:2020rvb} (the same upper limit as shown in Fig.~\ref{fig:bounds}). All of the experimental bounds shown here have been rescaled from the original sources. Firstly, they have been rescaled such that they all assume the same DM density of $\rho_0 = 0.45$~GeV~cm$^{-3}$. Secondly, we have rescaled them such that they all consider the fixed DP polarisation scenario. This relies on the factor $\excl$, which is different for each experiment. The method of deriving these factors is detailed in Sec.~\ref{sec:derivinglimits}, and the result for each experiment is listed in the final columns of Tables~\ref{tab:axionhaloscopes} and~\ref{tab:DPexperiments}. }
 	\label{fig:bounds_closeup}
\end{figure*}

\begin{figure*}
\centering
% 	%trim option's parameter order: left bottom right top
  \includegraphics[trim = 0mm 0mm 0mm 0mm, clip, width=0.49\textwidth]{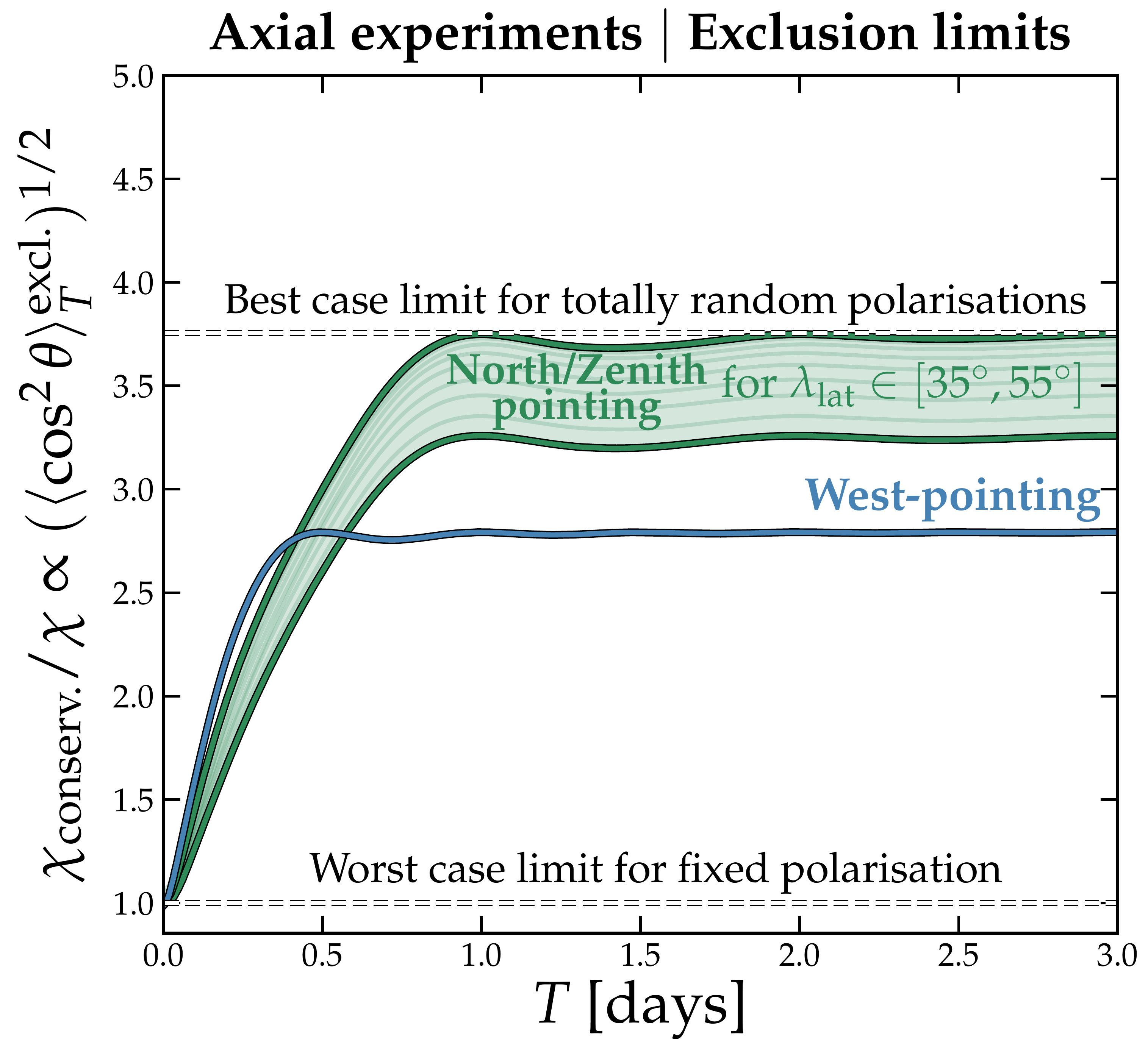}
    \includegraphics[trim = 0mm 0mm 0mm 0mm, clip, width=0.49\textwidth]{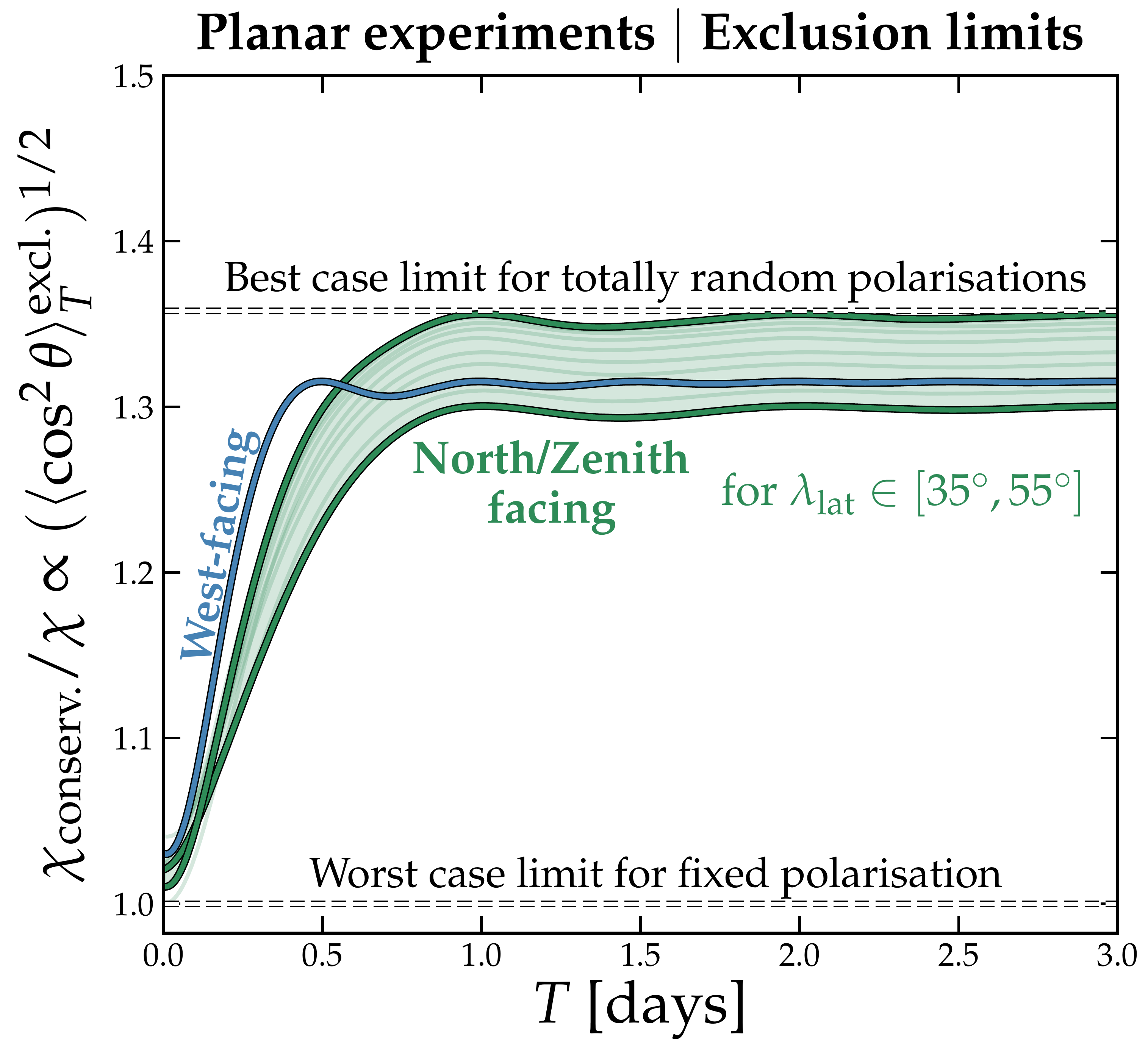}
      \includegraphics[trim = 0mm 0mm 0mm 0mm, clip, width=0.49\textwidth]{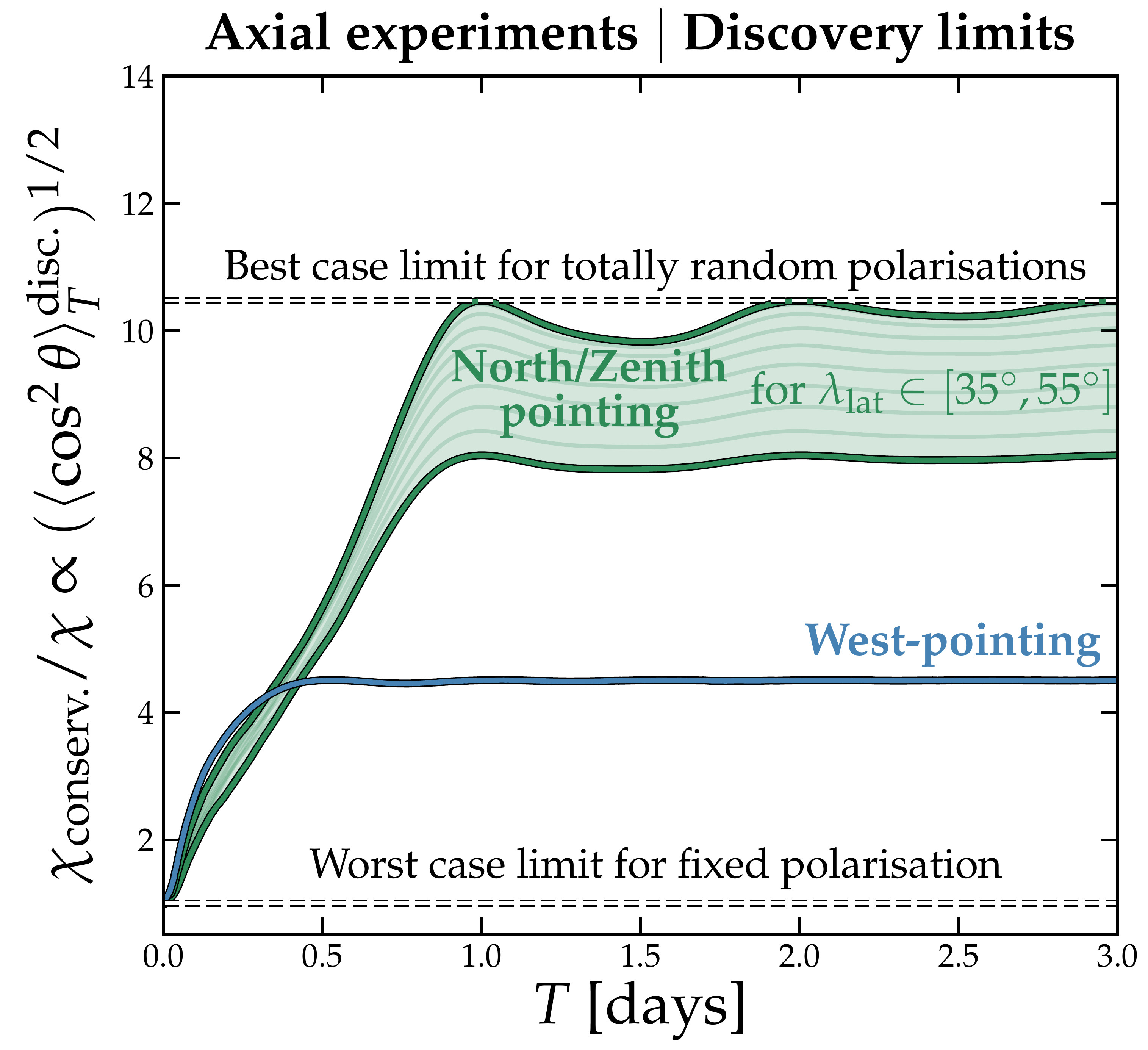}
    \includegraphics[trim = 0mm 0mm 0mm 0mm, clip, width=0.49\textwidth]{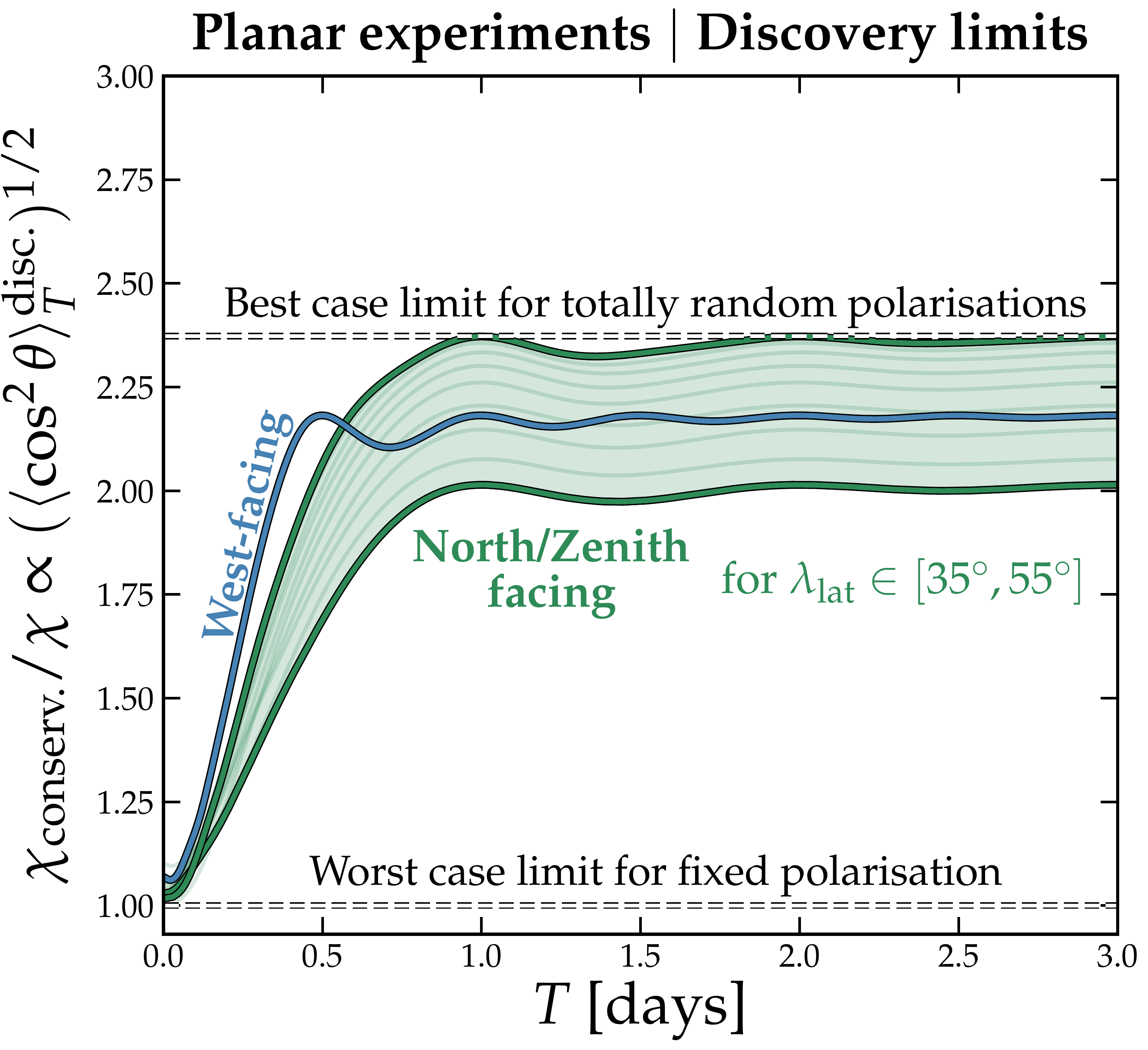}
\caption{We display the improvement that can be made to both exclusion limits (top row) and discovery limits (bottom row) on the DP's kinetic mixing parameter, $\chi$, as a function of the measurement time, $T$, and for several experiment geometries and orientations. The left-hand panels correspond to axial experiments, whereas the right-hand panels correspond to planar experiments. In all cases we display the improvement made to the limit relative to the most conservative possible assumption ($\chi_{\rm conserv.}$): ignoring daily modulation and taking the instantaneous results for $\excl$ and $\disc$ when $T\rightarrow 0$. In the scenario that the DP has a random polarisation every coherence time, $\langle\cos^2{\theta}\rangle_T$ effectively takes on a single value of $1/3$ (axial) or $2/3$ (planar), and is the best the experiment could do. What this figure shows is that by properly accounting for the daily modulation and how that changes the distribution of $\langle\cos^2{\theta}\rangle_T$, the sensitivity in the fixed-polarisation scenario can be significantly improved---and for certain cases it can even match the random-polarisation scenario. The potential improvement could be up to a factor of $\sim$3.75 or $\sim$1.36 in the exclusion limits, and $\sim$10.5 or $\sim$2.37 in the discovery limits.}
 	\label{fig:improvement}
\end{figure*}

Now that we know how to calculate the distributions of $\langle \cos^2{\theta}\rangle_T$, we can use this knowledge to estimate what $\excl.$ should be for past experiments. The results are listed in the final columns of Tables~\ref{tab:axionhaloscopes} and~\ref{tab:DPexperiments}. This means we can now do what we hinted at in Sec.~\ref{sec:DPdetection}, and reinterpret previous exclusion limits in the context of the fixed polarisation scenario. 

For a concrete example of how this is done, we consider the prototypical axion/DPDM experiment: the cavity haloscope. The power is often written in terms of the physical quantities of the cavity: volume $V$, quality factor $Q$, external applied $B$-field ${\bf B}$, and coupling factor $\kappa$,
\begin{subequations}
\begin{align}
\text{\bf Dark photons:} &\quad  P_{\rm cav} = \kappa {\cal G}_X V Q\rho_{\rm DM} \chi^2m_X \,, \\
\text{\bf Axions:} &\quad P_{\rm cav} = \kappa {\cal G}_a V \frac{Q}{m_a} \rho_{\rm DM} g_{a\gamma}^2B^2 \,, \label{eq:sikivieoverlap}
\end{align}
 \end{subequations}
where $g_{a\gamma}$ is the dimensionful axion-photon coupling, and $m_a$ is the axion mass. The geometry factors $\mathcal{G}_{X,a}$ are given by
\begin{subequations}
\begin{align}
\text{\bf Dark photons:} &\quad {\cal G}_X = \frac{\left(\int dV\,{\bf E}_\alpha\cdot {\bf \hat X}\right)^2}{V \frac{1}{2}\int dV\,\epsilon({\bf x}){\bf E}_\alpha^2+{\bf B}_\alpha^2},\\
\text{\bf Axions:} &\quad{\cal G}_a = \frac{\left (\int dV\,{\bf E}_\alpha\cdot {\bf  B}\right )^2}{V B^2\frac{1}{2}\int dV\,\epsilon({\bf x}){\bf E}_\alpha^2+{\bf B}_\alpha^2}. \label{eq:geometryfactor}
\end{align}
 \end{subequations}
Here we have denoted the field of the $\alpha$ mode of a cavity via ${\bf E}_\alpha,{\bf B}_\alpha$. It is simple to convert between the two expressions for the cavity power. So to recast a limit on $g_{a\gamma}$ to one on $\chi$ we just replace~\cite{Arias:2012az},
\begin{equation}
 \chi = g_{a\gamma}\frac{B}{m_X|\cos\theta|}\,,
\end{equation}
where $\cos\theta={\bf \hat X}\cdot{\bf \hat B}$. 

In fact, this statement holds generally for all the experiments we consider, as long as $\theta$ is defined with respect to the appropriate axis or plane. One caveat however is that this conversion assumes a constant magnetic field over the cavity, which may not always be the case. The differences in most experiments may be small, but designs such as the original Orpheus proposal~\cite{Rybka:2014cya}\footnote{Note that the name has been reused by the ADMX collaboration for a dielectric loaded resonator~\cite{Carosi:2020akt}.} would actually be very insensitive to DPs, as it would have employed an oscillatory magnetic field. We stress that deriving fully accurate limits requires dedicated calculations of the geometry factors---relying on this simple recasting may not always be sufficient.

Figure~\ref{fig:bounds_closeup} shows limits on DPDM in the fixed polarisation scenario that are the most accurate ones to date. We have zoomed in on the radio-microwave range where the majority of the experimental activity takes place. Though these limits are still not fully optimised for each experiment, they are consistent in their assumptions, which was not true prior to this. As we mentioned in Sec.~\ref{sec:derivinglimits}, the approach adopted for handling this scenario was to simply take the 5th percentile of the distribution: $\langle \cos^2{\theta}\rangle_{T\rightarrow 0}=0.0025$ or $0.0975$~\cite{Arias:2012az}.\footnote{We note that there was also subtle plotting error for the reinterpreted axion limits presented in Ref.~\cite{Arias:2012az}, which has been fixed in more recent work~\cite{Gelmini:2020kcu,Ghosh:2021ard}.} Looking at the final columns of Tables~\ref{tab:axionhaloscopes} and~\ref{tab:DPexperiments}, we can see that these values are certainly overly conservative for many experiments, especially those integrating for longer than a few hours. 

The factors of $\langle \cos^2{\theta}\rangle^{\rm excl.}_T = 1/3$ (axial) or $2/3$ (planar), obtained under the random polarisation scenario, would enhance these bounds by $\sim 3.75$ or $1.36$ respectively. As we discussed in Sec.~\ref{sec:DPcosmology}, this can only be a valid assumption for certain DP production mechanisms. We stress therefore, that if these values are chosen, an accompanying statement must be made about the requirement this places on the DP production mechanism. Due to the large differences in these values, such a statement would be more important than one about, say, the assumed DM density, which also varies between publications but whose differences lead to discrepancies of factors around $\sim$1.2.

In Fig.~\ref{fig:bounds_closeup}, we rescaled each limit so that they apply for $\rho_{\rm DM} = 0.45$~GeV~cm$^{-3}$. This value is what the axion direct detection community has adopted since Ref.~\cite{Asztalos:2001jk} in 2002.\footnote{This unusual choice seems to stem from the fact that Ref.~\cite{Asztalos:2001jk} presented their exclusion limits as a function the DM density for specific QCD axion models, as opposed to the other way around as is now convention. KSVZ axions were ruled as contributing more than 0.45~GeV~cm$^{-3}$ in 2002 and that appears to have been adopted in subsequent experimental analyses which presented limits for a fixed value of $\rho_{\rm DM}$, despite the fact they were at different frequencies. However, we are not certain if this interpretation of history is correct.} However a value of 0.3 has been the standard over many years in other direct detection communities. The other experiments shown here: Dark E-field, SHUKET, and WISPDMX all state (or at least imply) that they have chosen a value of 0.3 (with the exception of SQuAD who chose 0.4). 

To see even more clearly the difference in sensitivity between the fixed polarisation scenario, and the randomised polarisation scenario, we use Fig.~\ref{fig:improvement}. Since the sensitivity to $\chi$ scales as $\sim (\langle \cos^2{\theta} \rangle^{\rm excl.}_T)^{-1/2} $ we have plotted this against the measurement time for both axial and planar experiments, but rescaled relative to the instantaneous value $\langle\cos^2{\theta} \rangle^{\rm excl.}_{T\rightarrow0} = 0.025$ or 0.37. The difference between the most optimistic and most pessimistic assumptions is around a factor 3.75 for the axial case, and around a factor of 1.36 for the planar case. For the discovery limits however, because these factors are much more sensitive to the low tails of $f(\langle \cos^2{\theta}\rangle$, we reach improvement factors up to $10.5$ and $2.37$ for axial and planar experiments respectively. 

So in summary, simply using a certain value of $T$ can bring the fixed DP scenario limits closer, and even equal to, the randomised case. In particular, for the cases when the experiment is placed at the optimal latitude and integrates over integer-day-long measurement times, the sensitivity under the two scenarios is \emph{the same}.

\subsection{Making multiple measurements}

\begin{figure*}
\centering
% 	%trim option's parameter order: left bottom right top
  \includegraphics[trim = 0mm 0mm 0mm 0mm, clip, width=0.49\textwidth]{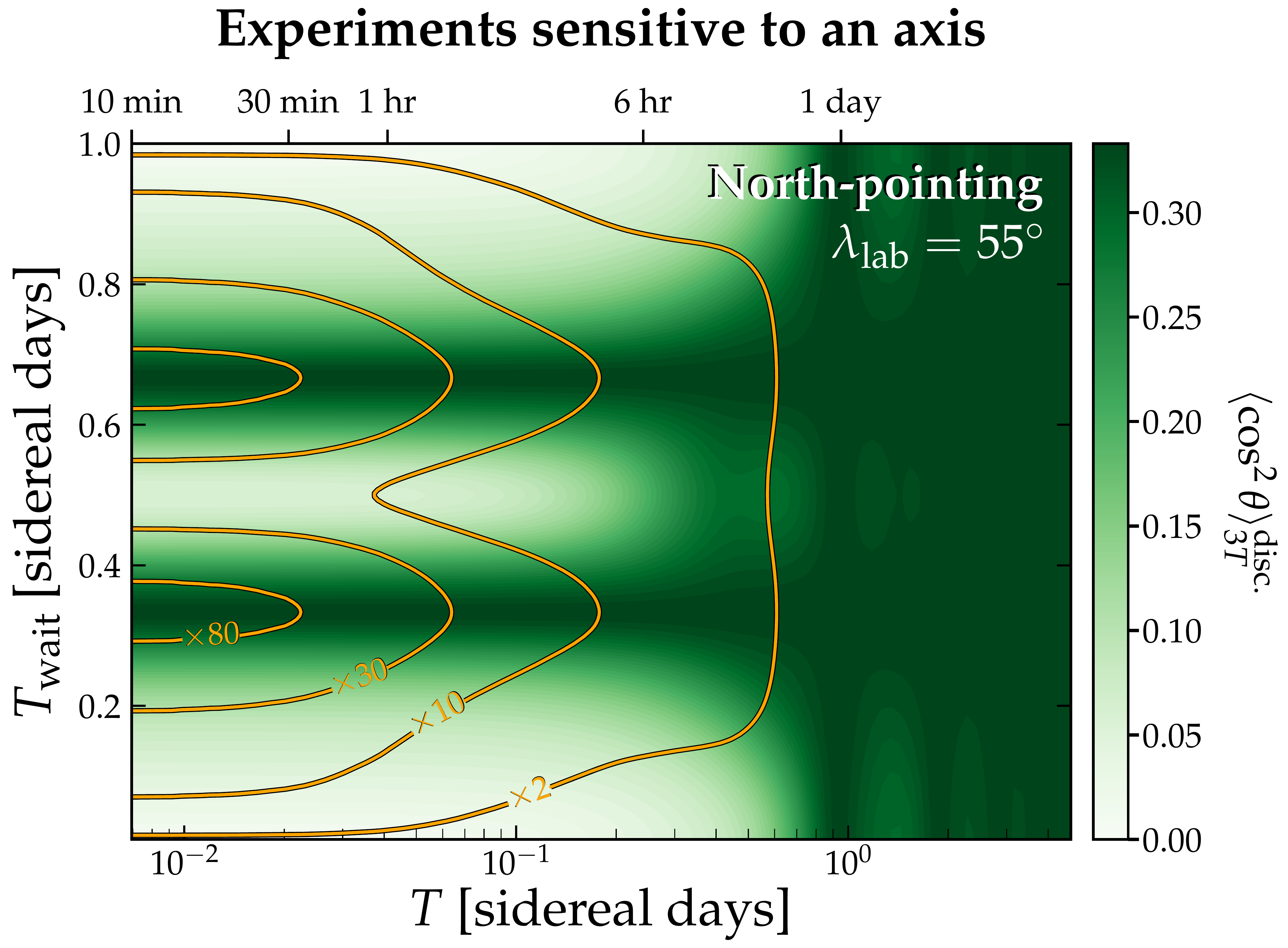}
  \includegraphics[trim = 0mm 0mm 0mm 0mm, clip, width=0.49\textwidth]{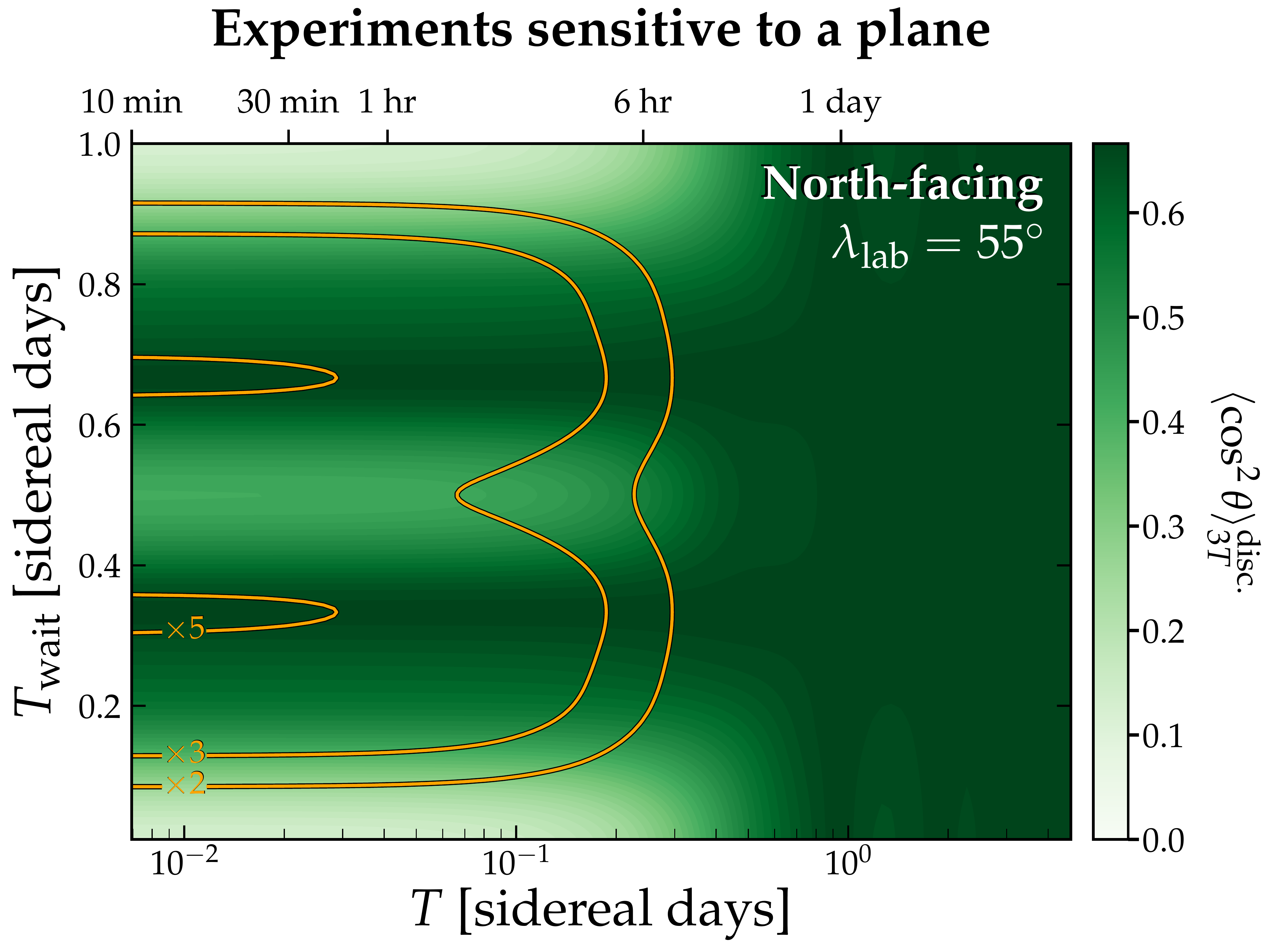}
  \includegraphics[trim = 0mm 0mm 0mm 0mm, clip, width=0.49\textwidth]{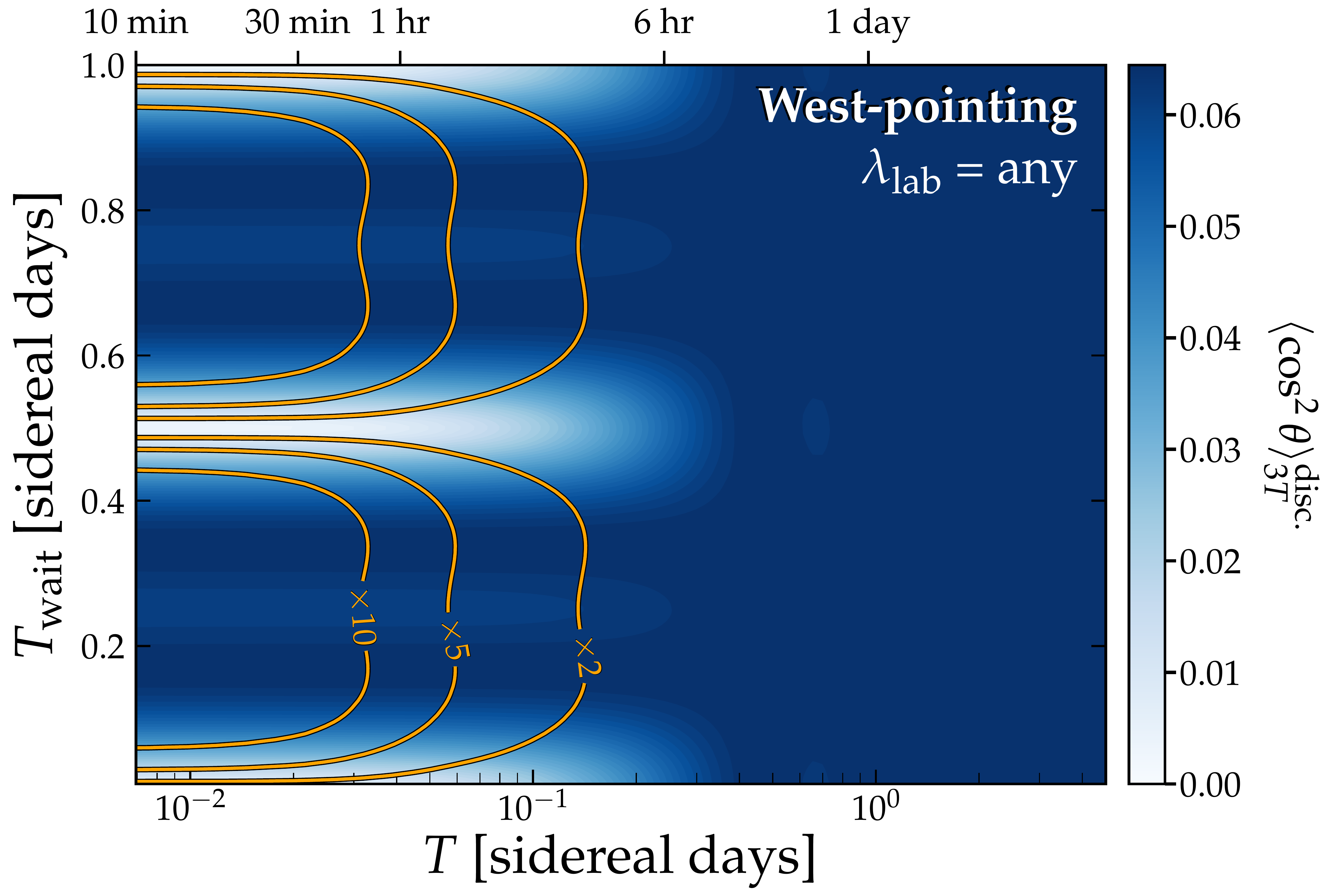}
  \includegraphics[trim = 0mm 0mm 0mm 0mm, clip, width=0.49\textwidth]{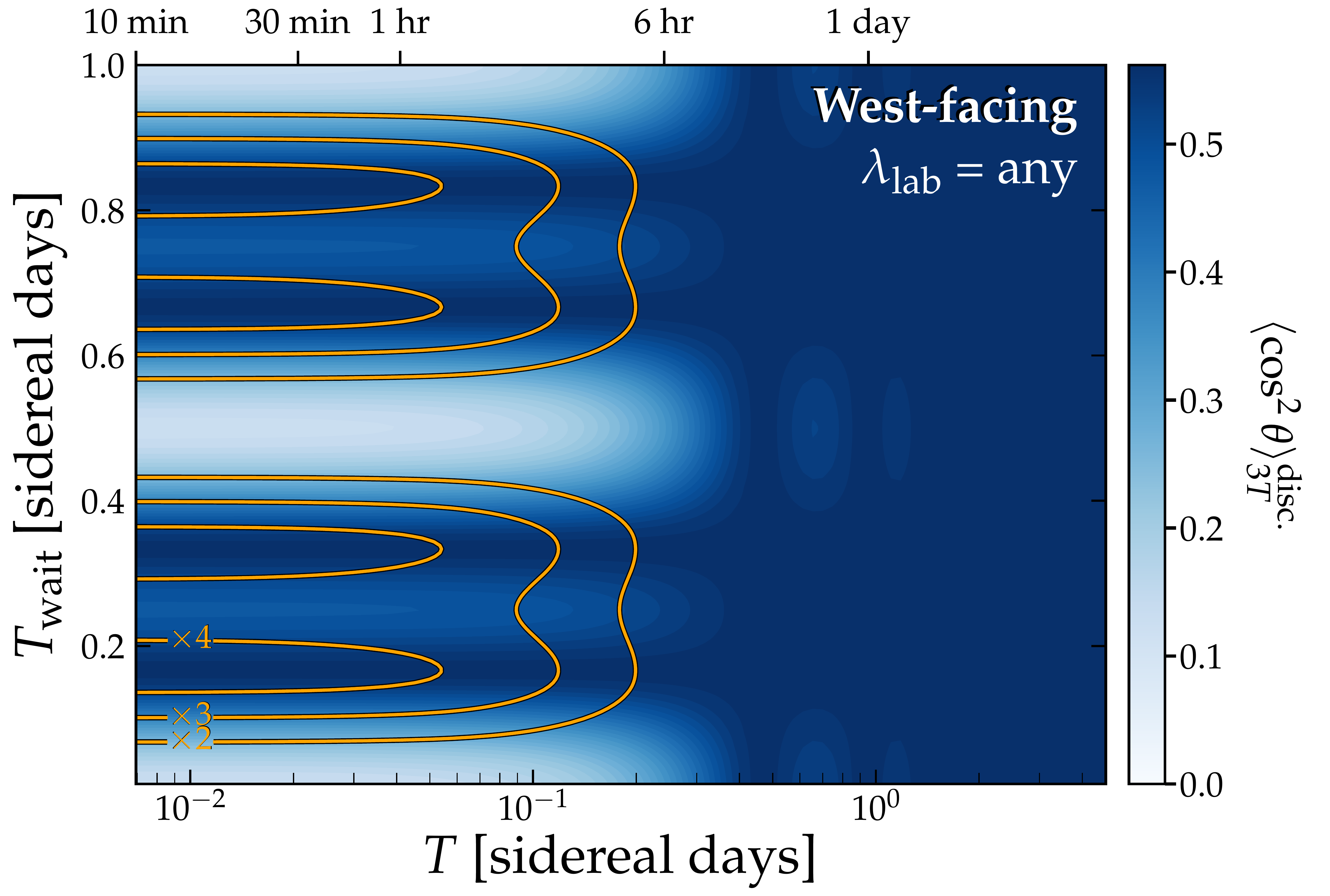}
  \includegraphics[trim = 0mm 0mm 0mm 0mm, clip, width=0.49\textwidth]{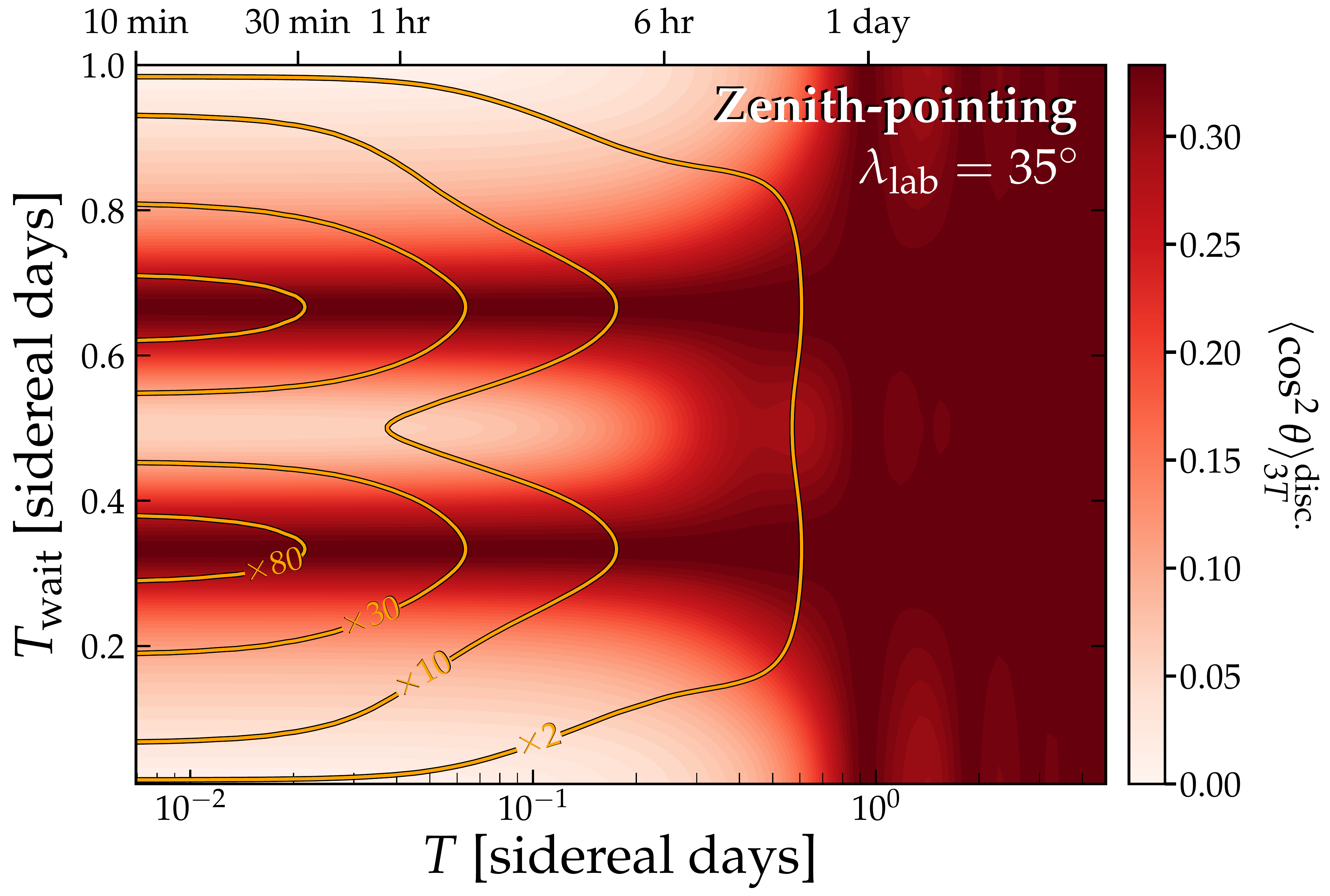}
  \includegraphics[trim = 0mm 0mm 0mm 0mm, clip, width=0.49\textwidth]{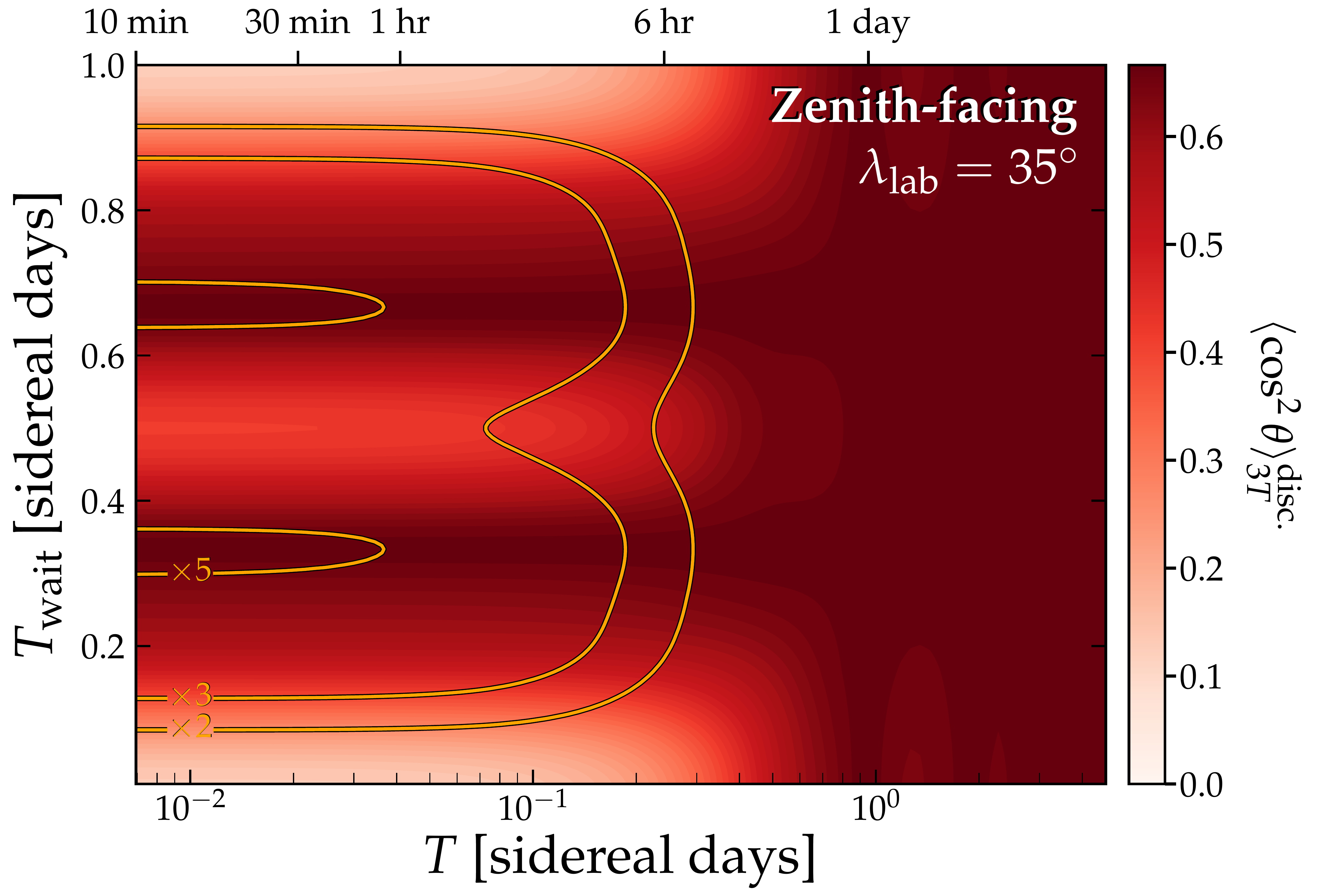}
\caption{In each panel the colour-scale refers to the value of $\langle \cos^2{\theta}\rangle^{\rm disc.}_{3T}$ obtained from \emph{three} measurements, each of duration $T$, and each separated by time $T_{\rm wait}$. The left column corresponds to axial experiments, and the right column to planar experiments. The top row is for North-pointing/facing experiments located at their optimal latitude of $\sim55^\circ$, the middle row is for West-facing/pointing experiments which can be located at any latitude, and finally the bottom row is for Zenith-pointing/facing experiments at the optimal latitude of $\sim35^\circ$. Note that in all cases the distribution is periodic in $T_{\rm wait}$ over one sidereal day, so there is no need to extend the plots vertically. The orange contours with labels enclose the region of the $T$-$T_{\rm wait}$ space where that choice of $T_{\rm wait}$ leads to an enhancement $\langle \cos^2{\theta}\rangle^{\rm disc.}_{3T}/\langle \cos^2{\theta}\rangle^{\rm disc.}_{T}$ by at least the amount shown.} 
 	\label{fig:costh_rescanned}
\end{figure*}

As we hinted at in Sec.~\ref{sec:S2N}, the sensitivity of a DP search can be greatly improved by a strategic choice of the time at which to do a repeat measurement, or rescan. For the same reason, the sensitivity can also be greatly \emph{harmed} if the time of rescan is poorly chosen. Fortunately, the rotation of the Earth is something we can plan for. While prior to now, no strategy of this kind of strategy seems to have not been adopted in any direct search for DPs. Nevertheless, we can detail some examples of better-practice here, should a future experiment wish to adopt a more strategic measurement schedule. Before we begin, it is worth emphasising that there are two distinct possibilities that are relevant here. One is if the experiment is repeating measurements to \emph{test} a candidate signal, and the other is if the repeat measurements are simply to obtain more data to enhance the \emph{sensitivity}. As we will see these two cases require very different strategies to be put in place.

We begin by picking up where the discussion in Sec.~\ref{sec:S2N} left off, namely the case where we have made a measurement of duration $T$, and we wish to add statistics to that measurement by repeating it, some time $T_{\rm wait}$ later. From what we have seen so far, there are scenarios where this measurement could be disastrous for the detection of the DP. Say we have a West-pointing experiment that happened to get very lucky and picked up the DP polarisation in perfect alignment with the instrument. If the experimentalists decided to check that measurement, say, 6 hours later, the polarisation would now be anti-aligned with the instrument and the signal will have disappeared. It would then be quite reasonable to chalk the event up to transient RF noise and toss the Nobel-prize winning signal in the bin with all the other radio stations.% i wouldn't blame you if you got rid of that last sentence

Of course, this is an extreme case, but the point is that the DP signal has large variations in time. These can benefit the experiment, but they can also cause signals to disappear when the alignment is bad. For this issue though the solution is simple: Test candidate DP signals at the same time of the sidereal day as they were originally measured.

Now, if we want to talk about maximising the potential to detect signals in the first place, we must flip this argument. The solution will be less straightforward, because it relies on the range of $\cos{\theta}$ swept about by a given experiment over a given time, but we can use our formulae derived in the previous subsections. We want to determine the best strategy for performing rescans in the \emph{absence} of any candidate signals. This issue is very relevant for many axion experiments which enhance their potential $S/N$ by means of rescans and subsequently stacking their data. In the context of DPs which are subject to strong daily modulations it can be even more relevant. With a judicious choice of $T_{\rm wait}$, it is possible to use the rotation of the Earth to game the statistics of $\langle \cos^2{\theta}\rangle_T$ in the experiment's favour. Note that this need not involve a substantial increase in the total time spent measuring a particular frequency.\footnote{The only increase in total measurement time comes from the additional time spent tuning. If such a time were short compared to the time spent measuring, then there is no real cost. Even if the tuning time is non-trivial, we will show that the gain is high enough that it is likely worth the additional time.} As long as the measurement is noise dominated (which is the case for all experiments considered here), the $S/N$ remains the same for any temporally constant signals, while the probability of a poor alignment can be significantly reduced for the temporally varying signals.

Imagine the following scenario: An experiment makes three measurements, each with a duration $T$. The start of each measurement is spaced out from the start of the previous one by a time $T_{\rm wait}$. We have already calculated the distribution $\langle \cos^2{\theta}\rangle_T$ for a single measurement of time $T$ (these are shown in Figs.~\ref{fig:costh_axis} and~\ref{fig:costh_plane}), but say we were to stack the data from the three measurements together and take it is one measurement, what is the distribution of $\langle \cos^2{\theta}\rangle_T$ then, and how does it depend on $T$ and $T_{\rm wait}$? Figure~\ref{fig:costh_rescanned} shows precisely this.

In Fig.~\ref{fig:costh_rescanned} we display $T$ on a logarithmic scale to show both the cases involving short sub-hour measurement times, as well as those lasting days. However, we show $T_{\rm wait}$ between 0 and 1 days, and on a linear scale, as these results are all periodic in that direction. This means that $T_{\rm wait}$ should be interpreted as the delay in the local time from the original measurement. For example if the original measurement took place at 00:00 then a value of $T_{\rm wait} = 0.5$~days, corresponds to any second measurement taking place at 12:00 (and then a third, the same time afterwards). It does not matter if that measurement is several days afterwards. Similarly, it does not matter what the exact time of the original measurement was, $T_{\rm wait} = 0.5$~days just means that there was 12 hours on the clock between them. Recall that we are considering sidereal days, i.e.~a shift of four minutes per calendar day is required to convert times of the sidereal day to the calendar day.

Since we are interested now in optimising future experiments, we need to ensure we are preparing them to have the best chance of discovering the DP, not just setting exclusion limits. Our discovery limit conversion factor $\disc$ is much more sensitive to the tails $f(\langle \cos^2{\theta}\rangle_T)$, so we can appreciate the effects of optimisation much more by focusing on this factor than $\excl$. The three panels on the left-hand side are for axial experiments, whereas the three on the right-hand side are for planar experiments. The colour-scale corresponds to the value of $\langle \cos^2{\theta} \rangle^{\rm disc.}_{3T}$. The orange contours enclose values of a different, but related, distribution: The enhancement in the value of $\langle \cos^2{\theta} \rangle^{\rm disc.}_{3T}$ that is gained from doing the repeat measurements. Keep in mind the quantity we are calculating is a time average, so any enhancement originates solely from the rotation of the Earth and not from the fact we are observing for longer.

Examining Fig.~\ref{fig:costh_rescanned}, we see that when $T\gtrsim1$~sidereal day, it does not matter when the next measurements start, since the experiment is already long enough that it samples all the DP polarisations it can. On the other hand, for very short $T$, the correct choice of $T_{\rm wait}$ makes a great difference. For instance, looking at the top-right panel, if the original measurement of the North-facing experiment was only 10 minutes, then simply choosing the next measurement to be at a time that was $\sim0.33$ or $\sim0.66$~sidereal days later would allow the power to be enhanced by a factor of 3, just from the factor of $\langle \cos^2{\theta} \rangle^{\rm disc.}_{3T}$ alone. 

The most dramatic cases are the North and Zenith-pointing experiments where a strategic timing of short measurements can lead to a power enhancement of over a factor of 100, leading to an enhancement in sensitivity to $\chi$ of a factor $\gtrsim$10. This is roughly the factor difference between the fixed and randomised polarisation scenarios that we saw in the lower panels of Fig.~\ref{fig:improvement}, which shows that three measurements is already enough to sample almost the whole distribution of possible polarisations. Since this factor is purely geometrical, the experiment gains an order of magnitude in sensitivity to $\chi$, with only a factor of three increase in measurement time. As a point of comparison, for an unmodulating DP signal, the sensitivity would scale as $\chi^{-1} \propto T^{1/2}$. Note again that for short measurement times, where such a technique is most appropriate, this could correspond to gaining a order of magnitude in sensitivity simply by dividing the measurement in three. 

Before moving on, we remark that we have also performed the same calculation for two measurements rather than three.\footnote{The equivalent figures for these cases can be found in the GitHub repository linked to this paper.} The qualitative trends are the same, however the enhancement factors are lower, gaining at most a factor of around 15 in $\langle \cos^2{\theta} \rangle^{\rm disc.}_T$. In the West-pointing and West-facing cases, the optimum time to do a rescan for sub-day-long measurements is around $\pm$6 hours, and does not improve for three measurements rather than two. Calling back to the discussion at the beginning of this subsection, this is the opposite strategy to what one would do to test for a candidate signal, however the reasoning is the same. With two short measurements 6 hours apart, the experiment is maximising its potential to capture a range of possible polarisation angles, but that is exactly what one should \emph{not} do to test for a signal seen in the first measurement.

It is unsurprising that the more individual measurements we make, the closer we can get to the optimal experiment which observes over the whole day and captures all possible DP polarisations. However, we found that three is already sufficient to get very close to the optimal case in the North and Zenith-pointing experiments, which represents a dramatic potential improvement in sensitivity for only a very minor reorganisation of the experimental data-taking. For planar and West-pointing experiments however, there is little difference between two and three measurements. A simple way to understand why is to consider that, when defining a volume, one needs either three axes or two planes. When an experiment is West-pointing, at most one can sweep out a disk, which requires only two axes to define.

\subsection{Optimising future experiments}\label{sec:optimising}
\begin{figure*}
\centering
% 	%trim option's parameter order: left bottom right top
  \includegraphics[trim = 0mm 0mm 0mm 0mm, clip, width=0.98\textwidth]{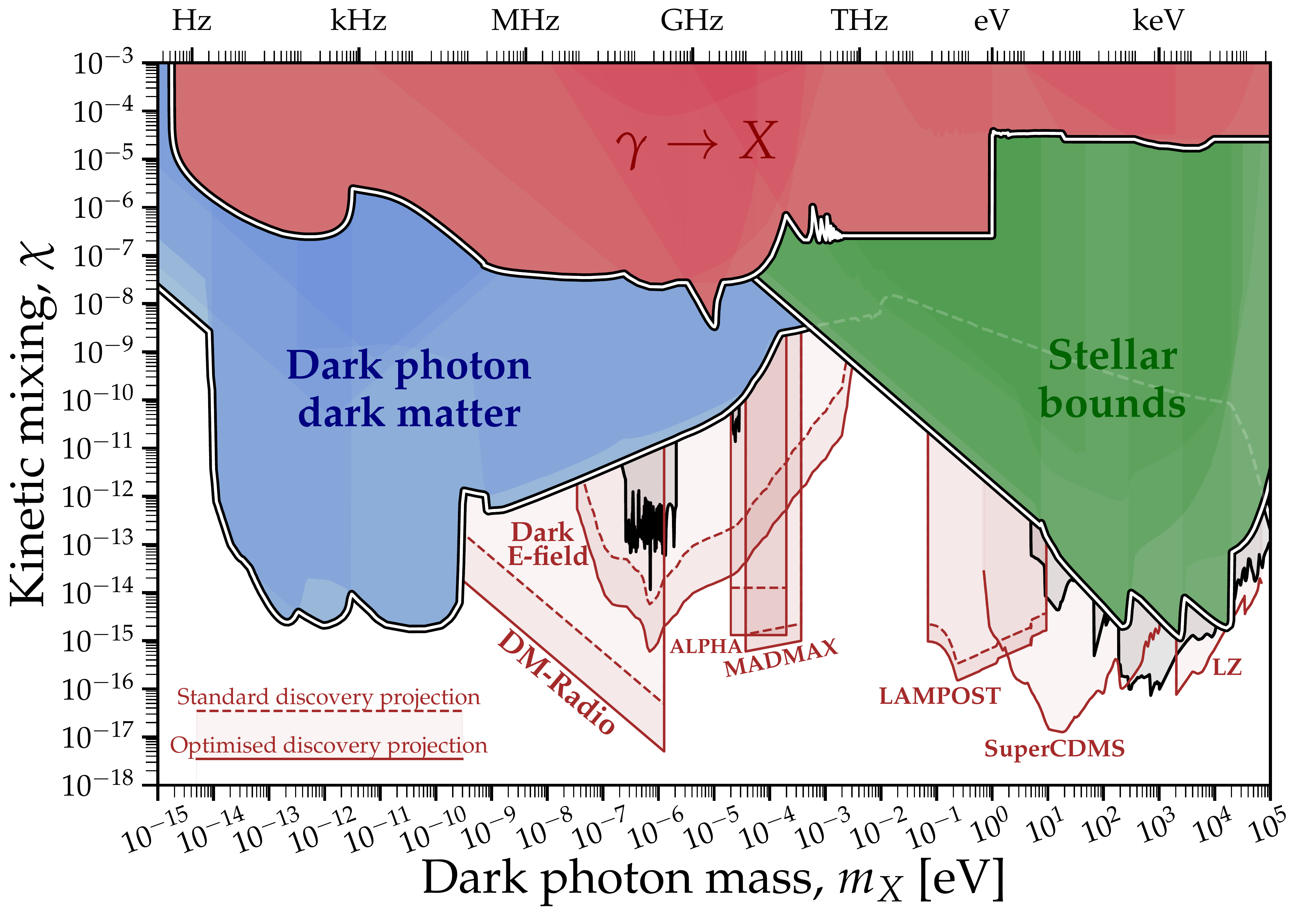}
\caption{Projected DP discovery limits for planned experiments, and future runs of existing experiments: DM-Radio~\cite{DMRADIO}, Dark E-field Radio~\cite{Godfrey:2021tvs}, ALPHA~\cite{Lawson:2019brd,Gelmini:2020kcu}, MADMAX~\cite{TheMADMAXWorkingGroup:2016hpc}, LAMPOST~\cite{Baryakhtar:2018doz}, SuperCDMS~\cite{Bloch:2016sjj} (assuming a Ge target), and LZ~\cite{Akerib:2021qbs}. We have plotted projections for two scenarios, the `standard' projection is the most pessimistic and assumes that no timing or directional information is taken into account when running the experiment, or calculating the limit. Our `optimised' projection would be obtained if the experiment followed some simple changes to the data-taking that are outlined in the text of Sec.~\ref{sec:optimising}. For the rest of the bounds, rather than show every existing limit individually as in Fig.~\ref{fig:bounds}, we have combined the limits into three categories based on the level of assumption involved. The red region encloses all constraints that are based on photon$\rightarrow$DP transitions (e.g.~light-shining through walls) which in a sense are the least model-dependent. In green, we show bounds invoking stellar cooling arguments, which also rely on photon-DP conversion, but could be circumvented through similar model-dependent arguments as those used for axions~\cite{Jain:2005nh, Masso:2005ym, Jaeckel:2006xm, Masso:2006gc, Budnik:2020nwz, DeRocco:2020xdt, Bloch:2020uzh}. Then in blue we show the bounds that rely on DPs comprising the majority of DM. In grey, we show the direct detection bounds from devoted DP experiments, namely, SHUKET, WISPDMX, and Dark E-field Radio. Note that the future projections vary in their levels of optimism about the future, hence this is not the fairest comparison, however it serves to demonstrate what might be possible.}
 	\label{fig:bounds_with_projections}
\end{figure*}

Now that we have discussed the best strategies for setting limits on DPs, we look towards the future and examine the extent to which these strategies could impact the sensitivities of upcoming experiments. Again we focus on the fixed polarisation scenario as this would require the greatest amount of care in orienting the experiment and would leave the DP sensitivity in any other polarisation scenario unchanged.

To do this, we first select a few of the major DP search proposals listed in Tables~\ref{tab:axionhaloscopes} and~\ref{tab:DPexperiments} that aim to cover the presently unconstrained DP parameter space. We then use their stated projections and apply scaling factors to improve their sensitivity in the fixed polarisation scenario. Since there are external factors that dictate what values of $T$ are possible, we will not attempt to alter anything about the total measurement times. Rather we suggest ways in which individual measurements could be divided, or in some cases, the orientation of the experiment itself could be changed. The strategies for each future experiment are as follows:
\begin{itemize}
    \item {\bf DM-Radio}: Each frequency is integrated over three $\mathcal{O}$(minute) measurements spaced 8 sidereal hours apart (the precise numbers of minutes make only minor differences to the resulting $\langle \cos^2{\theta}\rangle_T$). 
    \item {\bf Dark E-field Radio}: The readout antenna is rotated so that it is Zenith-pointing. The ${\cal O}({\rm month})$ measurement time is rounded to an integer number of sidereal days.
    \item {\bf ALPHA}: The experiment is placed at a latitude as close to $\lat = \pm35^\circ$ as possible. The $\mathcal{O}$(week)-long integration times are rounded to an integer number of sidereal days.
    \item {\bf MADMAX}: A two-polarisation sensitive antenna is used. The experiment is then aligned so that the disks face the North-South direction. The $\mathcal{O}$(week)-long integration times are rounded to an integer number of sidereal days.
    \item {\bf LAMPOST}: if a laboratory is available close to $\pm 35^\circ$ then the experiment should be rotated so its dielectric layers are parallel to the floor. If instead a latitude close to $\pm 55^\circ$ is more feasible then the layers should face the North-South direction. The $\mathcal{O}$(week)-long integration times are rounded to an integer number of sidereal days.
\end{itemize}
While integer-day-long measurements mildly improve the value of $\langle \cos^2{\theta}\rangle^{\rm excl.}_T$, in practice being able to measure for slightly longer may improve the total integrated power per cost of running. We stick to integer-day-long measurements purely for simplicity.   

In Fig.~\ref{fig:bounds_with_projections} we plot the rescaled experimental projections for the discovery of the DP at 5$\sigma$ in 95\% of experiments. We compare our optimised projections against the estimate one would make without incorporating any information about the timing or orientation of the experiment. This pessimistic scenario is, as we have already mentioned, overly conservative, and corresponds to setting our discovery conversion factor to the instantaneous result $\langle \cos^2{\theta}\rangle^{\rm disc.}_{T\rightarrow0} = 0.0036$ or 0.13.

We neglect to include any projections for axion experiments in this figure, whose DP sensitivity will likely rely on a post hoc data analysis (in the style of Fig.~\ref{fig:bounds_closeup}). We instead focus on those experiments that plan on a dedicated DP search. For comparison, we have also shown projections from SuperCDMS~\cite{Bloch:2016sjj} and LZ~\cite{Akerib:2021qbs} however these experiments rely on DP absorption and electronic emission, which is why their limits are insensitive to the DP polarisation scenario. Our point here is that some DP limits depend upon the production mechanism, and some do not---a fact that is commonly brushed over, or ignored.

In all cases we find that these strategies are sufficient to raise the limits in the fixed polarisation scenario up to a level very close to the randomised polarisation scenario. All this requires is some slight adjustments to the data-taking schedule, and a judicious choice in experimental alignment.

Much of this parameter space will therefore be constrained in the coming decades. As well as upcoming DM searches, several strategies used to set the bounds on Figs.~\ref{fig:bounds} and~\ref{fig:bounds_with_projections} will also see improvements over the next few years. For example, upgraded LSW experiments~\cite{Bahre:2013ywa,Spector:2016vwo,Inoue:2015cta,Kim:2020ask}, experiments using atomic transitions~\cite{Flambaum:2019cqi,Bhoonah:2019eyo,Jones:2019qny}, or Aharanov-Bohm experiments~\cite{Boulware:1989up,Arias:2016vxn} may improve the purely-laboratory bounds on DPs. Whereas searches using X-ray~\cite{Thorpe-Morgan:2020rwc} and radio~\cite{Lobanov:2012pt,An:2020jmf} telescopes, fast radio burst timing~\cite{Landim:2020ked}, or asteroseismology~\cite{Ayala:2019isl}, may improve upon existing astrophysical bounds.

\section{Using the daily modulation as a signal}\label{sec:postdetection}
The daily modulation of the DP signal in the fixed polarisation scenario represents one of the clearest smoking gun signals of DM that an experiments of this kind could hope to observe. This is fortunate, given that a DP signal candidate cannot be confirmed or rejected via the tuning of the magnetic field. At least one DP experimental collaboration has already suggested that they already will employ the daily modulation as a final stage in their signal confirmation decision trees~\cite{Godfrey:2021tvs}. It was also shown that daily modulation can be measured with an array of directionally sensitive detectors~\cite{Kopylov:2021ndl}. The most convincing property that the modulation will exhibit for confirming its DM origin will be a period equal to one sidereal day (23 hours and 56 minutes). If there were some way in which a source of terrestrial noise would modulate on the timescale of a day (and it is already hard to contrive of such a background), it would almost certainly have a period equal to one solar day. The way this would manifest in practice is that the phase of the daily modulation in local time would be seen to drift forward slowly over the year. Around 6 months later the phase would be 12 hours ahead of the time it was initially measured, but it would then cycle back around over the next 6 months.

The daily modulation would be a striking signal in the fixed polarisation scenario: it would dominate the signal when measured over timescales longer than a few hours. Yet the only scenario in which the DP signal will \emph{not} sidereally modulate\footnote{This statement only refers to the modulation due to the DP polarisation. In fact all DP signals will exhibit an extremely small ($v\sim 10^{-6}$) daily modulation in the lineshape due to the relative motion of the rotating Earth-bound laboratory and the DM halo. This is likely to be unobservable for the main DM halo which has a typical dispersion of $v\sim10^{-3}$. However it is conceivable that if the experiment encountered a strongly coherent configuration of the field like a Proca star~\cite{Brito:2015pxa}, or ultracold stream~\cite{Sikivie:1997ng,Ling:2004aj,Duffy:2006aa,Hoskins:2016svf}, then the daily modulation could be of comparable size to the signal's width in frequency space.} is when the polarisations are totally randomised in every coherence spacetime patch. One can imagine scenarios where only fraction of the DM is polarised to begin with, or the polarisation was mixed somehow during structure formation and hence varied over phase space, but in any of these scenarios a suppressed daily modulation would still be present. Thus one must be careful using daily modulation as a veto on possible DP signals as one could inadvertently exclude a potential discovery of a non-modulating or slightly modulating scenario. 

The natural question to ask is then: If we see a signal of DPs in one of the experiments we have discussed, can the DP polarisation axis be measured, and can this information be used to understand the production mechanism behind the creation the DP dark matter. Just as in the case of axions, any further scanning becomes unnecessary post-detection and the experiment can quickly accumulate very high $S/N$. This permits the fine-grained study of the spectral lineshape, in the process unravelling the velocity structure of the local DM halo. Note that the velocity distribution is annually modulated by the relative motions of the Earth and the halo, allowing the full \emph{velocity}-structure of $f(\mathbf{v})$ to be probed, as opposed to just the speed distribution, $f(v)$~\cite{OHare:2017yze}. However, one could also employ a multiplexed network of phase-linked detectors that rely on interferometry to further unravel this structure on much shorter timescales~\cite{Foster:2020fln}. These kinds of measurements would allow the experiment to measure properties about the DM halo and, via combination with simulations and galactic surveys, would allow us to probe the merger history of the Milky Way's halo~\cite{OHare:2018trr,Evans:2018bqy,OHare:2019qxc}. 

If the DP field is polarised in some way then the measurement of that polarisation may therefore offer an even more distant window into the early Universe. Without further intuition as to expected DP polarisation distribution in galaxies generated by different production mechanisms, we cannot be more precise than simply raising this as a tantalising possibility. However, even if the DP polarisation is of no fundamental interest at all, its measurement would be a crucial step in optimising the continuous study of the DP lineshape.

So in a similar spirit to Refs.~\cite{OHare:2017yze,Foster:2017hbq,Knirck:2018knd,Foster:2020fln}, we wish to understand the extent to which an experiment could reconstruct the true properties of the DP signal. In the case of the polarisation though, because we are less certain of what the distribution of $\bfhat{X}$ even is, it is not clear what we should assume the correct answer to be. As we have stressed, this subject is ripe for further exploration, but we can sketch a simple example to show how such a study would proceed.

\subsection{Measuring the daily modulation}
We will set up a toy statistical test that, while not accurately reflecting the details of a real experimental analysis, will at least resemble one, and incorporate some of the main features. Firstly we assume that our toy experiment reads in some time-series electromagnetic signal and takes the Fourier transform of it to measure a power spectrum. We then assume that many of these power spectra are stacked together, which eventually total the measurement time $T$. This has the effect of both amplifying any signal present, whilst mitigating against the exponential noise in the randomly drawn DP signal amplitudes in each frequency bin, and making the expected noise spectrum close to Gaussian. 

First let us assume that in this stacked data we have a signal $P_X \langle \cos^2{\theta}\rangle_T$ which is contained in the bandwidth of the experiment $\Delta \nu$ and is distributed according to the distributions shown in Fig.~\ref{fig:costh_axis}. Then we assume that we have some normally distributed noise with mean $P_N$ and standard deviation $\sigma_N = P_N/\sqrt{T\Delta \nu}$.

We adopt an Asimov dataset approach for analysing the daily modulation now. So we shift our definition of $\disc$ slightly since we are no longer interested in optimising an experiment for a guaranteed discovery, but rather we want to know what the typical experiment could do. In this case we wish to find the discovery threshold for when the median experiment seeing a signal could reject the noise at $3\sigma$. This can be straightforwardly calculated using the same technique as in Sec.~\ref{sec:derivinglimits},
\begin{equation}\label{eq:TS_excl}
    P_X \disc > (3+\Phi^{-1}[0.5])\sigma_N \approx 3\sigma_N \, ,
\end{equation}
where $P_X$ is the solution to\footnote{We have removed the superscript `0' to reduce clutter.},
\begin{align}
    \int_{0}^{1} \frac{f(\langle \cos^2{\theta}\rangle_T)}{2} \bigg[1+\text{erf}\bigg( \frac{3\sigma_N-P_X\langle \cos^2{\theta}\rangle_T}{\sqrt{2}\sigma_N} \bigg)\bigg] \textrm{d}\langle \cos^2{\theta}\rangle_T \nonumber \\ 
     = 0.5 \, .
\end{align}
Note that one can also arrive at the same results by writing down a profile likelihood ratio test statistic comparing the signal+noise and noise hypotheses, and this result will follow from the application of Chernhoff's theorem~\cite{Chernoff:1954eli} (see Refs.~\cite{Cowan:2010js,Algeri:2019arh} for further details).

Now we ask the following question: If the DP signal is strong enough that the median experiment can exclude the noise at 3$\sigma$, how much more time will it need to detect the daily modulation? To answer this, we need to write down a likelihood ratio that compares the modulated hypothesis with an unmodulated one,
\begin{align}\label{eq:llratio}
    \Lambda(\theta_X,\phi_X) = 2\bigg[\ln & \mathcal{L}\big(d|\mathcal{M}_t,\{\hat{P}_X,\theta_X,\phi_X,\hat{P}_N\}\big) \nonumber \\ 
    - \ln & \mathcal{L}\big(d|\mathcal{M}_0,\{\hat{\hat{P}}_X,\hat{\hat{P}}_N\}\big)\bigg] \, ,
\end{align}
where we use a single hat to refer to the maximum likelihood estimators (MLEs) under the modulating model, $\mathcal{M}_t$, and double hats to refer to the MLEs under the unmodulating model, $\mathcal{M}_0$. 

Say that we observe some power, $P^{\rm obs}$, from $t = 0$ to $t = T$, and split the data up into temporal bins of width $\Delta t$ centered on times $t_j$. Then the first likelihood in Eq.(\ref{eq:llratio}) is,
\begin{align}
    \ln\, \mathcal{L}(P^{\rm obs}|&\mathcal{M}_t,\{P^0_X,\theta_X,\phi_X,P_N\}) = \nonumber \\ 
    -\frac{1}{2\sigma^2_N} & \sum_{j =1}^{N_t} \bigg[P_j^{\rm obs} - P_X \langle c(\theta_X,\phi_X) \rangle_j - \frac{\Delta t}{T} P_N\bigg]^2,
\end{align}
where $N_t = T/\Delta t$, and we have used the shorthand,
\begin{equation}
    \langle c(\theta_X,\phi_X)\rangle_j \equiv \langle \cos^2{\theta}\rangle_j = \frac{1}{\Delta t} \int_{t_j-\Delta t/2}^{t_j+\Delta t/2} \cos^2{\theta} \, \mathrm{d}t \, .
\end{equation}
Then, the second likelihood in Eq.(\ref{eq:llratio}) is for the unmodulating case where we take the integrated signal over the full $T$, sacrificing any sensitivity to $(\theta_X,\phi_X)$,
\begin{align}\label{eq:unmodulatedL}
    \ln \mathcal{L}(P^{\rm obs}|\mathcal{M}_t,\{P_X,P_N\})  =  \\
    -\frac{1}{2\sigma^2_N} &\left(P^{\rm obs} - P_X - P_N \right)^2 \nonumber \, .
\end{align}

We can use the maximised ratio $\Lambda(\hat{\theta}_X,\hat{\phi}_X)$ as a test statistic (TS) for determining the detectability of daily modulation, but we can also use the unmaximised ratio to determine the accuracy with which the parameters $(\theta_X,\phi_X)$ could be measured (which we will do in the next subsection). Rather than doing a full Monte Carlo simulation of mock experiments, we can make progress with minimal effort by simply asking what the median experiment would be able to do. This can be answered with the Asimov dataset~\cite{Cowan:2010js}, because profile likelihood ratio test statistics obtained under this dataset often provide a very good estimate to the median of their full asymptotic distributions. The Asimov dataset is the case where we set the observation equal to the expectation for some set of ``true'' parameters. In our case when,
\begin{equation}
     P_j^{\rm obs} = P_j^{\rm Asi} \equiv P_X \langle c(\theta_X,\phi_X) \rangle_j + \frac{\Delta t}{T} P_N \, .
\end{equation}
for all $j$.

We now interpret $\Lambda$ as a test statistic for detecting modulations, ${\rm TS}_{\rm mod}$. Under the Asimov data, the first log-likelihood in Eq.(\ref{eq:llratio}) vanishes, leaving us with
\begin{equation}
    {\rm TS}_{\rm mod}(\theta_X,\phi_X) = \frac{1}{\sigma^2_N} \left(\sum_j P_X \langle c(\theta_X,\phi_X) \rangle_j-\hat{\hat{P}}_X\right)^2.
\end{equation}
The value of $P_X$ that maximises the unmodulated signal model likelihood will be $\hat{\hat{P}}_X = P_X \langle c \rangle_T$. Substituting that in and manipulating the subsequent expression we can find,
\begin{align}
    {\rm TS}_{\rm mod}(\theta_X,\phi_X) &= \left(\frac{P_X}{\sigma_N}\right)^2 \left(\sum_j \langle c \rangle_j-  \langle c \rangle_T\right)^2 \nonumber \\
    &= \left(\frac{P_X}{\sigma_N}\right)^2 \left( \frac{1}{\Delta t} \int_0^T c \mathrm{d}t - \langle c \rangle_T \right)^2 \, . \nonumber \\
    &= \left(\frac{T\langle c \rangle_T P_X}{\sigma_N}\right)^2 \left( \frac{1}{\Delta t} - \frac{1}{T} \right)^2 \nonumber \\
    &\approx \left(\frac{T \langle c \rangle_T P_X}{\Delta t \, \sigma_N}\right)^2 \, .
\end{align}
In the first step we converted the sum over $t$-bins to an integral, and in the final step we have assumed $\Delta t \ll T$. Note that we have suppressed the dependence on the DP angles, but they enter in via the dependence on $\langle c \rangle_T$. 

If we assume that the signal in one bin of interval $\Delta t$ was already large enough to detect, then we can substitute in our result from Eq.\eqref{eq:TS_excl} for the median 3$\sigma$ exclusion of the background,
\begin{equation}
    {\rm TS}_{\rm mod} \approx \left( \frac{3 T \langle c\rangle_T}{\Delta t\langle\cos^2{\theta}\rangle_{\Delta t}^{\rm disc.}} \right)^2 \, .
\end{equation}
Taking the typical value of $\langle c \rangle_T \sim 1/3$, we can see that if we needed a time $\Delta t$ to discover the DP, then we would need $(3\langle\cos^2{\theta}\rangle_{\Delta t}^{\rm disc.})^{-1}$ times that much data to exclude a non-modulating signal at the same significance. In the specific cases we discovered had $\langle\cos^2{\theta}\rangle_{\Delta t}^{\rm disc.} = 1/3$, (e.g. for Zenith-pointing experiments at $\lambda_{\rm lat} = 35^\circ$ observing for day-long measurement times) then this factor is equal to one. Meaning a DP signal discovered at 3$\sigma$ is already enough to make the same claim about the modulation. The results for the remaining experiments and for various values of $\Delta t$ can be inferred from Fig.~\ref{fig:improvement}.

\subsection{Measuring the DP polarisation}
\begin{figure}
\centering
% 	%trim option's parameter order: left bottom right top
    \includegraphics[trim = 0mm 0mm 0mm 0mm, clip, width=0.49\textwidth]{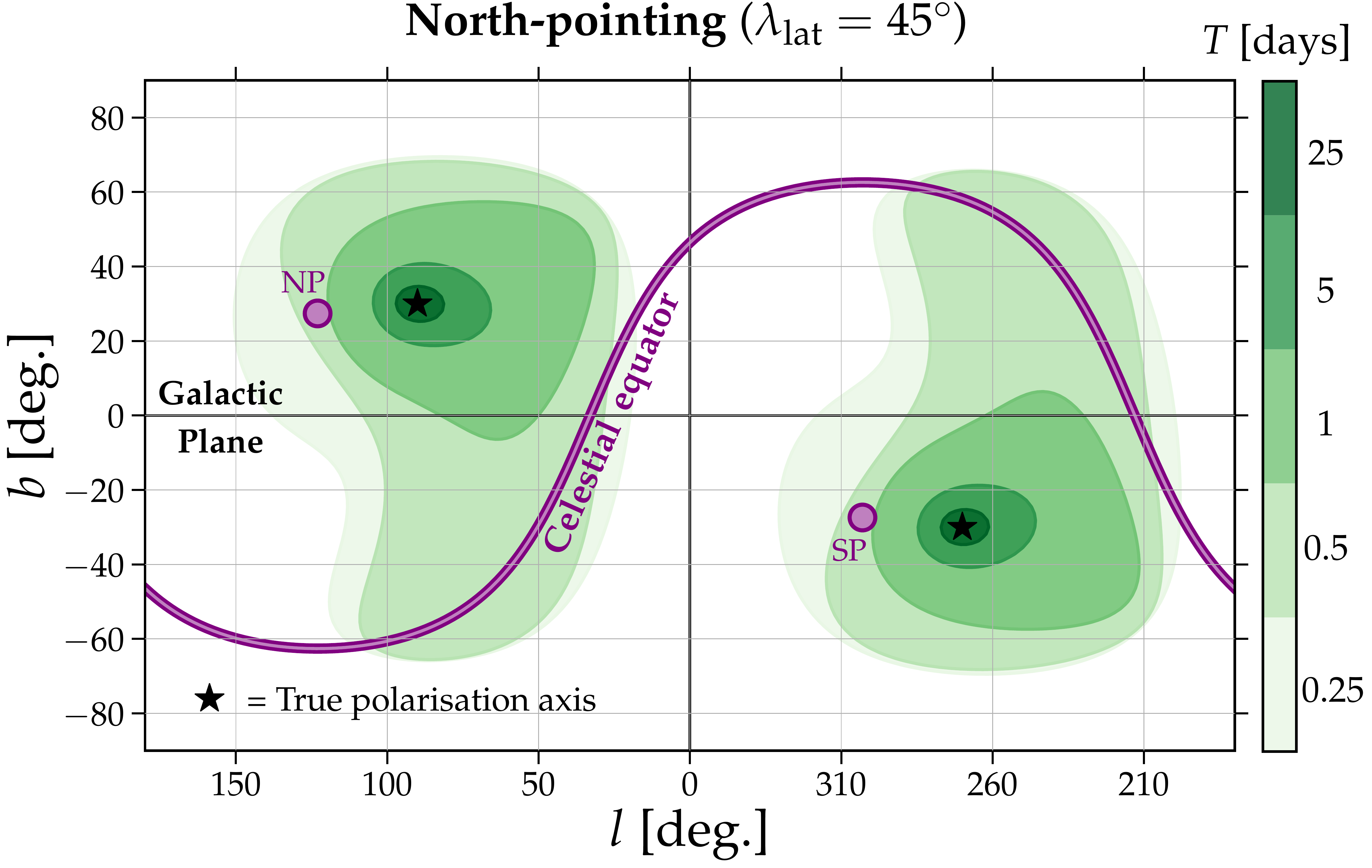}\vspace{1em}
  \includegraphics[trim = 0mm 0mm 0mm 0mm, clip, width=0.49\textwidth]{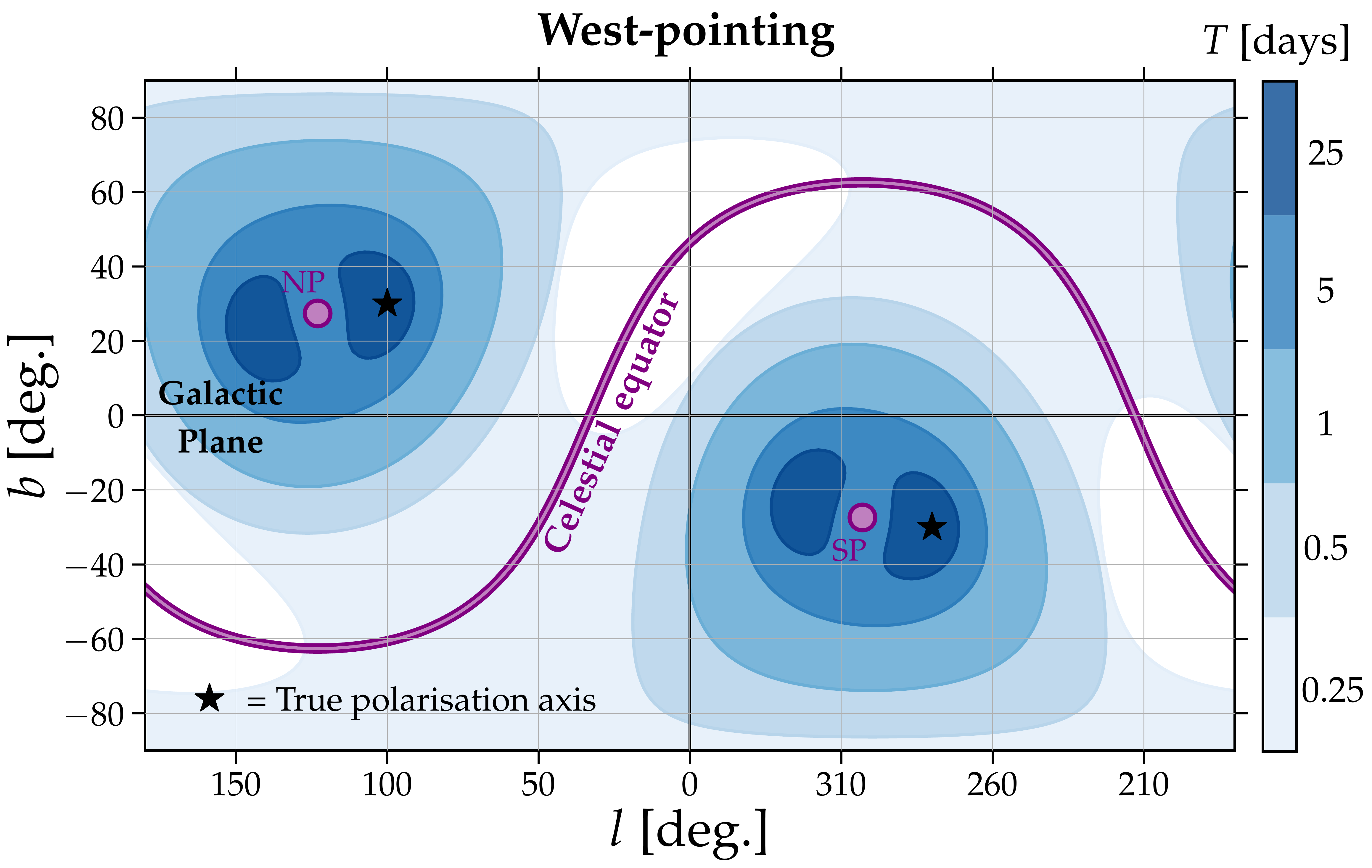}\vspace{2em}
    \includegraphics[trim = 0mm 0mm 0mm 0mm, clip, width=0.49\textwidth]{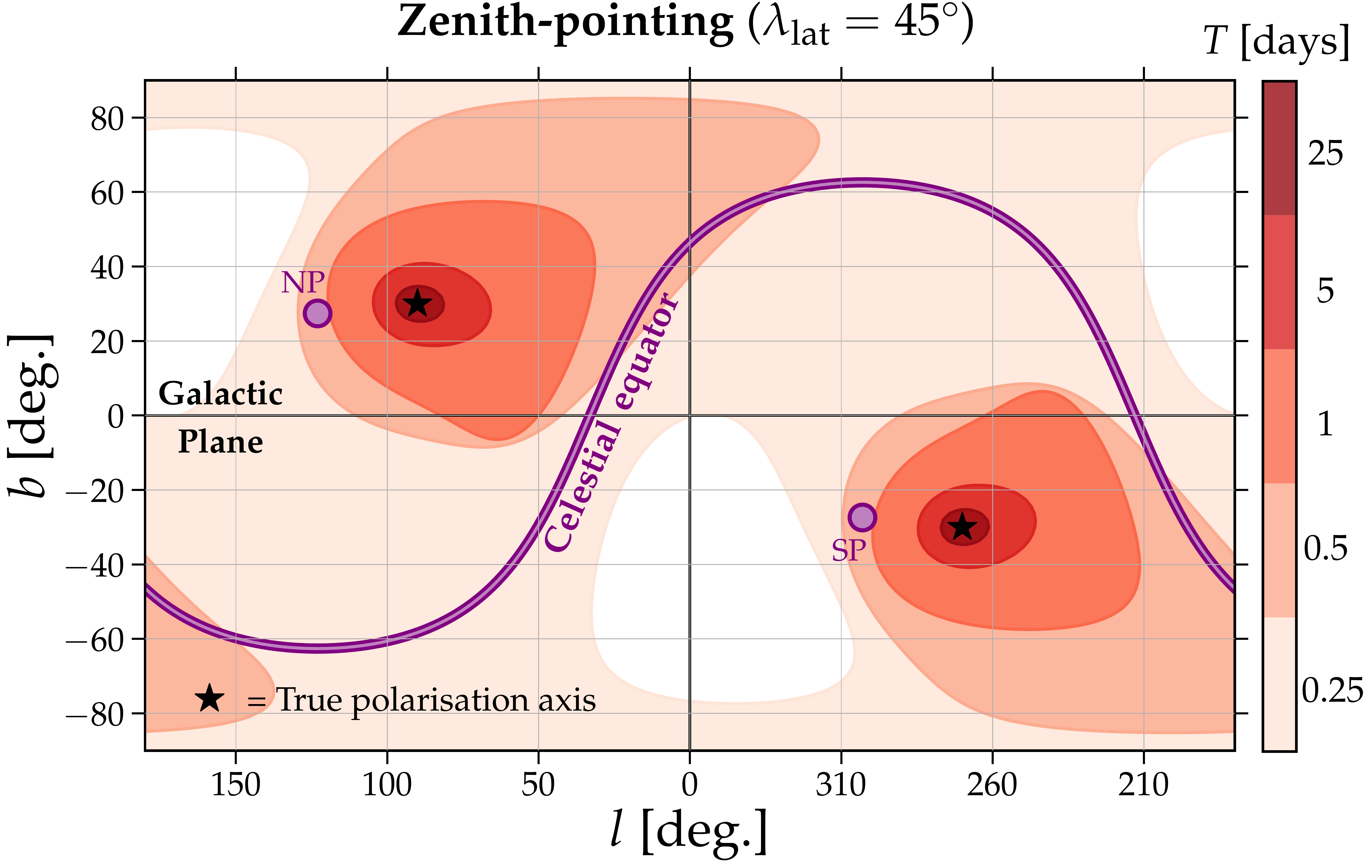}

\caption{Three mock examples of the reconstruction accuracy of the DP polarisation axis. The true polarisation axis is given by the black stars. In each case we assume that the experiment has been able to discriminate the signal from background at the 95\% C.L.~in one day, and the value of $T$ corresponds to the \emph{additional} amount of data used to test for the daily modulation. The size of each contour corresponds to the expected 2$\sigma$ contour around $(\theta_X,\phi_X)$ that the experiment could set given the measurement time $T$. We display these angles in galactic longitude and latitude $(l,b)$ with the galactic plane running horizontally through $b=0$, and the galactic centre at $(0,0)$. The celestial equator is shown with a purple line, and the projected directions of the North and South poles are shown with purple circles.}
 	\label{fig:polarisationmap}
\end{figure}

We can also use the same likelihood ratio to estimate how well some true values of the angles $(\theta^{\rm true}_X,\phi^{\rm true}_X)$ could be measured. We do this by taking the likelihood ratio, Eq.(\ref{eq:llratio}), under the Asimov dataset for the true angles, and then examine the dependence on $(\theta_X,\phi_X)$. The formula for this is very similar to the modulation test statistic we just derived, but now the first likelihood does not vanish since we are allowing $(\theta_X,\phi_X)$ to vary,
\begin{align}
    \Lambda(\theta_X,\phi_X) =& \left(\frac{3}{\langle\cos^2{\theta}\rangle_{\Delta t}^{\rm disc.}} \right)^2 \bigg[ \bigg(T\langle c_{\rm true}\rangle_T\bigg(\frac{1}{\Delta t} - \frac{1}{T}\bigg)\bigg)^2 \nonumber \\
    &- \frac{1}{\Delta t}\int_0^T \bigg(c_{\rm true} - c\bigg)^2 \mathrm{d}t \bigg] \, ,
\end{align}
where we write $c_{\rm true} = \cos^2{\theta}(\theta^{\rm true}_X,\phi^{\rm true}_X)$. This likelihood ratio will be asymptotically distributed according to a $\chi^2_2$ distribution since the models differ by two parameters~\cite{Wilks:1938dza}. Therefore we can draw contours for $\Lambda(\theta_X,\phi_X)-\Lambda(\theta^{\rm true}_X,\phi^{\rm true}_X) = -6.17$ to show the typical size of a 2$\sigma$ measurement of the DP polarisation axis. This is what is shown in Fig.~\ref{fig:polarisationmap}.

The darker shading in the three panels of Fig.~\ref{fig:polarisationmap} correspond to increasing the value of $T$. We assume that $\Delta t = 1$~day, i.e. the experiment required 1 day of observation to detect the DP signal. Then $T$ here corresponds to how many \emph{additional} days of observation are being assumed.

To make this result more evocative of a real scenario we have switched from the equatorial coordinates we defined $\bfhat{X}$ in, to galactic coordinates. We define the galactic longitude and latitude, $(l,b)$, using the coordinate rotation,
\begin{equation}
\left(\begin{array}{c}
\sin{\theta_X}\cos{\phi_X} \\
\sin{\theta_X}\sin{\phi_X} \\
\cos{\theta_X}
\end{array}\right) = R_{\rm gal}\begin{pmatrix}\cos l \cos b \\ \sin l \cos b \\ \sin b \end{pmatrix} \, ,
\end{equation}
where,
\begin{equation}
R_{\rm gal} =
\begin{pmatrix}
-0.05487556 & +0.49410943 & -0.86766615 \\
-0.87343709 & -0.44482963 & -0.19807637 \\
-0.48383502 & +0.74698225 & +0.45598378
\end{pmatrix}  \, , \nonumber
\end{equation}
with values assuming the International Celestial Reference System convention for the right ascension and declination of the North Galactic Pole, $(\alpha_{\rm GP},\delta_{\rm GP}) = (192^\circ.85948,\,+27^\circ.12825)$ as well as the longitude of the North Celestial Pole $l_{\rm CP} = 122^\circ.932$~\cite{BinneyGalacticAstronomy}.

%We can see from Fig.~\ref{fig:polarisationmap} that it appears as though the North and Zenith-pointing experiments take longer to measure the polarisation axis than the West-pointing experiment. However we have specifically chosen the initial time $\Delta t$ that the experiment needed to measure the signal to be equal to 1 day. Since the West-pointing experiments have a much lower value of $\langle \cos{\theta}\rangle^{\rm disc.}_{1\textrm{day}}$ compared to the North and Zenith cases (see the left hand panel of Fig.~\ref{fig:LocationDependence}), then it must have a stronger signal (i.e. larger $P_X$) to overcome it. As a consequence, this experiment can measure the DP polarisation axis much more quickly. However there is a downside: despite the fact that the West-pointing modulations are typically larger, they only allow the polarisation to be sampled along a single plane. So over longer times the North and Zenith pointing experiments, which sample the polarisation in three dimensions, can eventually refine their measurement of $(\theta_X,\phi_X)$ to a single axis. This can be seen by the fact that the contours on the top and bottom panels of Fig.~\ref{fig:polarisationmap} are centred around the black stars whereas the middle panel's contours are centered around four locations. The latter showing that there are two degenerate axes that the experiment is unable to tell apart. These two axes are separated by $180^\circ$ of rotation around the Earth's spin axis.

We can see from Fig.~\ref{fig:polarisationmap} that the North and Zenith-pointing experiments require a smaller amount of \emph{additional} data than the West-pointing experiments to obtain similar sized contours around the true polarisation axis. Unlike West-pointing experiments, the North and Zenith pointing experiments sample the polarisation in three dimensions, so they can eventually refine their measurement of $(\theta_X,\phi_X)$ to a single axis. This can be seen by the fact that the contours on the top and bottom panels of Fig.~\ref{fig:polarisationmap} are centred around the black stars whereas the middle panel's contours are centered around four locations. In the West-pointing case we find that there are two degenerate axes that the experiment is unable to tell apart. These two axes are separated by $180^\circ$ of rotation around the Earth's spin axis. This phenomenon would persist even with indefinite data-taking---a rotation of the experiment's antennae would be needed to lift the degeneracy.

Therefore, we conclude that the polarisation axis of the DP should be measurable very soon after the signal is detected. This is good because an experiment designed to \emph{study} the DP could then be rotated by a suitable amount to maximise the signal.

\section{Summary}\label{sec:summary}
We have highlighted some of the difficulties involved in setting limits on dark photons (DPs) as a dark matter (DM) candidate, focusing on the fact that the DP polarisation distribution around the Earth is unknown. The polarisation state of the DPDM is initiated by its production mechanism. However, as of now, the precise polarisation distributions generated by the several proposed production mechanisms, and the subsequent effects of structure formation, have not been rigorously studied. If it can be shown that structure formation completely randomises the polarisation then direct detection limits on DPs can be strengthened, sometimes significantly. However, a serious possibility that appears to be true for several mechanisms is for the DP to have a single polarisation over scales much larger than those probed by terrestrial experiments. This latter scenario presents the greatest challenge for detection and should be used as a conservative baseline. Most experiments are sensitive to either one or two (but not all) DP polarisations at a time. Thus a limit must take into account the probability of the experiment being misaligned with the DP polarisation. This probability varies by over two orders of magnitude depending on the measurement time, location, and alignment of the experiment.

We began our discussion by explaining how to reinterpret limits from axion haloscopes in the context of DPs. The first issue to reiterate is that the common approach of vetoing candidate axion signals by testing for their disappearance when the magnetic field is switched off means that a DP could have been discovered but would have been discarded as noise. This makes the RBF and UF axion haloscope bounds unsuitable for reinterpretation. For other haloscopes, a lack of information about the exact magnetic field employed for a given measurement also forbids a concrete reinterpretation in terms of DPs. 

The second issue is related again to the DP polarisation. Accurately accounting for the variation in the signal due to the rotation of the Earth requires precise timing and orientation information. In both axion and dedicated DP experiments, this information is usually not made available. Such omissions make it impossible to set robust DP bounds. Instead, we have demonstrated what would need to be done, by performing a conservative calculation based on the information that is available. The resulting bounds are shown in Fig.~\ref{fig:bounds_closeup}. As an example of what could be achieved in the future if the experimental data-taking were more strategic---and we have laid out the steps for doing so in this work---we have displayed some projections in Fig.~\ref{fig:bounds_with_projections}. 

To provide clarity for future experiments, we have outlined a recipe for calculating the DP polarisation geometrical factor $\cos^2{\theta}$ for any arbitrary measurement time, polarisation angle, and experiment orientation. This is detailed in Sec.~\ref{sec:DPscanning}. We then outlined the best practices to both maximise the sensitivity, and to avoid accidentally discarding DP signals. In general, by measuring for times close to an integer number of sidereal days, with well-aligned experiments, one can obtain close to (or even equal) to the best possible scenario. This represents a gain of over an order of magnitude in coupling for experiments polarised in a single direction, and a factor of three for experiments sensitive to a plane of polarisation. This result can be seen in Fig.~\ref{fig:improvement}. Even for experiments where measurement times are short, improvements in sensitivity of an order of magnitude are possible simply by splitting measurements into shorter ones spaced several hours apart---as is shown in Fig.~\ref{fig:costh_rescanned}. We have also found that the location of the experiment also impacts the sensitivity---as shown by Fig.~\ref{fig:LocationDependence}---with $\pm35^\circ$ and $\pm 55^\circ$ representing the optimal latitudes in most cases.

With the recent increase in the number of axion and DP-sensitive haloscopes, it is more important than ever to be sure that we are extracting robust and instructive limits. Rather than relying on others reinterpreting data with limited information, we strongly encourage experimental collaborations to perform dedicated DP analyses taking care of the role of magnetic field vetos, as well as orientation and timing information. Further, without any reduction to the experiment's axion sensitivity or incurring any additional run time, simple changes can result in substantial improvements in DP sensitivity. If all experiments adopted such strategies it would maximise our chances of discovering DPDM or, at the very least, would allow us to rule out vast swathes of unexplored parameter space.

\section*{Acknowledgements}
The authors thank Pierre Brun, Woohyun Chung, Akash Dixit, Stefan Knirck, Laura Manenti, Le Hoang Nguyen, Jonathan Ouellet, Tony Tyson, Darko Veberi{\v c} and SungWoo Youn for providing information about their experiments, and Gonzalo Alonso-Àlvarez, Raymond Co, Keisuke Harigaya, and Andrew Long for useful discussions about production mechanisms. 
CAJO thanks Masha Baryakhtar and Sam Witte for helpful information that improved Fig.~\ref{fig:bounds}. We also thank Amit Bhoonah, Priscilla Cushman, Divya Sachdeva, and Chang Sub Shin, for useful comments. AC is supported by the Foreign Postdoctoral Fellowship Program of the Israel Academy of Sciences and Humanities. CAJO is supported by The University of Sydney. AC also acknowledges support from the Israel Science Foundation (Grant 1302/19), the US-Israeli BSF (Grant 2018236) and the German-Israeli GIF (Grant I-2524-303.7). AM is supported by the European Research
Council under Grant No. 742104 and by the Swedish Research Council (VR) under Dnr
2019-02337 “Detecting Axion Dark Matter In The Sky And In The Lab (AxionDM)”.
The work of EV was supported in part by the U.S. Department of Energy (DOE) Grant No.~DE-SC0009937. AM dedicates this paper to Rosie Millar, born this day. May she enjoy all the wonders of the Universe.

\providecommand{\href}[2]{#2}\begingroup\raggedright\endgroup

\end{document}